\documentclass[prd,aps,superscriptaddress,nofootinbib,preprintnumbers,amsmath,amssymb,10pt,twocolumn,dvipsnames,floatfix]{revtex4-2}

\makeatletter 
    
\renewcommand\onecolumngrid{% <<<<<<
\do@columngrid{one}{\@ne}%
\def\set@footnotewidth{\onecolumngrid}% <<<<<<<<<<<<<<<<
\def\footnoterule{\kern-6pt\hrule width 1.5in\kern6pt}%
}

\renewcommand\twocolumngrid{% <<<<<<
        \def\footnoterule{% restore rule
        \dimen@\skip\footins\divide\dimen@\thr@@
        \kern-\dimen@\hrule width.5in\kern\dimen@}
        \do@columngrid{mlt}{\tw@}
}%

\makeatother    
\usepackage{physics}
\usepackage{braket}
\usepackage{tikz}
\usepackage{tensor}
\usepackage[caption=false]{subfig}
\usepackage{faktor}
\usepackage{mathrsfs}
\usepackage{orcidlink} 
\usepackage{ulem}
\usepackage{mathtools}

\usepackage[capitalize]{cleveref}    % Smart referencing
  \crefname{algorithm}{Alg.}{Algs.}
  \crefname{tab}{Table}{Tables}
  \crefname{fig}{Fig.}{Figs.}
  \crefname{section}{Sec.}{Secs.}
  \crefname{appendix}{App.}{Apps.}

\newcommand{\SU}{\mathrm{SU}(2)}
\newcommand{\su}{\mathfrak{su}(2)}
\newcommand{\SB}{\text{SB}}
\newcommand{\fields}{\mathcal{E}(\mathrm{G})}

\newcommand{\hilbert}{\mathcal{H}(\mathrm{G})}

\newcommand{\G}{\mathrm{G}}
\newcommand{\fieldsR}{\mathcal{E}(\mathrm{G}_R)}

\DeclareMathOperator{\Span}{span}

\DeclareMathOperator{\Ad}{Ad}

\DeclareMathOperator{\esup}{even sup}

\begin{document}

%%%%%%%%%%%%%%%%%%%%%%%%%%%%%%%%%%%%%%%%%%%%%%%%%%%%%%%%%%%%
%%%%%%%%%%%%%%%%%%%%%%%%%%%%%%%%%%%%%%%%%%%%%%%%%%%%%%%%%%%%

\title{Gauge Loop-String-Hadron Formulation on General Graphs and Applications to Fully Gauge Fixed Hamiltonian Lattice Gauge Theory}
\author{Ivan M. Burbano\,\orcidlink{0000-0002-3792-1773}}
\affiliation{Theoretical Physics Group, Lawrence Berkeley National Laboratory, Berkeley, CA 94720, USA}
\affiliation{Department of Physics, University of California, Berkeley, CA 94720, USA}
\author{Christian W. Bauer\,\orcidlink{0000-0001-9820-5810}}
\affiliation{Theoretical Physics Group, Lawrence Berkeley National Laboratory, Berkeley, CA 94720, USA}
\affiliation{Department of Physics, University of California, Berkeley, CA 94720, USA}

\begin{abstract}
We develop a gauge invariant, Loop-String-Hadron (LSH) based representation of $\SU$ Yang-Mills theory defined on a general graph consisting of vertices and half-links. Inspired by weak coupling studies, we apply this technique to maximal tree gauge fixing. This allows us to develop a fully gauge fixed representation of the theory in terms of LSH quantum numbers. We explicitly show how the quantum numbers in this formulation directly relate to the variables in the magnetic description. In doing so, we will also explain in detail the way that the Kogut-Susskind formulation, prepotentials, and point splitting, work for general graphs. In the appendix of this work we provide a self-contained exposition of the mathematical details of Hamiltonian pure gauge theories defined on general graphs.

\end{abstract}

\maketitle

\setcounter{tocdepth}{1}
\tableofcontents

\section{Introduction}

While traditional lattice simulations~\cite{Wilson:1974sk} have been tremendously successful at calculating a large variety of important observables in non-perturbative field theories such as QCD (for a review of the impact of lattice field theory, see~\cite{Gattringer:2010zz}), they rely on Monte-Carlo integration of the path integral, and therefore inherently limited to observables that do not exhibit a sign problem~\cite{Troyer:2004ge}. 
This puts some of the most interesting observables, most notably most dynamical observables, out of reach of classical lattice calculations. 
Hamiltonian lattice gauge theory (HLGT)~\cite{Kogut1975} is a formulation of gauge theories on spatial lattices, where time is kept continuous.
Time-dependent observables can be calculated directly by solving the Schr\"odinger equation, and since it does not rely on Monte-Carlo integration, systems with complex parameters in the action are accessible as well. 
The Hilbert space of an HLGT grows exponentially with the number of lattice points, and most problems are inaccessible to classical computers. 
On the other hand, it has been shown that HLGTs can be simulated on quantum computers with resources that scale only polynomially with the number of lattice points~\cite{Jordan:2012xnu}, which has resulted in many studies of HLGT formulations that are amenable to simulation on quantum computers. The first quantum field theory that was simulated on a quantum computer~\cite{Martinez:2016yna,Muschik:2016tws,Klco:2018kyo} was the Schwinger model, a $\mathrm{U}(1)$ field theory in (1+1) dimensions~\cite{Schwinger:1962tn,Schwinger:1962tp}. 
Since then, there has been much progress in the formulation of Abelian~\cite{Banerjee:2012pg,Zohar:2013zla,Kuhn:2014rha,Kasper:2015cca,Zohar:2015hwa,Yang:2016hjn,Kokail:2018eiw,Lu:2018pjk,Kaplan:2018vnj,Mil:2019pbt,Davoudi:2019bhy,Surace:2019dtp,Haase:2020kaj,Luo:2019vmi,Shaw:2020udc,Yang:2020yer,Ott:2020ycj,Paulson:2020zjd,Nguyen:2021hyk,Zhou:2021kdl,Riechert:2021ink,Bauer:2021gek,Kane:2022ejm,Grabowska:2022uos,fontana_reformulation_2022,zhang2023observation,Farrell:2023fgd,Nagano:2023uaq,Falcao:2024msg} and non-Abelian~\cite{Zohar:2012xf,Stannigel:2013zka,mathur_canonical_2015,Mezzacapo:2015bra,Raychowdhury:2018osk,Raychowdhury:2019iki,Klco:2019evd,Davoudi:2020yln,Atas:2021ext,ARahman:2021ktn,halimeh2022gauge,zache2023quantum,DAndrea:2023qnr,Turro:2024pxu,Anishetty:2009nh,Alexandru:2019nsa,Ciavarella:2021nmj,Farrell:2022wyt,Farrell:2022vyh,Atas:2022dqm,Ciavarella:2021lel,Ciavarella:2023mfc,hayata2023qdeformedformulationhamiltonian,Farrell:2024fit,Ciavarella:2024fzw,Cepollaro:2024qln,fontana_efficient_2024} gauge theories.
A broad overview of quantum computing for High-Energy Physics can be found in~\cite{Bauer:2022hpo}.

The lattice Hamiltonian of Yang-Mills theories contains two terms~\cite{Kogut1975}
\begin{equation}\label{eq:hamiltonian}
    H = \frac{g^2}{2} H_E + \frac{1}{2g^2} H_B
    \,,
\end{equation}
where the magnetic Hamiltonian $H_B$ contains the plaquette terms of the original action, restricted to spatial plaquettes, while the electric Hamiltonian $H_E$ contains the contributions from the original plaquettes that extended in the time direction. 
The parameter $g$ denotes the bare coupling constant.
The Hamiltonian can be constructed using the transfer matrix formalism from the Euclidean action of the lattice gauge theory by taking the lattice spacing of time to zero~\cite{Wilson:1974sk,Creutz:1976ch}. 
The magnetic Hamiltonian is written in terms of the Wilson line operator $U$, while the electric Hamiltonian is given in terms of the the conjugate electric operators $E$.
The transfer matrix formalism uses temporal gauge $A_0 = 0$, and the remaining spatial gauge transformations give rise to Gauss' law, which is generated by the so-called electric operators.

For any numerical simulation of the HLGT a basis of the Hilbert space needs to be chosen.
The operators $U$ and $E$ are conjugate to one another.
Consequently, the electric and magnetic Hamiltonians do not commute and can therefore not be diagonalized simultaneously. 
Two general classes of bases for the Hilbert space are generally considered. 
Basis states that diagonalize the electric Hamiltonian are called electric bases~\cite{Mathur:2004kr, mathur2006loop, mathur2007loop, Raychowdhury:2018osk, Raychowdhury:2019iki, Kadam:2022ipf, Kadam:2024zkj, Ciavarella:2021nmj, Klco:2019evd, Davoudi:2020yln, Byrnes:2005qx, Anishetty:2009nh, Banuls:2017ena}, while those that diagonalize the magnetic Hamiltonian are called magnetic bases~\cite{Alam:2021uuq, Alexandru:2021jpm, Gustafson:2022xdt, Gustafson:2023kvd, Gustafson:2024kym, Assi:2024pdn, Lamm:2024jnl, Muarari:2024dqx}. 
Most formulations of HLGTs are given in terms of electric bases, since 
for them Gauss' law is given in terms of local constraints. 

Since the electric Hamiltonian dominates for strong bare coupling, electric bases are best suited for simulations in this regime. 
At weak bare couplings, the magnetic Hamiltonian dominates and magnetic bases should be more suited.
Stated more precisely, at strong bare coupling the lowest lying energy states of the system are dominated by the lowest lying states of the electric Hamiltonian. 
This allows for a truncation of the nominally infinite dimensional Hilbert space of the system by keeping only a small number of electric basis states with smallest electric energy. 
At weak bare coupling, the lowest lying states are dominated by those of the magnetic Hamiltonian, while requiring higher energy states of $H_E$.
In this region of parameter space, truncation is therefore much more efficiently performed in a magnetic basis.
Given that asymptotic freedom requires simulations at small lattice spacing to be performed at small bare coupling values, it is important to understand the formulation of HLGTs in this limit.

Several formulations suitable to simulations at weak coupling have been proposed~\cite{Drell:1978hr,mathur_canonical_2015,Haase:2020kaj,Bauer:2021gek,Ji:2022qvr,Gustafson:2022xdt,Gustafson:2023kvd}.
One of the issues with magnetic bases, which are advantageous at weak coupling is ensuring gauge invariance after truncating the gauge fields. 
The approach taken in~\cite{Ji:2022qvr,Gustafson:2022xdt,Gustafson:2023kvd} is to perform the truncation using discrete subgroups of the full gauge group. 
An alternative to deal with the gauge invariance is to use gauge fixing, developed for Abelian U(1) in~\cite{Drell:1978hr,Haase:2020kaj} and non-Abelian gauge groups in~\cite{mathur_canonical_2015}. A gauge fixed formulation allowing for an efficient truncation of the resulting Hilbert space at weak coupling was developed for U(1) in~\cite{Bauer:2021gek}. 
In~\cite{DAndrea:2023qnr} a formulation for $\SU$ lattice gauge theory allowing for an efficient truncation at weak coupling was developed in~\cite{DAndrea:2023qnr}.
This work resulted in a mixed electric--magnetic basis, in which each $\SU$ rotation was parameterized by a rotation axis $(\theta, \phi)$ and rotation angle $\omega$. 
The wave function at weak coupling becomes localized at small values $\omega$, but can be represented by a small number of Fourier modes in defined in a region of small $|\omega|$, following a method originally introduced in~\cite{Bauer:2021gek} and implemented in~\cite{Grabowska:2022uos,Kane:2022ejm} for $\mathrm{U}(1)$ gauge theories. 
The dependence on $\theta$ and $\phi$ has no strong dependence on the value of $g$, and the wave functions can be well represented in terms of a small number of spherical harmonics~\cite{DAndrea:2023qnr}. 

In this work we apply a Schwinger boson representation~\cite{Schwinger:1952dse,Mathur:2004kr,Mathur:2005fb,Mathur:2007nu,Anishetty:2009nh,Mathur:2010wc,Anishetty:2014tta,mathur_canonical_2015} to the above construction and use the Loop-String-Hadron (LSH) formulation~\cite{Raychowdhury:2018osk,Raychowdhury:2019iki,Kadam:2022ipf} to obtain gauge invariant combinations of Schwinger boson operators. 
Even though the words ``String'' and ``Hadron'' refer to fermionic matter fields, which we do not include in this work, we will still refer to our formulation as LSH given the origin of the ideas.
While LSH has so far been utilized solely for electric bases, we will show that it can also be used to obtain a better understanding of magnetic bases defined in maximal tree gauge fixed theories. 
There are two main contributions the work in this paper will add to our theoretical understanding of HLGTs.
First, we introduce concepts from graph theory, and show how the formulation of HLGTs on maximal trees can be reformulated as a gauge theory formulated on general graphs. 
In particular we show how to use the technique of graph coarsening to morph one graph into one another, and how to relate the theories defined on the different graphs.
Both point splitting and maximal tree gauge fixing can be viewed as particular examples of graph coarsening.
In the main body we focus on these particular examples, while the appendix exhibits the general formalism and provides rigorous proofs.
Second, we develop a deeper understanding of the prepotential formulation of gauge theories, which will allow us to construct gauge invariant operators and states formulated on the general graphs that arose in the first part of this work.
This will give rise to a LSH formulation of the fully gauge fixed Hamiltonian of~\cite{mathur_canonical_2015, DAndrea:2023qnr}, with all operators required to construct both the states and the Hamiltonian as gauge invariant operators constructed out of creation and annihilation operators of Schwinger bosons. 
This therefore uses the same philosophy as the Loop-String-Hadron formulation of HLGTs, even though we work in a pure gauge theory, which does not include string or hadron operators.

This work is organized as follows: In~\cref{sec:Kinematics} we review the possible states in the Hilbert space of a HLGT on a general graph, and discuss how gauge invariance acts on these states. 
In~\cref{sec:point_splitting} we develop the technique of changing graphs into one another using maximal tree graph coarsening and point splitting. 
This is done in three steps. 

We first discuss in~\cref{sec:Lattices_To_Flowers} how after fixing almost all gauge redundancy, the theory is formulated in terms of Wilson lines forming closed loops. 
Since all closed loops originate from a single point, they can  be drawn as a flower, with each petal representing one of the closed loops.
This is the maximal tree coarsening.
In~\cref{sec:Petals_To_Leaves} we then consider a single such closed loop and show how one can use point splitting to localize the $\omega$ and $(\theta,\phi)$ on different locations of the graph.
The resulting graph manipulations transform the petals of the flower into leaves.
In~\cref{sec:Flowers_To_Branches} we show how to transform the graph of a flower, which has vertices of high valency, into a branch, which only has vertices of valency three. This is done through point splitting as well and will be an important step to be able to describe the theory using LSH operators.

In~\cref{sec:lsh_construction} we develop a deeper understanding of prepotential formulation, showing how one constructs gauge invariant loop operators on general graphs, making this formalism applicable to the scenarios developed before.
In~\cref{sec:LSH_on_branches} we use the techniques developed to construct the states and Hamiltonian of a fully gauge fixed $\SU$ gauge theory, and then give a few very simple explicit examples in~\cref{sec:examples}.
Our conclusions are presented in~\cref{sec:Conclusions}.
In the appendices, we provide many more details about our results, formulated in a rigorous mathematical language, as well as a general description of graph coarsening.

\section{Graph Theory for Hamiltonian Lattice Gauge Theory}
\label{sec:Kinematics}

\begin{figure}
    \centering
    \subfloat[]{\label{fig:single_link}\includegraphics[width = 0.35\textwidth]{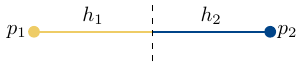}}
    \par\bigskip
    \subfloat[]{\label{fig:single_loop}\includegraphics[width = 0.2\textwidth]{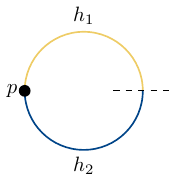}}
    \par\bigskip
    \subfloat[]{\label{fig:high_valency_vertex}\includegraphics[width = 0.2\textwidth]{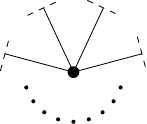}}
    \par\bigskip
    \subfloat[]{\label{fig:single_corner}\includegraphics[width = 0.3\textwidth]{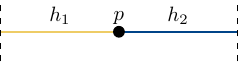}}
\caption{Possible combinations of vertices and half-links. Figure (a) on the top shows a typical ingredient of a hyper-cubic lattice, with two vertices connected by one link made out of two half-links. Below this in (b) there is a loop formed from a single vertex and two half-links. Underneath in (c), we show a single vertex, connected to many half-links. Finally, in (d) we show a corner, consisting of two half-links anchored at the same vertex.
Note that the loop in (b) is both a link and a corner. \label{fig:oneLinkDef}}
\end{figure}

The basic building blocks of a lattice are vertices and half-links, where one end of each half-link is anchored to a vertex, while the other end is attached to another half-link. 
This implies that pairs of half-links combine to form links, which connect vertices to one another. 
Each link is comprised of two half-links. 
Since in mathematical graph theory links are often referred to as edges, we use the symbol $e$ to denote a link, and $h$ to denote half-links. 
This is illustrated in~\cref{fig:single_link}. 

Spatial graphs have very particular arrangements of half-links and vertices with the structure of a lattice. 
For a spatial lattice in $d$ dimensions each vertex not at the boundary of the lattice has $2d$ half-links connected to it. 
It also contains plaquettes, which are combinations of four vertices and four links forming a closed loop in the lattice.
However, other graphs can be formed out of half-links as well. 
One example is a closed loop as shown in~\cref{fig:single_loop}, consisting of two half-links attached to each other and a common vertex.
Furthermore, each vertex can have in principle an arbitrary number of half-links anchored to it, as shown in~\cref{fig:high_valency_vertex}, and we will call the number of anchored links the valency of a vertex. 
A final structure we will highlight here is a {\emph{corner}}, constructed out of a single vertex and two anchored half-links. 
This is shown in~\cref{fig:single_corner}.
Such closed loops, high valency vertices and corners will be an important ingredient in this work.

An oriented link defines one of the two half-links as the start of the link and the other as the end of the link, and defines the corresponding vertices as the starting and target vertex $s$ and $t$, respectively. 
Multiple links can join to form a trajectory $\gamma$, where the target vertex of one link has to be the same as the starting vertex of the next link. 
A trajectory has again a starting vertex $s$ and target vertex $t$. 
Precise definitions for all of the graph-theoretic notions we will use in this paper can be found in~\cref{app:graph_theory}.

\begin{figure}
    \centering
    \includegraphics[width=0.3\textwidth]{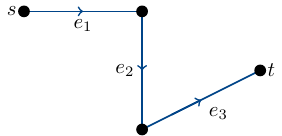}
    \caption{Single trajectory $e_3e_2e_1$. If the links $e_i$ have Wilson lines $u_i$, we assign to this trajectory the Wilson line $u_3u_2u_1$.}
    \label{fig:single_trajectory}
\end{figure}

Classically, a field configuration of Yang-Mills theory is determined by assigning to each oriented link $e$ a Wilson line.
Let us now specify informally what these Wilson lines are.
For a more precise discussion, we refer the reader to~\cref{app:classical_theory}.
If the theory has a structure group $G$, such a Wilson line is an element $u\in G$ that transports the color information from the starting vertex $s$ of $e$ to its target vertex $t$. 
For consistency we must then assign to $e^{-1}$, the link obtained by reversing the orientation of $e$, the Wilson line $u^{-1}$. 
Furthermore, once we have a prescription to transport along links, it is clear how to transport over trajectories using the group law of $G$. 
For example, given two links $e_1$ and $e_2$, respectively supporting Wilson lines $u_1$ and $u_2$, and such that $e_2$ starts at the endpoint of $e_1$, we can assign to the trajectory $e_2e_1$ that first goes through $e_1$ and then to $e_2$ a Wilson line given by $u = u_2u_1$. 
This Wilson line then transports color information from the start of $e_1$ to the endpoint of $e_2$ along the trajectory defined by these two links\footnote{In this paper we will adopt a somewhat unusual convention in which the composition described above is described by $u_2u_1$ instead of $u_1u_2$. 
This is under the interpretation that the color degrees of freedom a Wilson line will transform sit on its right.
They thus follow the standard convention of operators acting on the variables to their right.}. 
This generalizes to trajectories formed out of multiple links, and each trajectory from starting point $s$ to endpoint $t$ can be represented by an ordered set of links along this trajectory, and a Wilson line along this trajectory is given by the product of the Wilson lines on each of the links. 
This Wilson line will transport the color information from $s$ to $t$. 
An example of a trajectory with three links is shown in~\cref{fig:single_trajectory}.

The Wilson lines however are themselves not observable.
They are subject to a gauge redundancy identifying equivalent configurations of Wilson lines. 
Such a redundancy is parameterized by the gauge group, which is obtained by placing the structure group at every vertex of the graph. 
Under a transformation which has $g_t\in G$ at $t$ and $g_s\in G$ at $s$, the Wilson line transforms as\footnote{Observe that the placement of $g_t$ on the left and $g_s$ on the right on this formula is completely determined by the Wilson line composition convention above. 
The same goes for asking that one of them is inverted while the other one is not. 
The fact that $g_s$ is inverted comes from the convention that $g_s$ acts by left multiplication on the color degrees of freedom at its corresponding vertex.}
\begin{equation}\label{eq:gauge_transformation}
    u\mapsto g_tug_s^{-1}
    \,.
\end{equation}
Only configurations which are invariant under these transformations are physically observable. 

Moving to a quantum theory (for a rigorous approach to this subject, see~\cref{app:quantum_theory}) one defines states $\ket{u}$ characterized by a complete knowledge of the Wilson line $u$ at $e$. 
A generic state is defined by a quantum superposition of such states
\begin{equation}
    \ket{\psi}=\int\!\dd{u}\ket{u}\psi(u)
    \,,
\end{equation}
and the wave function is described by 
\begin{align}
    \psi(u) = \ip{u}{\psi} \,.
\end{align}
Tensor products of the wave functions at each link span the full Hilbert space. 
Therefore a general state will be a wave function over the space of Wilson lines on all of the oriented links of the graph, and the wave function is given by
\begin{equation}
    \psi(u_1,u_2,\dots) = \ip{u_1,u_2,\dots}{\psi}
    \,.
\end{equation}

The gauge redundancy mentioned above means that only those wave functions that are insensitive to gauge transformations of the form \eqref{eq:gauge_transformation} are physically meaningful. 
This can be expressed by noting that these transformations naturally lift into the quantum theory through a representation
\begin{equation}
    \ip{u}{\psi}\mapsto\ip{g_t^{-1}ug_s}{\psi}
    \,.
\end{equation}
Physical wave functions should be invariant whenever such a transformation is applied to all the Wilson lines. 
This defines the physical subspace of of the full Hilbert space. 
The condition that wave functions are in this subspace will be called Gauss' law. 

The discussion so far has shown how one can define an HLGT on a general graph by defining the Wilson lines on links or trajectories containing multiple links of the graph.
One can transform graphs into one another by taking trajectories containing multiple links and mapping them onto trajectories containing less links and vice versa. 
Consider a graph $\G$ containing a set $R$ of trajectories on $\G$ (where $R$ is a subset of all possible trajectories). Then one can define a new graph $\G_R$ through a coarsening procedure:
For each starting point $s(\gamma)$ and endpoint $t(\gamma)$ of the trajectories in $\G$ we identify a vertex in $\G_R$. 
In other words, the vertices in $\G_R$ are given by
\begin{align}
	p \in \bigcup_{\gamma\in R} \{s(\gamma), t(\gamma)\}\,.
\end{align}
For each trajectory in $R$ we now identify two half-links $h_\gamma^{(s)}$ and $h_\gamma^{(t)}$ in $\G_R$, with $h_\gamma^{(s)}$ attached to the vertex corresponding to the starting point $s(\gamma)$, while $h_\gamma^{(t)}$ attached to the vertex corresponding to the end point $g(\gamma)$.
The half-links in $\G_R$ are therefore 
\begin{align}
	h \in \bigcup_{\gamma\in R} \{h_\gamma^{(s)}, h_\gamma^{(t)}\}\,.
\end{align}

Now assume that a theory is defined by specifying Wilson lines on all trajectories $\gamma \in R$ on the original graph $\G$.  
To obtain the equivalent theory on the new graph, one needs to define the Wilson lines on the links formed in the graph $\G_R$ such that they agree with the Wilson lines on the trajectories $\gamma$ in the original graph $\G$. 
As we will see later, the theory also requires electric operators defined on the half-links of the graph $\G_R$ to be related in a very particular way to the electric operators defined on the original lattice.
The precise conditions under which we claim the equivalence of the theories on $\G$ and $\G_R$, as well as its proof, are given in~\cref{app:classical_graphs} for the classical setting, and ~\cref{app:quantum_graphs} for the quantum one.
In the next section, we will apply this concept to our example of choice, namely a theory defined on a hypercubic lattice, gauge fixed using a maximal tree procedure. 

\section{Maximal Trees from the Persepective of Graph Coarsening}
\label{sec:point_splitting}

\subsection{From Lattices to Flowers: formulation of HLGT in maximal tree gauge fixing}
\label{sec:Lattices_To_Flowers}
\begin{figure}
    \centering
    \includegraphics[width = 0.20\textwidth]{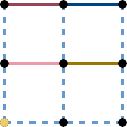}
    \caption{A maximal tree on a $2\times 2$ square latticehis shown.
    The marked vertex on the bottom left denotes the origin of the tree. The four physical links remaining are in the middle left, middle right, top left, and top right.}
    \label{fig:maximal_tree}
\end{figure}

We begin by reviewing the maximal tree construction studied in~\cite{DAndrea:2023qnr}.
In the main body of the text, we will keep our review at an informal level which will suffice to communicate our main results.
A mathematically rigorous treatment together with the proofs provided by our general graph coarsening formalism is presented in~\cref{app:maximal_tree}.

Consider an HLGT with structure group $\SU$.
On such a lattice we can define a maximal tree, defined by a maximal set of links that do not admit the construction of a closed loop. 
A choice for such a maximal tree is show in~\cref{fig:maximal_tree} in two spatial dimensions with open boundary conditions, where all blue links are part of the maximal tree. 
One can choose one vertex $o$ on this maximal tree as the origin of the tree, which we choose to be the lower left vertex, denoted in orange. 
A maximal tree also defines a unique path between any two vertices on the lattice, which is obtained by only traversing links on the maximal tree without backtracking.

Gauge invariance can be used to fix all Wilson lines on the maximal tree to the identity. 
To see this, pick a vertex $q$ at the end of the maximal tree, such that it only attaches to a single link in it and consider the path going from $o$ to $q$ along the tree.
Use the gauge freedom at the first vertex along this path after $o$ to gauge fix the first link of the trajectory to the identity.
This can of course be done by applying a gauge transformation equal to the inverse of the Wilson line at this first link.
Due to the uniqueness of the path between any two vertices, one can see that this uses the gauge invariance of all vertices except for the one at $o$.

\begin{figure}
    \centering
    \subfloat[]{\label{fig:flower_trajectories}\includegraphics[width = 0.22\textwidth]{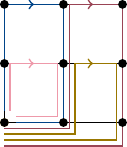}}
    \par\bigskip
    \subfloat[]{\label{fig:flower_complete}\includegraphics[width = 0.22\textwidth]{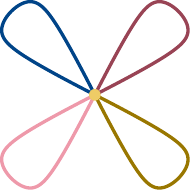}}
    \caption{Loop construction the $2\times 2$ lattice of\cref{fig:maximal_tree}, containing 4 physical links. For each of these an orientation has been chosen and a corresponding loop has been drawn through it. These are the loops in $R$ used for the maximal tree construction.
    On the bottom, the resulting flower $\G_R$ is shown.}
    \label{fig:flower}
\end{figure}

Using the maximal tree, for each physical link one can now construct a unique  trajectory originating and ending at the origin of the lattice $o$, and which traverses only links on the maximal tree apart from the physical link chosen\footnote{At this stage we would like to remark that one could have replaced the trajectories we constructed out of the maximal tree by trajectories that go through each plaquette and are otherwise restrained to be along the maximal tree. 
Indeed, this is the choice made in~\cite{mathur_canonical_2015}. 
The techniques developed in this paper can be applied to either of these (or any other) choices.}.
An example of these trajectories is shown on the top of~\cref{fig:flower}. 
Given that each trajectory only goes through a single physical link, and the links on the maximal tree have been gauge-fixed to the identity, the physical links support Wilson lines equal to the Wilson line supported on the corresponding closed trajectory. 
Furthermore, since they start and end at the vertex $o$, each such Wilson line $u$ transforms under a gauge transformation $g_{o}$ at said vertex as
\begin{equation}
    u\mapsto g_{o}ug_{o}^{-1}
    \,.
\end{equation}

One can now use the coarsening technique discussed in the previous section to construct a new graph that contains the same information as the original point split lattice. 
Let $R$ be the set of trajectories just discussed.
Since the vertices on the new graph $\G_R$ are given by the start and end points of the trajectories on the original lattice, the graph $\G_R$ contains only a single vertex.
The half-links forming the links in the new graph $\G_R$ therefore all connect to a single vertex, and are connected on the other side to each other. 

This can be summarized by replacing our lattice by a ``flower,'' like the one shown on the bottom of~\cref{fig:flower}. 
The single vertex in the flower corresponds to the origin of the tree. 
Each loop in the flower, which we will refer to as ``petals,'' corresponds to one of the trajectories we pointed out on the lattice. 
There is therefore one petal for each physical link, and the Wilson line supported on each petal will be equal to the Wilson line on each physical link.
Such a flower is an example of more general structures that can be formed out of vertices and half-links, with a flower consisting of a single vertex and $N$ links connected as loops to this vertex. 
As long as one equips the flower with an appropriate Hamiltonian, the theories on the original lattice and on the flower are completely equivalent.

The maximal tree construction in this paper has therefore left us with a much simpler representation of the theory, exemplified by the fact that the remaining gauge group has been reduced to a single $\SU$ at the remaining vertex. 
The price we have to pay however is that the dynamical information is now non-local, for each petal is generically probing extended parts of the original lattice, where they have non-trivial overlaps. 

\subsection{From Petals to Leaves: understanding the Hilbert space of a single loop}
\label{sec:Petals_To_Leaves}

\begin{figure}
    \centering
    \includegraphics[width = 0.12\textwidth]{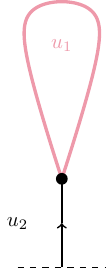}
    \caption{Generic leaf. The loop at the top will still be referred to as the petal, while the straight line at the bottom, which connects it to the rest of the graph will be called the stem. Conventions for Wilson lines $u_1$ and $u_2$ used throughout the text are set. In particular, for definiteness, the stem will be oriented towards the petal.}
    \label{fig:leaf}
\end{figure}

Let us now focus on a single petal of the flower.
By pinching the base of the petal, we form a ``leaf,'' like the one shown in~\cref{fig:leaf}. 
As discussed in the previous section, the petal supported a single Wilson line $u$.
The leaf on the other hand supports a Wilson line $u_1$ on its loop and $u_2$ on its stem.
The theories are related by identifying
\begin{equation} 
    u=u_2^{-1}u_1u_2
    \,.
\end{equation}
In other words, the petal can be recovered from the leaf by graph coarsening through said leaf.
The leaf also contains an additional vertex joining the loop to the stem.
As we will explain below, the theories on petals and leaves are equivalent as long as one is considering states on the leaf that are gauge invariant with respect to this new vertex.
This is an example of the general concept of point splitting, which is discussed more thoroughly in~\cref{app:point_splitting}. 

The gauge invariance on the new vertex acts by
\begin{equation} 
    u_1\mapsto gu_1g^{-1}\qand u_2\mapsto gu_2
    \,, 
\end{equation} 
which of course keeps $u$ fixed. 
Taking $g_a=u_2^{-1}$, we see that we can always gauge fix the leaf such that the loop supports the same Wilson line as the original petal, while the Wilson line on the stem is gauge-fixed to the identity
\begin{equation} 
    u_1\mapsto g_au_1g_a^{-1} = u\qand u_2\mapsto g_au_2 = 1
    \,.
\end{equation} 

Starting from a configuration satisfying this gauge-fixing condition, however, we can obtain a more interesting configuration. 
Indeed, let us start with a configuration that has been gauge fixed such that $u_1=u$ and $u_2=1$.
If we use an axis-angle representation for the Wilson line
\begin{equation}
    u_1=u=e^{i\vb*{\omega}\cdot\vb{J}}
    \,,
\end{equation}
a gauge transformation then acts by
\begin{equation}
    gu_1g^{-1}=e^{ig\vb*{\omega}\cdot\vb{J}}
    \,,
\end{equation}
where on the right hand side $g$ is acting by the vector representation. 
We can therefore choose a gauge transformation $g_b$ such that $g_b\vb*{\omega}$ is running along the $z$-axis. 
This gauge-fixing condition now leaves us with
\begin{equation}
    u_1=e^{i\omega J_z}\qand u_2=g_b
    \,,
\end{equation}
thereby localizing the $\omega=\|\vb*{\omega}\|$ information, which characterizes the angle of rotation, to the loop of the leaf, while the $\hat{n}$ ($\vb*{\omega}=\omega\hat{n}$) information, which determines the axis of rotation, is left on the stem. 
As shown in~\cite{DAndrea:2023qnr}, the former is localized around $\omega=0$ for low-energy states in the low coupling limit, while the later is slowly varying.
The two gauge transformations $g_a$ and $g_b$ can of course be combined into a single transformation $g = g_b g_a$, so that this gauge fixing condition can always be attained.

We now have a construction of each loop formed in the maximal tree (corresponding to the petals of the flower) in terms of a leaf, which supports a Wilson line $u_1$ aligned with the $z$ direction, and only depending on a single variable $\omega$, and a stem, which encodes the angular dependence of the axis of rotation. 
As we will show later, all relevant information can be captured through simple integer quantum numbers by employing a formulation in terms of Schwinger bosons and prepotentials.  

\subsection{From Flowers to Branches: enforcing the final gauge constraint}
\label{sec:Flowers_To_Branches}

\begin{figure}
    \centering
    \includegraphics[width=0.25\textwidth]{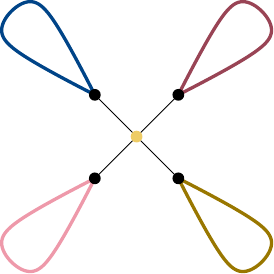}
    \caption{Flower on which the petals have been replaced by leaves.}
    \label{fig:flower_leaves}
\end{figure}

Having transformed our petals to leaves, our flower now looks like the one shown in~\cref{fig:flower_leaves}.
The origin of the original lattice, which contained the last remaining gauge redundancy, has been mapped to the center of this flower.
This vertex however has a high valency of $2N$. This poses a difficulty for the implementation of the LSH formalism due to the presence of Mandelstam constraints \eqref{eq:mandelstam}. Thus, in order to set the stage for the introduction of prepotentials, let us show how one can reduce the valency of this vertex. This will be yet another example of the point splitting method developed more thoroughly in~\cref{app:point_splitting}.

\begin{figure}
    \centering
    \includegraphics[width=\hsize]{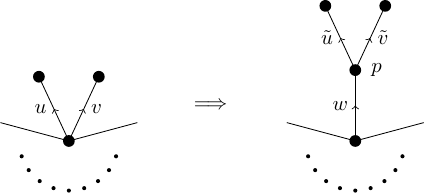}
    \caption{On the left we have a vertex of valency greater than 3 with Wilson lines $u$ and $v$ coming out of it. On the right we show the same vertex after having point split along these two links.}
    \label{fig:point_splitting}
\end{figure}

The procedure for point splitting is shown in~\cref{fig:point_splitting}. 
The physical intuition behind this procedure is that we have pinched the base of the links supporting the Wilson lines $u$ and $v$ together, thereby creating a new link. 
This new link supports the Wilson line $w$ on the figure. 
The remainder of the original links now supports $\tilde{u}$ and $\tilde{v}$. 
As shown in~\cref{app:point_splitting}, the theory before and after point splitting are physically equivalent. 
For now it will be enough to note that this equivalence is obtained via the identification
\begin{equation}
    u=\tilde{u}w \qand v=\tilde{v}w
    \,,
\end{equation}
as well as the gauge symmetry on the new trivalent vertex $p$
\begin{equation}
    \tilde{u}\mapsto\tilde{u}g^{-1}_p\qc\tilde{v}\mapsto\tilde{v}g^{-1}_p\qand w\mapsto g_pw
    \,,
\end{equation}
under which $u$ and $v$ are invariant. 
In other words, the vertex on the left of~\cref{fig:point_splitting} can be recovered from the graph on the right by graph coarsening through the trajectories supporting the Wilson lines $\tilde{u}w$ and $\tilde{v}w$.
Choosing $g_p=w^{-1}$ on the latter, we can obtain a gauge-fixing condition on which
\begin{equation}
    u=\tilde{u}\qc v=\tilde{v}\qand w=1
    \,.
\end{equation}

\begin{figure}
    \centering
    \subfloat[]{\includegraphics[width=0.4\textwidth]{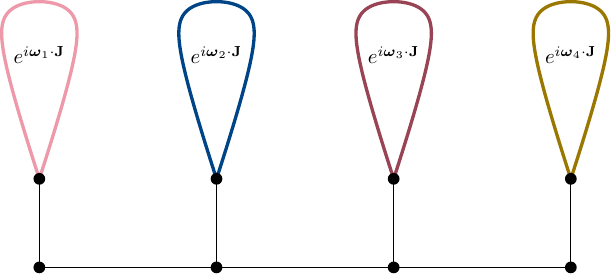}}
    \par\bigskip
    \subfloat[]{\includegraphics[width=0.4\textwidth]{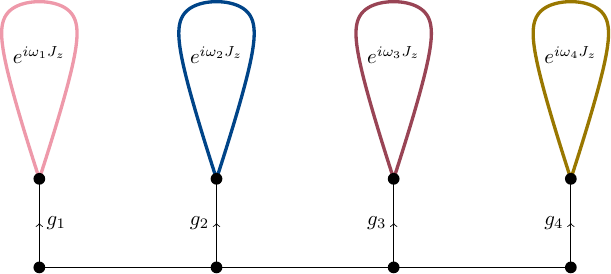}}
    \par\bigskip
    \subfloat[]{\includegraphics[width=0.4\textwidth]{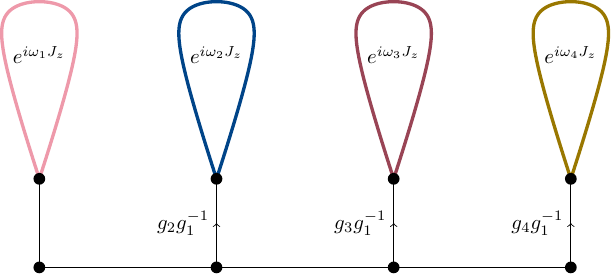} }
    \caption{Generic branch obtained through the point splitting of a flower. The three figures show equivalent classical configurations of Wilson lines which have the same amplitude on physical states. Unlabeled undirected links support the trivial Wilson line.}
    \label{fig:branch_noLSH}
\end{figure}

Using repeated applications of this point splitting, we can transform the flower of leaves in~\cref{fig:flower_leaves} to a branch like the one shown in~\cref{fig:branch_noLSH}. 
This was first considered in~\cite{mathur_canonical_2015} and gives a graph-theoretic interpretation for the result found there. 
Although our main objective is to study the physics of this branch in the LSH formalism, it is interesting to mention the magnetic possibilities of this formulation.
We first note that gauge transformations of the kind described in this section and in the discussion of~\cref{sec:Petals_To_Leaves} show that, as far as physical states are concerned, we can always restrict to configurations like the one shown at the top of~\cref{fig:branch_noLSH}.
These have trivial Wilson lines in all of their links except for their petals, which have general rotations. If we enumerate the petals with indices $i$, these rotations can be labeled as $e^{i\vb*{\omega}_i\cdot\vb{J}}$.

In~\cref{sec:Petals_To_Leaves} we also showed that we can go further and consider configurations where each petal has rotations $e^{i\omega_i J_z}$ around the $z$ direction. 
The simplification on the petals come at the cost of having to add group elements $g_i$ to the $i$-th stem. These were defined to be the elements of $\SU$ which rotate $\vb*{\omega_i}$ to the $z$ direction. 
They are of course only defined up to the $\mathrm{U}(1)$ group of rotations around the $z$-axis. 
This is shown in the middle of~\cref{fig:branch_noLSH}.
This remaining redundancy sits at the vertex joining the petals with their corresponding stem.
This separation of the axis-angle degrees of freedom throughout the branch will be useful to us when we attempt to understand the magnetic content of the LSH basis in~\cref{sec:LSH_on_leaves}.

But now that we have this branch structure, we can further exploit the remaining degrees of freedom to reduce the configurations we consider even more.
For example, we can apply $g_1$ on all of the bottom vertices to obtain the configurations shown in the bottom of figure~\cref{fig:branch_noLSH}.
These are configurations viewed ``from the perspective of $\vb*{\omega}_1$.''
Indeed, they are of the form we would have obtained using the previous gauge-fixing if $\vb*{\omega}_1$ happened to lie on the $z$-axis.
In this gauge, the stems carry the Wilson lines $g_ig_1^{-1}$.
These correspond to rotations mapping $g_1\vb*{\omega}_i$, which is the vector $\vb*{\omega}_i$ viewed from the reference frame of $\vb*{\omega}_1$, to the $z$-axis. 
These are then of course conjugate to the rotations $g_1^{-1}g_i=g_1^{-1}g_ig_1^{-1}g_1$ mapping $\vb*{\omega}_i$ to $\vb*{\omega}_1$.
Indeed, we can perform an additional gauge transformation of $g_1^{-1}$ on all of the vertices to leave such rotations on the stems.
This however would rotate the petals away from the $z$ direction and is therefore undesirable from the point of view of the LSH formulation. Let us turn to this now.

\section{LSH Construction on General Graphs}
\label{sec:lsh_construction}
In this section we will use the construction discussed in~\cref{sec:Kinematics}, in which each link is comprised of two half-links. 
The importance of half-links stems from the Peter-Weyl theorem, which states that the Hilbert space of each link can be written as a direct sum of the tensor product of the irreducible representations $V_r$ and its dual $V_r^*$
\begin{equation}
    L^2(G)\cong\bigoplus_{r\in \hat{G}}(V_r\otimes V_r^*)
    \,,
\end{equation}
with $\hat{G}$ an exhaustive collection of the inequivalent irreducible unitary representations of $G$. 
This localization suggests an enlargement of the Hilbert space to
\begin{equation}\label{eq:abelian_enlargement_main}
    L^2(G)\subseteq \mathcal{H}_{\SB}\otimes\mathcal{H}_{\SB}^*\qc \mathcal{H}_\SB\coloneq \bigoplus_{r\in\hat{G}}V_r
    \,, 
\end{equation}
where each of the two $\mathcal{H}_\SB$ factors is thus supported on a single half-link.
In this extended Hilbert space, each half-link can support a different representation of the gauge group.
However, only states in which every two half-links forming a link have the same representation are physical. 
The original Hilbert space is therefore recovered by the introduction of new \emph{Abelian} Gauss' laws demanding that physical vectors are spanned by tensor products of vectors in the same irreducible representation.

\begin{figure}
    \centering
    \includegraphics[width = 0.1\textwidth]{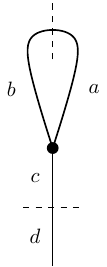}
    \caption{Generic leaf with named half-links.}
    \label{fig:half_edge_leaf}
\end{figure}

On can now use the prepotential formulation, where each half-link (like the ones labeled $a$, $b$, $c$ and $d$ in~\cref{fig:half_edge_leaf}) is equipped with a pair of Schwinger bosons $a^\dagger_A$, $A=1,2$ satisfying the usual canonical commutation relations\footnote{Some more details of this formulation can be found in~\cref{app:prepotential}, but we provide enough background to understand the claims of the paper.}.
The gauge structure of the theory is reflected in the fact that these bosons form an $\SU$ doublet. 
Such Schwinger bosons can be used to create spin $j$ states
\begin{equation}\label{eq:sb_electric}
    \ket{j,m}\coloneq \frac{(a_1^\dagger)^{j+m}(a_2^\dagger)^{j-m}}{\sqrt{(j+m)!(j-m)!}}\ket{0}
\end{equation}
on their associated half-link.
In particular, $2j$ counts the total number of bosons on the half-link. 

As already discussed, each link is composed of two half-links, and the resulting state on a link is a tensor product of two such spin states
\begin{equation}
    \ket{j,m;j',m'}\coloneq \ket{j,m}\otimes\ket{j',m'}
    \,,
\end{equation}
with $j \neq j'$ in general. 
One can use the Peter-Weyl theorem to relate these states to the ones of Yang-Mills theory. 
Indeed, if $j=j'$, such a state induces a wave function which is defined by the matrix element
\begin{equation}\label{eq:electric_to_magnetic}
    \ip{u}{j,m;j,m'}\coloneq (-1)^{j+m}\sqrt{2j+1}\mel{j,-m}{r_j(u)}{j,m'}
    \,,
\end{equation}
with $r_j(u)$ the spin $j$ representation of $u$.
The Peter-Weyl theorem then guarantees that these wave functions form a basis of all the wave functions of the theory. 

A major advantage of the prepotential formulation is that gauge invariance at a vertex is simply ensured by contracting all of the indices of the Schwinger bosons on the half-links originating from it. 
As already discussed, one side of each half-link will be connected to a vertex, while the other side will be joined with another half-link to form a link. 
These half-links therefore transform under the $\SU$ gauge group on one side, and under the $\mathrm{U}(1)$ Abelian gauge symmetry on the other. 
As discussed in more detail in~\cref{sec:LSHOps} these transformation properties can be captured by defining a single operator
\begin{equation}\label{eq:wilsonmain}
    \begin{aligned}
        A\indices{^A_a}(h)&\coloneq \begin{cases}
            a^{\dagger A}(h) & a=+\\
            a^{A}(h) & a=-
            
        \end{cases}\\
        &=\begin{pmatrix}
            a^{\dagger 1}(h) & a^1(h) \\
            a^{\dagger 2}(h) & a^2(h)
        \end{pmatrix}\indices{^A_a}\\
        &=\begin{pmatrix}
            a_2^\dagger(h) & a^1(h) \\
            -a_1^\dagger(h) & a^2(h)
        \end{pmatrix}\indices{^A_a}\,.
    \end{aligned}
\end{equation}
Here we have used that $\SU$ indices $A$ can be raised and lowered 
using the totally antisymmetric symbol $\epsilon^{AB}$. 
We will define $\epsilon_{AB}=-\epsilon^{AB}$ so that $\epsilon_{AB}\epsilon^{BC}=\delta^{C}_A$.
By convention we will take $\epsilon^{12}=1$. 
Note also that since creation and annihilation operators are conjugates of one another, one is naturally indexed using covariant indices, while the other using contravariant ones.
The metric to raise the indices corresponding to the Abelian gauge symmetry is given by
\begin{align}
g=\mqty(0 & -1\\ -1 & 0)
\,.
\end{align}
The alert reader will recognize this matrix as the metric of $\mathrm{SO}(1,1)$ written in terms of light-cone coordinates.
Since we are working with complex representations, this is equivalent to the vector representation of $\mathrm{SO}(2)=\mathrm{U}(1)$, which is the Abelian gauge symmetry associated with the Abelian Gauss law.
Note that this operator satisfies 
\begin{equation}
\label{eq:normA}
    A(h)\indices{^A_a}A^\dagger(h)\indices{^a_B}=\delta\indices{^A_B}(N(h)+1)\,,
\end{equation}
where $N(h)$ denotes the usual number operator build out of the Schwinger bosons
\begin{align}
\label{eq:NOpdef}
    N(h) = a^\dagger(h) a(h)
    \,.
\end{align}

We now proceed to construct gauge invariant operators out of the $A\indices{^A_a}(h)$. The first is the number operator already defined in~\eqref{eq:NOpdef}. 
Next, we construct a corner operator, built out of two half-links, connected by a single vertex. 
To make this operator invariant under gauge transformations at this vertex, it should be constructed out of two operators $A\indices{^A_a}(h)$, with the $\SU$ index contracted. 
This leads to the so-called loop operators~\cite{Raychowdhury:2019iki}, involving two half-links that are connected to the same vertex
\begin{equation}
    \mathcal{L}_{ab}(h_1,h_2)=A\indices{_{Aa}}(h_1)A\indices{^A_b}(h_2)
    \,.
\end{equation}
Choosing for the two indices $a = b = +$, one finds the gauge invariant operator 
\begin{equation}\label{eq:loop_creation_main}
    \mathcal{L}_{++}(h_1,h_2)=a_A^\dagger(h_1)a^{\dagger A}(h_2)=-\mathcal{L}_{++}(h_2,h_1)
    \,,
\end{equation}
which generates units of chromoelectric flux along the corner between half-links $h_1$ and $h_2$. 
In the same way, one can construct the operators $\mathcal{L}_{--}(h_1,h_2)$ and $\mathcal{L}_{+-}(h_1,h_2) = - \mathcal{L}_{-+}^\dagger(h_1,h_2)$.
These operators agree (up to a minus sign for the operator $\mathcal{L}_{-+}$) with the results in~\cite{Raychowdhury:2019iki}.
With this formulation the commutation relations between loop operators can be succintly described by the expression
\begin{equation}
    \begin{aligned}
        &[\mathcal{L}_{ab}(h_1,h_2),\mathcal{L}_{cd}(h_3,h_4)] = \\
        &g_{ad}\delta_{h_1,h_4}\mathcal{L}_{cb}(h_3,h_2) -g_{bc}\delta_{h_2,h_3}\mathcal{L}_{ad}(h_1,h_4)\\
        &+g_{bd}\delta_{h_2,h_4}\mathcal{L}_{ac}(h_1,h_3)-g_{ac}\delta_{h_1,h_3}\mathcal{L}_{db}(h_4,h_2)\,,
    \end{aligned}
\end{equation}
which reproduces all of the results contained in Table I of their work.
This can also be viewed as a precursor of the $\mathrm{SO(4,2)}$ algebra found in~\cite{mathur_canonical_2015}.

Another important operator is the Wilson line operator on a single link. Since each link is composed of two half-links, the Wilson line operator depends again on two half-links $h_1$ and $h_2$, which are however connected to each other, rather than to a common vertex. 
Since links start and end on vertices, this operator needs to transform under $\SU$ on both ends, making it a tensor with two $\SU$ indices $U\indices{^A_B}(h_1, h_2)$. 
This operator needs to return the Wilson line between the two endpoints of the link when acting on the state $\ket{u}$. 
This requires the operator to be unitary, satisfying the  condition
\begin{align}
U\indices{^A_B}(h_1, h_2) U\indices{^{\dagger B}_C}(h_1, h_2) = \delta\indices{^A_C}
\,,
\end{align}
as well as having unit determinant. 
One might attempt to compose this operator by combining two of the half-link operators $A\indices{^A_a}(h)$ by contracting the Abelian index. This standard contraction, however, does not give the correct Abelian Gauss law. 
As discussed in~\cref{sec:Wilson_line_operators}, the Abelian Gauss law is instead recovered by contracting the Abelian indices with a matrix 
\begin{equation}
    t\indices{^a_b}(h_1,h_2)=\mqty(0 & 1\\
    -1 & 0)\indices{^a_b}
\end{equation}
for links ordered in the positive orientation, and the transpose (negative) for negative orientations.
As shown in more detail in~\cref{sec:Wilson_line_operators}, combining this with the normalization of the of the single link operators given in~\cref{eq:normA}, one finds
\begin{align}
\label{eq:UABdef}
        U\indices{^A_B}(h_1,h_2)=U\indices{^A_a}(h_1) t\indices{^a_b}(h_1, h_2) U\indices{^{\dagger b}_{B}}(h_2)\,,
\end{align}
where we have defined the normalized single link operator as
\begin{align}
\label{eq:UNormdef}
    U(h)\indices{^A_a} = \frac{1}{\sqrt{N(h)+1}}A(h)\indices{^A_a}\,.
\end{align}
We can again compare these results to those of~\cite{Raychowdhury:2019iki}. 
The left-hand side of~\cref{eq:UABdef} is identical to the the operator $\hat U$ in (24) of their paper. 
Our operator $U\indices{^A_a}(h_1)$ is equal to their operator $\hat U_L$, while their operator $\hat U_R$ corresponds to $t\indices{^a_b}(h_1, h_2) U\indices{_{B}^b}(h_2)$. 

The Wilson line operators of a single link can be combined into Wilson line operators for trajectories over multiple links, using
\begin{align}
    U\indices{^A_B}(h_1,h_2)U\indices{^B_C}(h_3,h_4)= U\indices{^A_C}(h_1,\ldots, h_4)
    \,,
\end{align}
where $U\indices{^A_C}(h_1,\ldots, h_4)$ denotes the Wilson line along a trajectory going from $h_4$ to $h_1$. 
Using this, the Wilson line operator for any trajectory between two vertices can be written by stringing together Wilson operators for the links $\gamma_1, \ldots, \gamma_n$ traversed by the trajectory $\gamma$, contracting all indices corresponding to the intermediate vertices
\begin{align}
    U\indices{^A_B}(\gamma) = U\indices{^A_{A_1}}(\gamma_n) \ldots U\indices{^{A_{n-1}}_B}(\gamma_1)
    \,.
\end{align}

Using~\cref{eq:UABdef,eq:UNormdef,eq:wilsonmain} the Wilson line operator can be written in terms of Schwinger boson creation and annihilation operators. 
This can be simplified by rearranging and regrouping the relevant terms to write
\begin{align}
    U\indices{^A_C}(h_1,\ldots, h_4) =&\, U(h_1)\indices{^A_a} t\indices{^a_b}(h_1, h_2) U\indices{^b_c}(h_2, h_3) \nonumber\\
    & \times t\indices{^c_d}(h_3, h_4) U\indices{^{\dagger d}_C}(h_4)
    \,,
\end{align}
introducing the LSH Wilson line operator defined on a corner
\begin{align}
\label{eq:UCorner}
        &U\indices{^a_b}(h_2,h_3)=U\indices{^{\dagger a}_A}(h_2)U\indices{^A_b}(h_3)
        \,,
\end{align}
depending on two half-links connected to a single vertex.
The LSH Wilson line operator can be written in term of the loop operators as
\begin{equation}\label{eq:corner_op_to_vertex}
    \begin{aligned}
        U\indices{^a_b}&(h_1,h_2)=\\
        &\begin{cases}
            a\frac{1}{\sqrt{N(h_2)+1}}\mathcal{L}\indices{^a_b}(h_1,h_2)\frac{1}{\sqrt{N(h_1)+1}} & h_1\neq h_2\,\\
            \delta^a_b & h_1 = h_2\,.
        \end{cases}
    \end{aligned}
\end{equation}
These operators are known as vertex factors in~\cite{Raychowdhury:2019iki}.
Just as was the case for the Wilson line operators, one can string together multiple LSH Wilson line operators by contracting indices with $t\indices{^a_b}$ to write, for example,
\begin{align}
    U\indices{^a_d}(h_2, \ldots, h_5)&=U\indices{^a_{b}}(h_2, h_3)t\indices{^{b}_{c}}(h_3, h_4)U\indices{^{c}_{d}}(h_4, h_5)\,.
\end{align}
Again, this can be extended to a larger corner trajectory $\gamma_c$, traversing the corners $\gamma_{c,1}$, $\gamma_{c,2},\ldots,$ up to $\gamma_{c,n}$, by
\begin{align}
\label{eq:UCornerTrajectory}
    U\indices{^a_c}(\gamma_c)&= U\indices{^a_{b_{1}}}(\gamma_{c,n})t\indices{^{b_{1}}_{b_{2}}}(\gamma_{n-1}) \ldots U\indices{^{b_{2n-2}}_{c}}(\gamma_{c,1})
    \,.
\end{align}
In here $\gamma_{n-1}$ is the link common to the corner $\gamma_{c,n}$ and $\gamma_{c,n-1}$.

\begin{figure}
    \centering
    \includegraphics[width=0.25\textwidth]{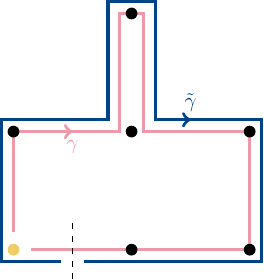}
    \caption{A trajectory $\gamma$ that starts and ends at a marked vertex. The last link $\gamma_n$ that the trajectory goes through is highlighted by a dashed line that separates its two half-links. A corresponding corner trajectory $\tilde{\gamma}$ is shown on the outside. It starts at a corner at the marked vertex. The first link that it goes through is only traversed once by these trajectories. This is in contrast to what would have happened if we had chosen $\tilde{\gamma}$ to start at the corner in the upper most vertex.}
    \label{fig:trajectory_v_corner_trajectory}
\end{figure}

Note that the LSH Wilson line operators no longer carry any explicit $\SU$ indices and are therefore invariant under $\SU$ gauge transformations.
Any gauge invariant operator, in particular those present in the Hamiltonian of the theory, will be constructed out of these LSH Wilson line operators. 
For example, one can construct gauge invariant Wilson loop operators. Indeed, consider a closed trajectory $\gamma$ whose last link is $\gamma_n$. While this can be thought of as a sequence of links, it can be equivalently be thought of a sequence of corners. Let us denote the corresponding corner trajectory by $\tilde{\gamma}$. The Wilson loop on $\gamma$ can be equivalently expressed in terms of the LSH Wilson line on $\tilde{\gamma}$ as
\begin{equation}\label{eq:loops_lsh_main}
    \begin{aligned}
        &\tr U(\gamma)=t\indices{^a_b}(\gamma_n)U\indices{^b_a}(\tilde{\gamma})\,,
    \end{aligned}
\end{equation}
as long as the last half-link that $\gamma$ goes though (which is the first one $\tilde{\gamma}$ goes through) is only traversed once.
This is the situation depicted in~\cref{fig:trajectory_v_corner_trajectory}.
Such Wilson loop operators will be used to construct the magnetic part of the Hamiltonian of the system in~\cref{sec:Dynamics}.

\begin{figure}
    \centering
    \includegraphics[width=0.2\textwidth]{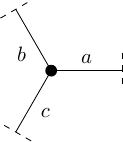}
    \caption{We show a standard trivalent vertex with half-links $a$, $b$, and $c$.}
    \label{fig:trivalent_vertex}
\end{figure}

What we have shown so far is that any gauge invariant combination of Schwinger boson creation and annihilation operators can be written in terms of the loop operators $\mathcal{L}_{ab}(h_1,h_2)$, depending on two half-links $h_1$ and $h_2$ which are connected to the same vertex and therefore form a corner. 
We have also seen in~\cref{sec:point_splitting} that using point splitting any graph can be transformed into a graph constructed only out of trivalent vertices which do not contain any Mandelstam constraints. 
The gauge invariant Hilbert space of such a trivalent vertex is therefore spanned by states that count the gauge invariant combination of Schwinger bosons for each of the three corners of a trivalent vertex. 
Labeling the three half-links connected to a trivalent vertex by $a$, $b$ and $c$, as indicated in~\cref{fig:trivalent_vertex}, the trivalent vertex is therefore characterized by three quantum numbers $n^{ab}$, $n^{bc}$ and $n^{ca}$, and the Hilbert space of any graph constructed out of such trivalent vertices is given by the tensor product of such states 
\begin{align}
    \ket{\psi} = \ket{n_1^{ab}, n_1^{bc}, n_1^{ca}} \otimes \ket{n_2^{ab}, n_2^{bc}, n_2^{ca}}\otimes\cdots
    \,.
\end{align}
Since the operators $\mathcal{L}_{++}(h_1, h_2)$ create excitations on the corner made out of half-links $h_1$ and $h_2$, the state $\ket{n^{ab}, n^{bc}, n^{ca}}$ is proportional to
\begin{align}
    \ket{n^{ab}, n^{bc}, n^{ca}} \propto &\, \mathcal{L}_{++}(a,b)^{n_{ab}}\mathcal{L}_{++}(b,c)^{n_{nc}}\nonumber\\
    & \times \mathcal{L}_{++}(c,a)^{n_{ca}} \ket{0, 0, 0}
    \,,
\end{align}
where $\ket{0,0,0}$ denotes the state with no excitations. 
To simplify the calculation of the normalization factor we introduce the normalized operators of~\cite{Raychowdhury:2019iki} (see~\cref{app:point_splitting} for details)
\begin{equation}\label{eq:ladder_creation}
    \mathcal{L}_{++}(a,b) = \sqrt{(N(a,b,c)+1)(N(a,b)+1)}\Lambda_+(a,b)
    \,,
\end{equation}
with
\begin{align}
    N(a,b)&=\frac{1}{2}(N(a)+N(b)-N(c))\\
    N(a,b,c)&=N(a,b)+N(b,c)+N(c,a)+1\,,
\end{align}
and $N(h)$ defined in~\cref{eq:NOpdef}. 
Note that the normalization factor depends on the number of excitations at all three half-links of the corner. 
These operators are normalized raising operators
\begin{align}
    \Lambda_+(h_1,h_2) \ket{n^{h_1h_2}} = \ket{n^{h_1h_2}+1}
    \,,
\end{align}
allowing us to write
\begin{align}
    \ket{n^{ab}, n^{bc}, n^{ca}} = &\,\Lambda_+(a,b)^{n_{ab}}\Lambda_+(b,c)^{n_{nc}}\nonumber\\
    & \times \Lambda_+(c,a)^{n_{ca}} \ket{0, 0, 0}
    \,.
\end{align}
Similarly, one can define the lowering operator
\begin{align}
    \Lambda_-(h_1,h_2) \ket{n^{h_1h_2}} = \ket{n^{h_1h_2}-1}
    \,.
\end{align}
The remaining loop operators can then be written as
\begin{align}\label{eq:rest_ladder}
    \mathcal{L}_{--}(a,b)=&\,\Lambda_{-}(a,b)\sqrt{N(a,b)N(a,b,c)}\\
    \mathcal{L}_{-+}(a,b)=&\,\Lambda_-(c,a)\Lambda_+(b,c)\sqrt{N(c,a)(N(b,c)+1)}\nonumber\\
    \mathcal{L}_{+-}(a,b)=&-\Lambda_-(b,c)\Lambda_+(c,a)\sqrt{N(b,c)(N(c,a)+1)}\nonumber\,.
\end{align}

\section{LSH Formulation on Branches}
\label{sec:LSH_on_branches}
\subsection{The Hilbert space on a single leaf}\label{sec:LSH_on_leaves}

We begin by considering a single petal, point split into a leaf and a stem, such as the one illustrated in~\cref{fig:half_edge_leaf}.
This leaf consists of one trivalent vertex where two of the half-links $a$ and $b$ are connected to each other, and the third half-link $c$ is attached to another half-link $d$. 
From the discussion of~\cref{sec:lsh_construction} the Hilbert space on this leaf is characterized by the excitations on the three corners, supplemented by Schwinger boson excitations on the remaining half-link $d$.

While for a general trivalent vertex there are three independent corner operators, the fact that two of the half-links are joined together gives rise to an additional Abelian Gauss law constraint. 
This constraint ensures that the flux flowing between half-links $a$ and $c$ and that between half-links $b$ and $c$ has to be equal, for otherwise there would be more bosons on $a$ than on $b$. 
The two corresponding operators can therefore be bundled into the loop operator
\begin{align}
    \mathscr{L}_{++} &= a^\dagger_A(a) a^{\dagger A}(c) \,\, a^\dagger_B(b) a^{\dagger B}(c)\nonumber\\
    \ell_{++} &= a^{\dagger A}(a) \,\, a^\dagger_A(b)
    \,.
\end{align}
One can see that $\ell_{++}$ creates chromoelectric flux within the loop, while $\mathscr{L}_{++}$ creates chromoelectric flux that goes in and out of the loop.
Appending the most general Schwinger boson configuration supported on the half-link $d$, subject to the Abelian Gauss law constraint between half-links $c$ and $d$, we then obtain a basis of states of the form
\begin{equation}
\label{eq:LSHStates}
    \ket{n,l,m}\propto\ell_{++}^n\mathscr{L}_{++}^l(a_1^\dagger(d))^{l+m}(a_2^\dagger(d))^{l-m}\ket{0}
    \,,
\end{equation}
where $n$ counts the internal chromoelectric flux while $l$ the external.
The possible values for $m$ must satisfy
\begin{equation}
    -l\leq m\leq l
    \,.
\end{equation}
These are similar to the Hydrogen atom or tadpole states discussed in~\cite{mathur_canonical_2015}. They however do not correspond to each other exactly. For example, the $\SU$ representation on the leaf corresponding to our states has Casimir
\begin{equation}
    \frac{n+l}{2}\qty(\frac{n+l}{2}+1)\,.
\end{equation}
The tadpole states there instead have Casimir
\begin{equation}
    \frac{n^2-1}{4}.
\end{equation}

Due to the geometry of the petal, $\mathscr{L}_{++}^{l}$ always has $2l$ bosons on $c$. 
Accordingly, the state on the stem always has integer spin $j=l$, i.e. is in a representation of the rotation group $\mathrm{SO}(3)$.
Therefore $l$ and $m$ can be thought of as orbital angular momentum quantum numbers.
As shown in~\cref{app:wavefunction}, using the gauge-fixing condition discussed in~\cref{sec:Petals_To_Leaves} which aligns the Wilson line $u_1$ along the $z$-direction, one can write 
\begin{equation}\label{eq:petal_function}
    \ip{u}{n,l,m} = \phi_{n,l}(\omega) Y^l_m(\hat{n})
    \,,
\end{equation}
with
\begin{align}\label{eq:wave_function}
        \phi_{n,l}(\omega)=& i^n\sqrt{n+l+1}\sum_{m=-\frac{n+l}{2}}^{\frac{n+l}{2}}e^{i(\omega-\pi) m}\nonumber\\
        &\times\ip{\frac{n+l}{2},m;\frac{n+l}{2},-m}{l,0}
        \,.
\end{align}

\begin{figure}
    \centering
    \includegraphics[width=0.35\textwidth]{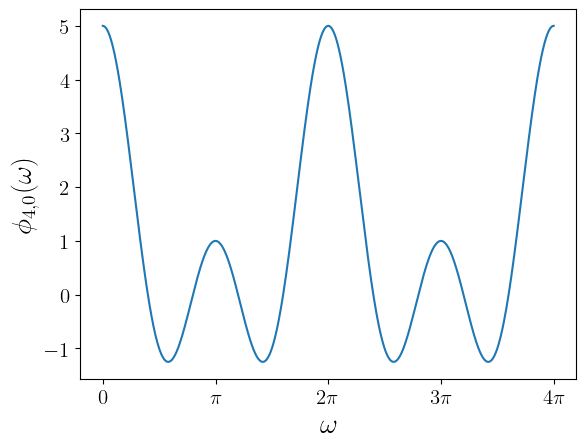}
    \centering
    \includegraphics[width=0.35\textwidth]{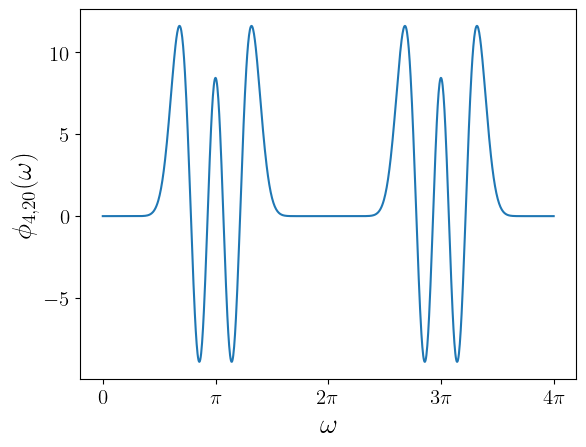}
    \centering
    \includegraphics[width=0.35\textwidth]{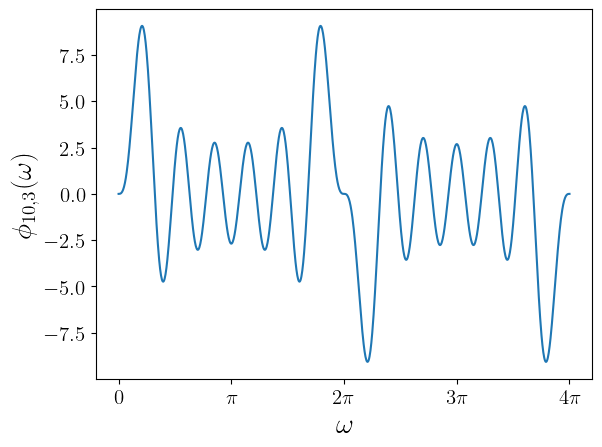}
    \centering
    \includegraphics[width=0.35\textwidth]{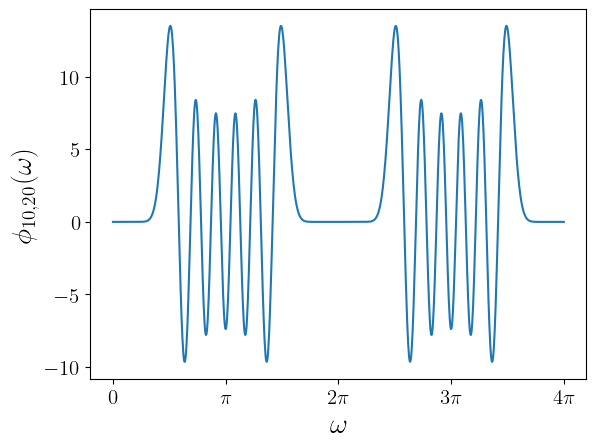}
    \caption{The functions $\phi_{n,l}$ for different values of the quantum numbers.}
    \label{fig:special_function}
\end{figure}

Using the results of~\cite{DAndrea:2023qnr,mathur_canonical_2015} and~\cref{sec:two_plaquette} these wave functions satisfy the differential equations
\begin{equation}\label{eq:differential_equation}
    \begin{aligned}
        &\qty(-\dv[2]{\omega}-\cot(\frac{\omega}{2})\dv{\omega}+\frac{l(l+1)}{4\sin(\frac{\omega}{2})^2})\phi_{n,l}\\
        &\qquad =\frac{n+l}{2}\qty(\frac{n+l}{2}+1)\phi_{n,l}\,.
    \end{aligned}
\end{equation}
There are several interesting properties of this function which are showcased in~\cref{fig:special_function}.
The more obvious one is that, being a function on $\SU$, it is $4\pi$ periodic.
In fact, when $n+l$ is an even integer, $m$ only takes integer values, which makes it $2\pi$ periodic.
In other words, it descends to a function on $\mathrm{SO}(3)$.
Along the same vein, the shift $\omega\mapsto\omega+2\pi$, produces an overall factor of $e^{i2\pi m}$, which is always $1$ for $n+l$ even and $-1$ for $n+l$ odd.
Therefore, the function is completely determined by its behaviour in $\omega\in[0,2\pi)$, to which we will restrict from now on.

According to the results of~\cite{Bauer:2021gek}, its behaviour around $\omega = 0$ is of high importance for small coupling investigations, where low energy wave functions are expected to be highly localized around these points.
Quite interestingly, we have
\begin{equation}
    \left.\dv[k]{\phi_{n,l}}{\omega}\right|_{\omega=0}=0 
\end{equation}
for all $k < l$.
In other words $\phi_{n,l}=\order{\omega^l}$.
For $k=0$ this is easy to see, for $e^{-i\pi m}$ is proportional to the CG coefficient of $\ket{0,0}$ in the tensor product.
One then sees that the function is proportional to $\ip{0,0}{l,0}=\delta_{l,0}$.
We have further checked this numerically for $0\leq n,l\leq 10$.
This result is shown in~\cref{fig:special_function} by the fact that the higher $l$ is, the flatter the function is around $\omega\in\{0,2\pi\}$, which in turn squishes the function at $\omega\in\{\pi,3\pi\}$. 

Turning this result around, one finds that wave functions receiving most of of their support from small $\omega$ can be described by low $l$ values.
It was shown in~\cite{DAndrea:2023qnr} that low energy states at small coupling are dominated by small values of $\omega$. 
Our result therefore proves one of the results of that paper, namely that low lying states at small coupling can be described with a small cutoff in the $l$ quantum numbers.
In fact, the results of this paper show that this small cutoff is directly compatible with the localization at $\omega=0$ in this basis.

A final interesting property of this function is that $n$ controls the number of nodes of the function.
For $l=0$ it is $2n$.
For $l>0$ it is $2(n+1)$.

\subsection{The Hilbert space of the fully gauge-fixed theory}
\label{sec:hilbert_fully_gauge_fixed}
The construction of the gauge invariant states at each of the leaves has not yet enforced the gauge invariance at the origin of the lattice, where all stems are connected to one another.
In this section we will use the graphs of~\cref{sec:Flowers_To_Branches}, where the stem of each leaf is attached to a branch using only trivalent vertices, to construct a fully gauge invariant theory described by gauge LSH operators. 

\begin{figure}
    \centering
    \includegraphics[width=\hsize]{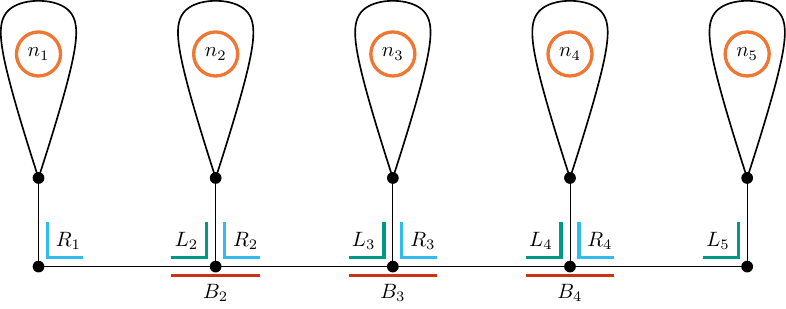}
    \caption{We show a generic branch with a spatial indicator of its associated quantum numbers. Each quantum number corresponds to the number of times the loop operator on its corresponding highlighted corner has been excited.
    The rest of the loop operators corresponding to corners that weren't highlighted are obtained uniquely through Gauss' law.
    Of the $B$s, only the leftmost is independent in our basis, and its minimum value is determined by the differences amongst the $L$s and the $R$s.}
    \label{fig:branch}
\end{figure}

The configuration of stems we consider is shown in~\cref{fig:branch}. The number of loops $n_i$ internal to each leaf remains just as for the single leaf discussed in the previous section, but now we need to construct gauge invariant combination of the Schwinger boson operators at each of the stems.
Each stem connects to the branch at a trivalent vertex, and at each trivalent vertex in the branch, we can now build three LSH operators corresponding to the three corners at that vertex. 
The number of excitations at each of them is labeled by $L_i$, $R_i$ and $B_i$ in~\cref{fig:branch}. 
Note that the first and last vertex are only bivalent, such that only a single corner exists\footnote{In practice, these corners can be flattened out removing the associated vertex. We retain the corners here for conceptual clarity.}. 
We take this into account by the convention 
\begin{align}
    L_1 = R_N = B_1 = B_N = 0
    \,.
\end{align}

The corner operators on the left and right corners will create Schwinger bosons on the stem of the leaf.
One immediately finds the relation
\begin{align}
    L_i + R_i &= 2l_i
    \,.
\end{align}
This implies that $L_i$ and $R_i$ need to have the same parity.

The Abelian gauss law on the links connecting the leaves imposes
    \begin{equation}
    R_i+B_i=L_{i+1}+B_{i+1}
    \,,
\end{equation}
where one needs to remember the convention $B_1=B_N=0$. 
This allows to solve for the $B_i$ quantum numbers
\begin{equation}
    \begin{aligned}
        B_2&=R_1-L_2,\\
        B_3&=R_1-L_2+R_2-L_3,\\
        B_4&=R_1-L_2+R_2-L_3+R_3-L_4,\\
        \vdots&
    \end{aligned}
\end{equation}
and obtain an overall constraint on the $L_i$ and $R_i$ values
\begin{equation}
    \sum_{i = 1}^{N-1} (R_i - L_{i+1}) = 0
    \,.
\end{equation}
Each relationship is simply counting the total number of chromoelectric flux pushed to the right by the $R$'s minus the one pushed up by the $L$'s. 

Another constraint comes from the fact that all loop quantum numbers $L_i$, $R_i$ and $B_i$ have to be positive. 
This can be ensured by requiring that $R_1$ is big enough
\begin{align}
\label{eq:R1constraint}
    R_1&\geq \esup\{L_2,L_2+L_3-R_2,\dots,L_2+\dots\nonumber\\
        & \hspace{2cm} + L_{N-1}-R_2-\dots-R_{N-1}\}
        \,,
\end{align}
where $\esup$ is the smallest even number greater or equal to the set.

Combining all the above information gives us a completely gauge-fixed basis of orthogonal gauge invariant states
\begin{align}
    &\ket{n_1,\dots,n_N;R_1; L_2,R_2;\dots;L_{N-1},R_{N-1}} = \\
    &\quad\left(\Lambda_+^{(n, 1)}\right)^{n_1}
    \left(\Lambda_+^{(R, 1)}\right)^{R_1}
    \left(\Lambda_+^{(n, N)}\right)^{n_N}\left(\Lambda_+^{(L, N)}\right)^{L_N}\nonumber\\
    &\quad \prod_{i=2}^{N-1}
    \left(\Lambda_+^{(n,i)}\right)^{n_i}
    \left(\Lambda_+^{(L, i)}\right)^{L_i}
    \left(\Lambda_+^{(R, i)}\right)^{R_i} 
    \left(\Lambda_+^{(B, i)}\right)^{B_i} 
     \ket{0}\nonumber
    \,,
\end{align}
where $\Lambda_+^{(n, i)}$ denoted the normalized raising operator adding one internal flux to the $i$'th leaf, while $\Lambda_+^{(L, i)}$, $\Lambda_+^{(R, i)}$ and $\Lambda_+^{(B, i)}$ denotes the normalized raising operator adding flux on the $L$, $R$ or $B$ corner of the $i$'th stem. These operators are only well-defined up to a sign depending on the ordering of the corners.
Thus, for explicit computations, setting up the Hilbert space requires choosing an ordering of the corners in the branch.

The resulting states depend on a set of integer quantum numbers greater or equal to zero. 
First, there are $N$ integer quantum numbers $n_i$ counting the number of internal loops of each leaf. Second, for each of the $N-2$ leaves $2, \ldots , N-1$, we require two quantum numbers $L_i$ and $R_i$, corresponding to the the number of left and right corner excitations of the stems (by the Abelian Gauss law, the parity of both of these numbers has to be equal). 
Finally, we need to give $R_1$, which satisfies the constraint given in~\cref{eq:R1constraint}.
Observe that this basis can easily be written in terms of unrestricted positive integers or enumerated in base $2$ using some simple classical precomputations.
\begin{figure}
    \centering
    \includegraphics[width = 0.8\hsize]{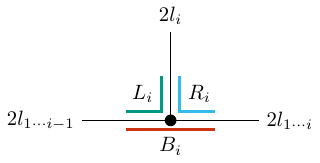}
    \caption{We show a generic trivalent vertex. The relationship between the orbital angular momentum and the LSH quantum numbers is depicted.}
    \label{fig:singlebranch}
\end{figure}

We can furthermore relate this gauge invariant basis to the one discussed in~\cite{DAndrea:2023qnr} based on angular momentum quantum numbers. 
This is done by defining
\begin{equation}
    l_{12\cdots i}=(R_i+B_i)/2
    \,.
\end{equation}
Indeed, focusing on a trivalent vertex, like the one shown in~\cref{fig:singlebranch}, we have three relations
\begin{equation}
\label{eq:LSH_relations}
    \begin{aligned}
        2l_i &= L_i+R_i,\\
        2l_{1\cdots i-1} &= L_i + B_i,\\
        2l_{1\cdots i} &= R_i + B_i
        \,.
    \end{aligned}
\end{equation}
We can use these relations to solve for the LSH quantum numbers $L_i$, $R_i$ and $B_i$.
The fact that the LSH quantum numbers must be positive then implies the three inequalities 
\begin{equation}
    |l_{1\cdots i-1} - l_i|\leq l_{1\cdots i}\leq l_{1\cdots i-1} + l_i
    \,.
\end{equation}
These are precisely the constraints arising from the addition of angular momentum quantum numbers.
As already discussed, the $l_i$ quantum numbers denote the total orbital angular momentum of the leaf $i$. The inequalities just proven suggest that the quantum numbers $l_{1\cdots i}$ determine the sum of the angular momentum of the leaves $1$ through $i$. 
The fact that the system can be characterized through the quantum numbers $l_1$, $l_2$, $l_{12}$, $l_3$, $l_{123}$, $l_{1\ldots (N-1)}$, $l_N$ represents the well known fact that this set of total angular momentum operators form a complete set of commuting obserbables (CSCO).

\subsection{The Dynamics of Branches: obtaining the Hamiltonian of fully gauge fixed SU(2) HLGT}
\label{sec:Dynamics}

Having obtained the states of the fully gauge fixed Hamiltonian, we now proceed to discuss how to construct the Hamiltonian of this theory. 
We will begin by constructing the magnetic Hamiltonian, and then move on to the electric Hamiltonian. 
Both constructions however have the same two steps. 
First, the Hamiltonian has to be translated from the spatial graph to the flower. 
This has already been done in~\cite{mathur_canonical_2015,DAndrea:2023qnr}, but we will briefly recall the construction in here for the sake of completeness. 
Second, it has to be translated from the flower to the branch.
These are of course specific examples of translating a Hamiltonian through the graph coarsening procedure. 
Indeed, we provide a more general perspective towards this matter in~\cref{app:dynamics}, where we also provide rigorous proofs.
In here we restrict to citing the results as they apply to the specific case of the branches we have been discussing so far.
It may be useful for the reader to refer to~\cref{sec:examples} for explicit examples of these constructions as well.

The magnetic Hamiltonian of a spatial lattice is given by
\begin{equation}
    H_B=\sum_{\gamma}\tr U(\gamma)\,,
\end{equation}
where the sum is taken over all plaquettes of the lattice.
In practice, since the the Wilson loop operators of $\gamma$ and $\gamma^{-1}$ are related to one another, it is often useful for computations to choose a fixed orientation for each plaquette. 
In theories based on unitary structure groups they are conjugates of one another, so that one can write
\begin{equation}
    H_B=2\sum_{\gamma}\Re\tr U(\gamma)\,,
\end{equation}
where the sum is now taken only over plaquettes with a chosen orientation.
In fact, in $\SU$ the Wilson loops are real so we can skip the $\Re$ in this formula.
Let us then focus on a single plaquette $\gamma$ with a definite orientation and study the Wilson loop $\tr U(\gamma)$ supported on it.

Let us recall that in the maximal tree construction each physical link $e$ is associated to a trajectory $\gamma_e$ as shown in~\cref{fig:flower_trajectories}. 
Let us in particular focus on those physical links that touch our plaquette of interest $\gamma$. 
As it turns out, one can find a unique composition of the trajectories $\gamma_e$ associated to these physical links, which supports the same Wilson loop as $\gamma$. 
For example, the plaquette oriented clockwise on the top right of~\cref{fig:flower_trajectories} supports the same Wilson loop as the trajectory that first goes through the trajectory associated to the physical link in the top right, and then goes through the inverse of the trajectory associated to the physical link in the middle right.
More generally, if we set $\sigma_e=1$ when the plaquette goes through $e$ in the positive orientation or $\sigma_e=-1$ when it traverses it in the opposite direction, we will have
\begin{equation}
    \tr U(\gamma)=\tr U\qty(\prod_{e}\gamma_e^{\sigma_e})\,,
\end{equation}
where the product is over all the physical links which are part of the plaquette.

So far, our discussion has been restricted to the spatial lattice. The translation to the flower is simply obtained by identifying the trajectory $\gamma_e$ with its associated trajectory $(h_{\gamma_e}^{(t)},h_{\gamma_e}^{(s)})$ on the flower. So, in our ongoing example, the trajectory $\prod_{e}\gamma_e^{\sigma_e}$ would go first through the top right petal in the flower~\cref{fig:flower_complete}, and then through the lower right petal. The precise orientation in which these are traversed depends on the user's choice of which half-links are of the type $h^{(t)}_\cdot$ and which are of the type $h^{(s)}_\cdot$.

\begin{figure}
    \centering
    \subfloat[]{\label{fig:marked_trajectory_branch}\includegraphics[width = 0.4\textwidth]{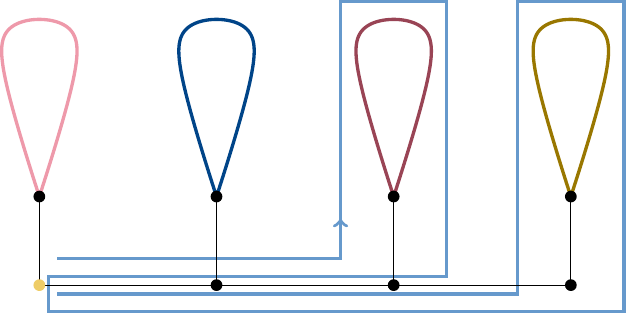}}
    \par\bigskip
    \subfloat[]{\label{fig:simplified_marked_trajectory_branch}\includegraphics[width = 0.4\textwidth]{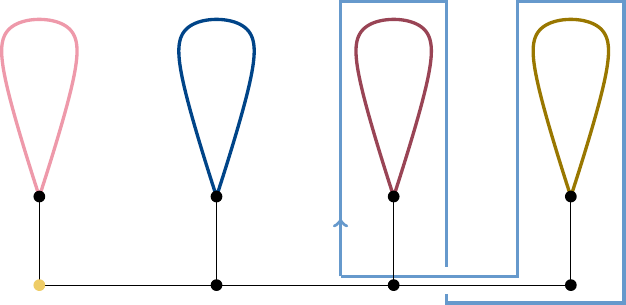}}
    \caption{We show a trajectory on the branch that corresponds to the upper right plaquette of~\cref{fig:flower_trajectories}. In here we have assumed the leaves, when studying them from left to right, correspond to the middle left, top left, top right, and middle right physical links. We are further assuming that the orientation of the leaves corresponding to the orientation of the trajectories in~\cref{fig:flower_trajectories} is from left to right. This is a convention.}
    \label{fig:magnetic_branch}
\end{figure}

Finally, we note that the trajectory $\gamma_e$, which starts and ends on the marked vertex of the flower, gets identified by the point splitting procedure to a stretched trajectory on the branch. Said trajectory starts from the marked vertex of the branch, traverses the leaf corresponding to the petal, and then comes back to the marked vertex. After this identification $\prod_{e}\gamma_e^{\sigma_e}$ becomes a trajectory on the branch. In the case of our example, this trajectory is shown in~\cref{fig:marked_trajectory_branch}. Note however, that at this stage, the Wilson lines on several sections of this trajectory cancel. Thus, one can in general replace this trajectory by a simpler trajectory $\gamma_\text{branch}$, such as the one shown in~\cref{fig:simplified_marked_trajectory_branch}. Such a trajectory loops through all the leaves associated to physical links traversed by our original plaquette $\gamma$. The precise orientation with which it does so depends on the user. In $d=2$, there can at most be two physical links traversed by each plaquette. In $d=3$, some plaquettes can also traverse four physical links.

In this way we have completed our program of transforming the Wilson loop on the original lattice to one on the branch
\begin{equation}
    \tr U(\gamma)\mapsto \tr U(\gamma_\text{branch})\,.
\end{equation}
At this stage this plaquette operator can be written in terms of LSH operators using~\cref{eq:loops_lsh_main,eq:UCornerTrajectory,eq:corner_op_to_vertex}, and further in terms of normalized raising and lowering operators using~\cref{eq:ladder_creation,eq:rest_ladder}.

Let us now consider the electric Hamiltonian.
On a spatial lattice it is given by
\begin{equation}
    H_E=\sum_{h\in H_\text{dyn}}\vb{E}(h)^2\,.
\end{equation}
In here the sum is taken over a single half-link of each of the links of the lattice. For notebooking purposes, we will call these ``dynamical'' links. At this stage which of the two half-links of a given link is chosen is immaterial for both will give the same result. However, as we will explain shortly, the maximal tree construction restricts the choice for the links on the maximal tree. The operator $\vb{E}(h)$, known as the electric operator, is the generator of gauge transformations on the half-link $h$. More details on this operator are given in~\cref{app:quantum_theory}.

The operators $\vb{E}(h)$ do not have a simple interpretation in terms of the trajectories supplied by the maximal tree construction~\cref{fig:flower_trajectories}.
However, there are closely related operators that do. 
Indeed, consider the trajectory that goes along the maximal tree from the marked vertex to the vertex at which $h$ is anchored. 
If said trajectory does not touch $h$, it can be used to parallel transport $\vb{E}(h)$ back to the marked vertex. 
Every half-link not in our maximal tree satisfies this condition.
For the links in the maximal tree, one of its two constituent half-links satifies it.
This assumption thus restricts $H_\text{dyn}$.

The resulting operator has the same square as $\vb{E}(h)$ and has a simple interpretation in terms of the trajectories supplied by the maximal tree. 
Whenever such a trajectory $\gamma$ traverses $h$ in the negative direction (meaning when it passes through the link associated with $h$, it does so in a way that $h$ is the starting half-link), it generates on it a gauge transformation equivalent to the one $h_\gamma^{(s)}$ would generate on the flower.
We will denote the set of such $\gamma$ by $R_{h,\leftarrow}$. 
Conversely, whenever $\gamma$ traverses $h$ in the negative direction, the gauge transformation generated on it is equivalent to the one generated by $h_\gamma^{(t)}$ on the flower.
Let us denote the set of such trajectories by $R_{h,\rightarrow}$. With this we conclude that the translation of the electric Hamiltonian on the spatial lattice to the flower is obtained by
\begin{equation}\label{eq:electric_hamiltonian_flower}
    \vb{E}(h)^2\mapsto\qty(\sum_{h\in R_{\gamma,\leftarrow}}\vb{E}(h_\gamma^{(s)})+\sum_{h\in R_{\gamma,\rightarrow}}\vb{E}(h_\gamma^{(t)}))^2
    \,.
\end{equation}

Now, we need to translate the electric operators on the flower to electric operators on the branch. 
After point splitting, every half-link of the flower, whether it is of the form $h_\gamma^{(s)}$ or $h_\gamma^{(t)}$ will correspond to a half-link $h$ on a loop of the branch.
The corresponding electric operator $\vb{E}(h)$ will be equivalent to its corresponding electric operator on the flower once it has been parallel transported back to the marked vertex.
For terms in \eqref{eq:electric_hamiltonian_flower} which are squares of electric operators, such a parallel transport is immaterial. For such terms we simply have
\begin{equation}
    \vb{E}(h)^2=\frac{N(h)}{2}\qty(\frac{N(h)}{2}+1),
\end{equation}
which is the casimir of the $\SU$ representation living at the half-link $h$ of the branch.

\begin{figure}
    \centering
    \includegraphics[width=0.45\textwidth]{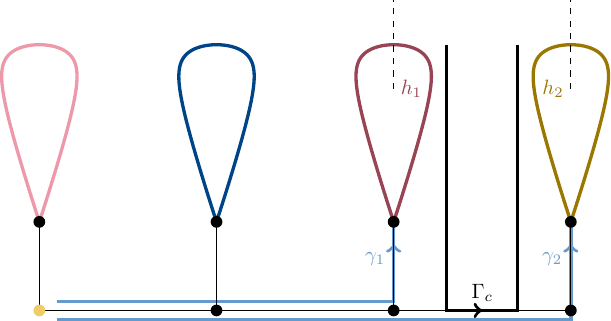}
    \caption{Two half-links $h_1$ and $h_2$ are shown with trajectories $\gamma_1$ and $\gamma_2$ which start at a marked vertex and end at the vertex where they are anchored. A corner trajectory $\Gamma_c$ joining $h_1$ and $h_2$ is also shown.}
    \label{fig:electric_corner_trajectory}
\end{figure}

The mixed inner products coming from \eqref{eq:electric_hamiltonian_flower} are less trivial. Indeed, consider 
\begin{align}
    \vb{E}(\gamma_1, h_1) \cdot \vb{E}(\gamma_2, h_2)
    \,,
\end{align}
where $\gamma_i$ denotes the path connecting the marked vertex to the half-link $h_i$, and $\vb{E}(\gamma_i, h_i)$ denotes the parallel transport of the electric operator at $h_i$ back to the marked vertex. 
As shown in~\cref{app:lsh_dof}, the overlapping contributions of $\gamma_1$ and $\gamma_2$ cancel, so that only the trajectory joining the anchors of $h_1$ and $h_2$ is needed. In fact, one can extend said trajectory to a corner trajectory $\Gamma_c$ joining these two half-links. This situation is depicted in~\cref{fig:electric_corner_trajectory}. 
As shown in~\cref{app:lsh_dof}, putting these considerations together one can show that the product of two electric operators in terms of LHS Wilson line operators can then be written as
\begin{equation}\label{eq:electric_inner_product}
    \begin{aligned}
        &\vb{E}(\gamma_1,h_1) \cdot \vb{E}(\gamma_2,h_2)_i=\\
        &U\indices{^-_-}(\Gamma_c^{-1})U\indices{^-_-}(\Gamma_c)\frac{N(h_1)}{2}(N(h_2)+1)\\
        &-\frac{N(h_1)}{2}\qty(\frac{N(h_2)}{2}+1)
        \,.
    \end{aligned}
\end{equation}
As was the case for the plaquette operator, this operator can be written in terms of LSH operators using~\cref{eq:loops_lsh_main,eq:UCornerTrajectory,eq:corner_op_to_vertex}, and further in terms of normalized raising and lowering operators using~\cref{eq:ladder_creation,eq:rest_ladder}.

At this stage we have succeeded in developing an algorithm which produces a fully gauge-fixed Hamiltonian on the branch in terms of LSH operators. In fact, said Hamiltonian is now a product of ladder operators and functions of number operators. Operating with such expressions is simple, for one has
\begin{equation}
f(N(h_1,h_2))\Lambda_q(h_1,h_2)=\Lambda_q(h_1,h_2)f(N(h_1,h_2)+q)\,.
\end{equation}
Thus, these expressions can be manipulated algebraically into a normal ordering in which all ladder operators are at the left of the expression and all functions of number operators are on the right. When this is done, the ladder operators have simple geometric interpretations in terms of creation and annihilation of loops of flux across the branch.
Fortunately, this is the kind of operations that symbolic computer algebra programs are good at doing.
Results of this are shown in the examples in~\cref{sec:examples}.

\section{Examples}
\label{sec:examples}

\subsection{One plaquette universe}\label{sec:one_plaquette}

\begin{figure}
    \centering
    \subfloat[]{\includegraphics[width=0.15\textwidth]{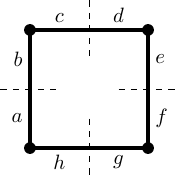}}\qquad
    \subfloat[]{\includegraphics[width=0.15\textwidth]{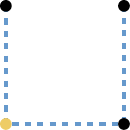}}
    \par\bigskip
    \subfloat[]{\includegraphics[width=0.12\textwidth]{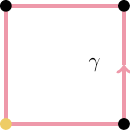}}\qquad
    \subfloat[]{\includegraphics[width=0.12\textwidth]{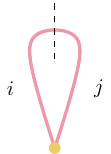}}
    \caption{A one plaquette universe. The top left corner shows the associated spatial graph and labels its eight half-links. On the top right, we make a choice of a maximal tree, which is shown in brown dashed lines, and a marked vertex, depicted in orange. On the bottom left we show the Wilson line $\gamma=(a,b)(c,d)(e,f)(g,h)$ associated to the link $(c,d)$ left unfixed by the maximal tree. On the bottom right we show the graph obtained by coarsening the original graph through $\gamma$. In this petal the link $(i,j)$ now corresponds to $\gamma$. We will orient this graph so that $(i,j)$ is positively oriented.}
    \label{fig:one_plaquette}
\end{figure}

Let us now start with the simplest system consisting of a single plaquette, like the one shown in~\cref{fig:one_plaquette}.
This system has already been explored in a mixed basis~\cite{DAndrea:2023qnr} as well as in the prepotential formulation in~\cite{mathur_canonical_2015, Raychowdhury}, but it will be useful to review it to set the stage for a second, more complicated example.
The maximal tree construction can be used to gauge fix all but one of the links in this system to the identity.
Accordingly, in this gauge the Wilson line on said link would be equivalently represented as the Wilson line on the loop $\gamma$.
Thus, we can equivalently think of this system as comprised of a single petal.

The kinematics of this problem are completely fixed by noting that there is a single LSH creation operator up to orientation
\begin{equation}
    \mathcal{L}_{++}(i,j)\equiv\mathcal{L}_{++}\,.
\end{equation}
Therefore, all of the states are of the form
\begin{equation}
    \ket{n}=\Lambda_+^n \ket{n} = \frac{1}{\sqrt{n!(n+1)!}}(\mathcal{L}_{++})^n\ket{0}\,.
\end{equation}
This vectors have unit norm when supplemented by this normalization, as explained in~\cref{sec:lsh_construction}, which agrees with~\cite{Raychowdhury:2019iki}.

Now, let us construct the electric Hamiltonian following \ref{sec:Dynamics}. We note that~\cref{fig:one_plaquette} has eight half-links. 
Of these, $a$, $c$, $d$, $f$, and $h$ are potential candidates for dynamical in our formulation (in~\cref{sec:Dynamics} and~\cref{app:dynamics} we explain that $b$, $e$ and $g$ are not good candidates because the path along the maximal tree connecting them to the marked vertex goes through them). 
We can then choose
\begin{equation}
    H_\text{dyn}=\{a,c,f,h\}\,.
\end{equation}
We then have
\begin{equation}
    R_{a,\leftarrow}=R_{c,\leftarrow}=R_{f,\rightarrow}=R_{h,\rightarrow}=\{\gamma\}\,,
\end{equation}
with the rest of the sets empty.
We conclude that all of the squares of the electric operators on the original half-edges can be replaced by $\vb{E}(i)^2=\vb{E}(j)^2$ and
\begin{equation}
    \begin{aligned}
        &H_E = 4\vb{E}(i)^2= 2N\qty(\frac{N}{2}+1)\,,
    \end{aligned}
\end{equation}
with 
\begin{equation}
    N=N(j)=N(i)
\end{equation}
the number operator.

For the magnetic Hamiltonian, we have
\begin{equation}
    H_B=2\Re\tr U(\gamma)\,,
\end{equation}
with \eqref{eq:loops_lsh_main}
\begin{equation}
    \begin{aligned}
        \tr U(\gamma)=t\indices{^a_b}(i,j)U\indices{^b_a}(j,i)\,.
    \end{aligned}
\end{equation}
A quick computation then shows that
\begin{equation}
    \tr U(\gamma)=-\Lambda_+-\Lambda_-\,.
\end{equation}

\begin{figure}
    \centering
    \includegraphics[width=0.4\textwidth]{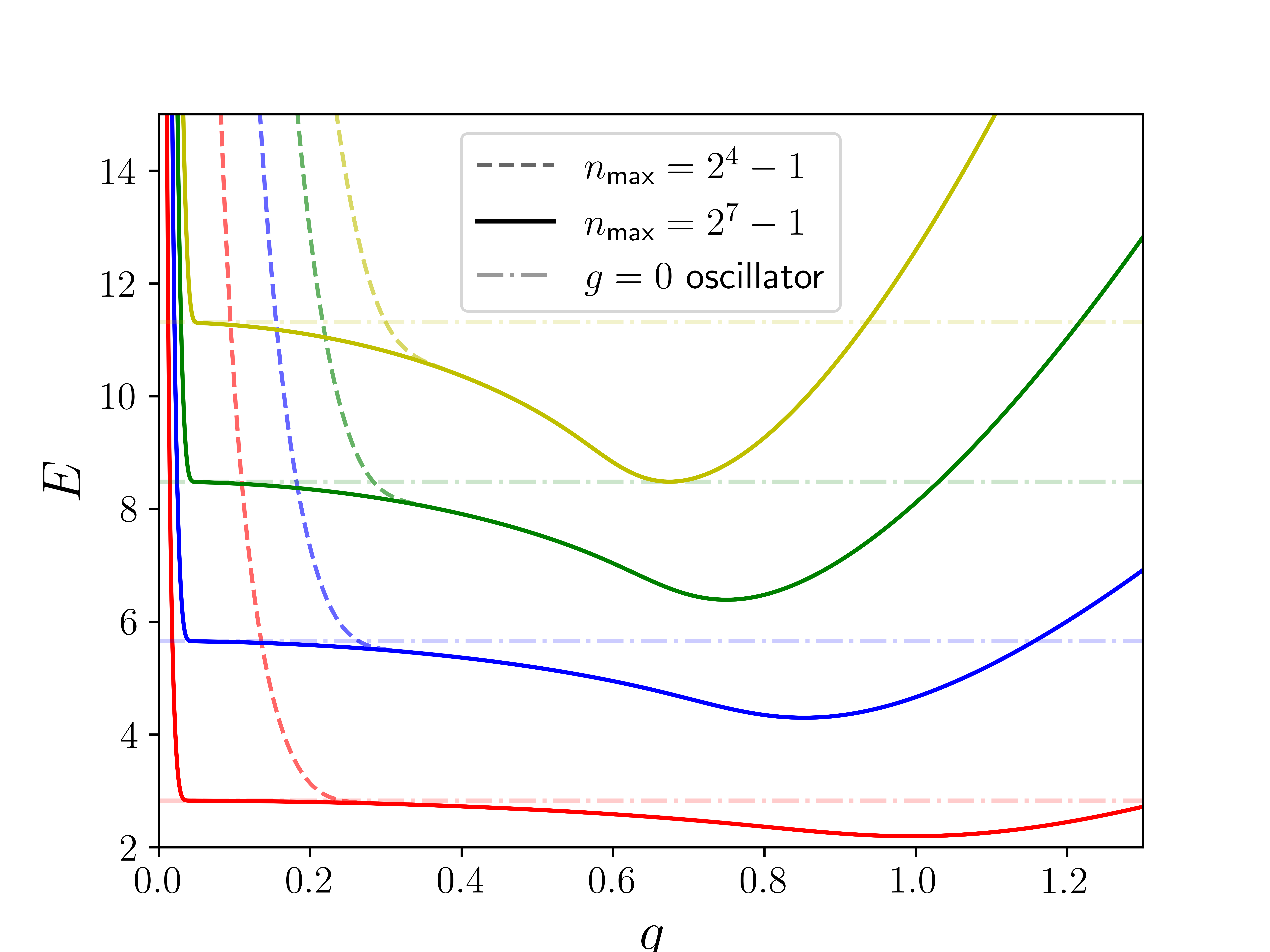}
    \caption{Energy levels for the one plaquette system.}
    \label{fig:energy_levels}
\end{figure}

The LSH states are then clearly eigenstates of the electric part
\begin{equation}
    H_E =\sum_{n=0}^\infty 2n\qty(\frac{n}{2} + 1)\op{n}\,,
\end{equation}
while for the magnetic Hamiltonian we have
\begin{equation}
    H_B=-2\sum_{n=0}^\infty (\op{n}{n+1}+\op{n+1}{n})\,.
\end{equation}
The energy levels can then be recovered by cutting off the sums at some $n_\text{max}$. The results from this are shown in~\cref{fig:energy_levels}.

It is quite interesting to note that this Hamiltonian has an interpretation in terms of a quantum random walk.
Indeed, consider a particle on a one dimensional half-infinite lattice indexed by $n$.
Thus, in this interpretation, $N$ is the position operator in lattice units.
As shown in~\cite{silva_two-particle_2024}, the kinetic Hamiltonian for such a particle is given by $H_B/(2g^2)$ up to an additive constant.
In this interpretation $g/\sqrt{2}$ is the lattice spacing on this walk. 
$g^2H_E/2$, being diagonal in position space, is the potential Hamiltonian.
Defining the position variable $x=gn/\sqrt{2}$, it corresponds to a potential
\begin{equation}
    V(x)=\sqrt{2}x\qty(\frac{x}{\sqrt{2}}+g)=\qty(x+\frac{g}{\sqrt{2}})^2-\frac{g^2}{2}\qc x\geq 0\,.
\end{equation}

In particular, this leads to an interesting way of understanding the $g\rightarrow 0$ limit.
In this limit the lattice becomes continuous and we are left with the theory of a quantum particle moving on a continuous line in a displaced harmonic oscillator potential for $x\geq 0$
\begin{equation}
    H=\frac{P^2}{2}+X^2\,,
\end{equation}
and an infinite wall at $x=0$. 
The eigenstates of this system are the antisymmetric states of this oscillator which has frequency $\sqrt{2}$.
Therefore we recover the result that the energy levels of this system, when measured with respect to the ground state are
\begin{equation}
    \sqrt{2}\qty(k+\frac{1}{2})-\sqrt{2}\frac{3}{2}=\sqrt{2}\qty(k-1)\qc k=1,3,5,\dots\,.
\end{equation}

Let us end by noting that the magnetic expression for the electric Hamiltonian obtained in~\cite{DAndrea:2023qnr} and the fact that our states are eigenstates of said Hamiltonian by construction, implies that the special function \eqref{eq:petal_function} satisfies \eqref{eq:differential_equation} for $l=0$.

\subsection{Two plaquette system}\label{sec:two_plaquette}

\begin{figure}[h!]
    \centering
    \subfloat[]{\includegraphics[width = 0.3\textwidth]{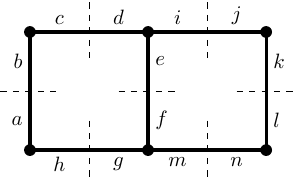}}
    \par
    \subfloat[]{\includegraphics[width = 0.25\textwidth]{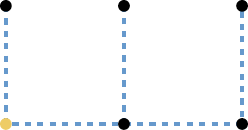}}
    \par\bigskip
    \subfloat[]{\includegraphics[width = 0.25\textwidth]{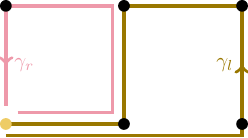}}
    \par
    \subfloat[]{\includegraphics[width = 0.25\textwidth]{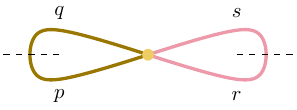}}\par
    \subfloat[]{\includegraphics[width = 0.35\textwidth]{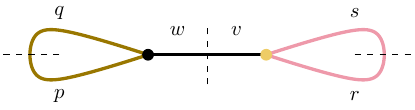}}
    \caption{A two plaquette universe. The top figure (a) shows the associated spatial graph and labels its fourteen half-links. Below this, we show in (b) a selection of a maximal tree along with a marked vertex. Below this in (c), we show trajectories $\gamma_r$ and $\gamma_l$ corresponding to the physical links $(c,d)$ and $(i,j)$, respectively. Below this in (d), we show on the left the associated flower. In it the links $(p,q)$ and $(r,s)$ correspond to $\gamma_l$ and $\gamma_r$ respectively. Finally in (d), on the right we consider the point split version of this flower leading to our branch.}
    \label{fig:two_plaquette}
\end{figure}

Now let us consider the two plaquette system shown in figure~\cref{fig:two_plaquette}. The three creation operators satisfying the Abelian Gauss law are
\begin{equation}
    \begin{aligned}
        \ell_{++}(l)&\coloneq \mathcal{L}_{++}(p,q),\\
        \ell_{++}(r)&\coloneq \mathcal{L}_{++}(r,s)\qand\\
        \mathcal{L}_{++}&\coloneq \mathcal{L}_{++}(p,w)\mathcal{L}_{++}(q,w)\mathcal{L}_{++}(r,v)\mathcal{L}_{++}(s,v)\,.
    \end{aligned}
\end{equation}
Associated to these are their normalized versions $\lambda_+(l)$, $\lambda_+(r)$, and $\Lambda_+$, respectively, and the Hilbert space of the system is given by
\begin{equation}
    \ket{n_l,n_r,l}=\lambda_{+}(l)^{n_l}\lambda_{+}(r)^{n_r}\Lambda_{+}^l\ket{0}\,.
\end{equation}
Therefore $n_l$ and $n_r$ counts the number of fluxes on their respective loops, while $l$ counts the number of fluxes flowing across the whole branch. We will denote the corresponding number operators by $N_l$, $N_r$ and $L$ .

For the magnetic operator, we note that one of the plaquettes is $\gamma_r$ while the other has a Wilson loop equivalent to that of $\gamma_l$. Furthermore, when translated to the branch, the $(w,v)$ contributions of the latter vanish. With these in mind, we conclude that 
\begin{equation}
    H_B=2\Re\qty(\tr U(l) + \tr U(r))\,,
\end{equation}
where
\begin{equation}
    U(l)=U(p,q)\qand U(r)=U(r,s)\,.
\end{equation}
Explicitly, we have
\begin{equation}
    \begin{aligned}
        \tr U(l)&=-\lambda_-(l)\sqrt{\frac{N_l(2L+N_l+1)}{(L+N_l+1)(L+N_l)}}\\
        &\hphantom{{}={}}-\lambda_+(l)\sqrt{\frac{(N_l+1)(2L+N_l+2)}{(L+N_l+2)(L+N_l+1)}}\,.
    \end{aligned}
\end{equation}
The formula for $\tr U(r)$ is the same under the exchange $u\mapsto d$.

We can verify this result by comparing to the ones found in ~\cite{mathur_canonical_2015,DAndrea:2023qnr}. Indeed, we can use the wave functions~\eqref{eq:wave_function} to show that
\begin{equation}
    \mel{\omega_l,\hat{n}_l}{\tr U(l)}{n_l,l,m}=2\cos(\omega/2)\ip{\omega_l,\hat{n}_l}{n_l,l,m}\,,
\end{equation}
numerically.
In this computation the degrees of freedom on the right petal are irrelevant and $m$ acts as a place holder for them.

For the electric part of the Hamiltonian, we need an admissible half-link from each link in the graph.
We choose
\begin{equation}
    H_\text{dyn}=\{a, c, i, h, m, f, l\}\,.
\end{equation}
The non-empty sets associated to each of these are
\begin{alignat}{2}
        R_{a, \rightarrow}=\{\gamma_r\}\qc & R_{c,\rightarrow}=\{\gamma_r\}\qc && R_{i, \rightarrow}=\{\gamma_l\}\,,\nonumber\\
        R_{l,\leftarrow}=\{\gamma_l\}\qc & R_{m,\leftarrow}=\{\gamma_l\}\qc && R_{f,\leftarrow}=\{\gamma_r\}\,,\\
        R_{f,\rightarrow}=\{\gamma_l\}\qc & R_{h, \rightarrow}=\{\gamma_l\}\qc &&
        R_{h,\leftarrow}=\{\gamma_l,\gamma_r\}\nonumber
        \,.
\end{alignat}
We also identify the links on the flower as 
\begin{align}
    h_{\gamma_r}^{(t)} = r\,, \quad h_{\gamma_r}^{(s)} = s\,, \quad h_{\gamma_l}^{(t)} = p\,, \quad h_{\gamma_l}^{(s)} = q
    \,.
\end{align}

Therefore, the electric Hamiltonian is given by
\begin{equation}
    H_E=\sum_{h\in H_\text{dyn}}\vb{E}(h)^2\,,
\end{equation}
with
\begin{equation}
    \begin{aligned}
        \vb{E}(a)^2&=\vb{E}(c)^2
        =\vb{E}(r)^2\,,\\
        \vb{E}(i)^2&=\vb{E}(p)^2\,,\\
        \vb{E}(l)^2&=\vb{E}(m)^2=\vb{E}(q)^2\,,\\
        \vb{E}(f)^2&=\qty(\vb{E}(s)+\vb{E}(p))^2\,,\\
        \vb{E}(h)^2&=\qty(\vb{E}(q)+\vb{E}(s)+\vb{E}(p))^2\,.
    \end{aligned}
\end{equation}
In these formulae we have suppressed the parallel transport on the left-hand side for simplicity of reading.
We can therefore write
\begin{align}
    H_E=&\,2\left(\vb{E}(r)^2+\vb{E}(s)^2\right)+3\left(\vb{E}(p)^2+\vb{E}(q)^2\right)\\
    &+2 \left(2\vb{E}(p)\cdot\vb{E}(s)+\vb{E}(q)\cdot\vb{E}(s)+\vb{E}(p)\cdot\vb{E}(q)\right)\nonumber
    \,.
\end{align}

The terms corresponding to the square of single electric operators are given by
\begin{align}
    \vb{E}(r)^2&=\vb{E}(s)^2
    =\frac{N_r+L}{2}\qty(\frac{N_r+L}{2} + 1)\,,\\
    \vb{E}(p)^2&=\vb{E}(q)^2=\frac{N_l+L}{2}\qty(\frac{N_l+L}{2} + 1)
    \,.
\end{align}
Note that there is an asymmetry in the coefficients of the terms involving the petal involving half-links $p$ and $q$ and the petal involving half-links $r$ and $s$. 
However, including the inner product of electric operators on a single petal 
\begin{equation}
    \vb{E}(p)\cdot\vb{E}(q)=\frac{L}{2}(L+1)-\frac{N_l+L}{2}\qty(\frac{N_l+L}{2}+1)
    \,,
\end{equation}
there is a manifest $u\leftrightarrow d$ symmetry. 
Furthermore, all electric contributions for which we have given explicit LSH formulae thus far are functions of number operators, and are therefore diagonal on the LSH basis.

This is in contrast with combination of electric field operators on the two different petals. 
These terms include a parallel transport to ensure gauge invariance. 
We will suppress this dependence again for notational convenience. 
But once it has been taken into account, one obtains
\begin{widetext}
\begin{align}
    \vb{E}(p)\cdot\vb{E}(s)&=-\frac{L}{4}(L+1)+\lambda_{-}(l)\lambda_{-}(r)\Lambda_+\frac{L+1}{4}\sqrt{\frac{N_lN_r(N_l+2L+2)(N_r+2L+2)}{4(L+1)^2-1}}\\
    &\hphantom{{}={}}+\lambda_+(l)\lambda_+(r)\Lambda_-\frac{L}{4}\sqrt{\frac{(N_l+1)(N_r+1)(N_l+2L+1)(N_r+2L+1)}{4L^2-1}}\nonumber\\
    \vb{E}(q)\cdot\vb{E}(s)&=-\frac{L}{4}(L+1)-\lambda_{-}(l)\lambda_{-}(r)\Lambda_+\frac{L+1}{4}\sqrt{\frac{N_lN_r(N_l+2L+2)(N_r+2L+2)}{4(L+1)^2-1}}\\
    &\hphantom{{}={}}-\lambda_+(l)\lambda_+(r)\Lambda_-\frac{L}{4}\sqrt{\frac{(N_l+1)(N_r+1)(N_l+2L+1)(N_r+2L+1)}{4L^2-1}}\nonumber\,.
\end{align}
\end{widetext}
Our LSH basis is no longer comprised eigenstates of for these terms in the Hamiltonian. 
While they conserve the overall loops running through each leaf, such that $n+l$ is preserved, they can move internal loops to the external loop and vice versa.
Putting all this information together, one finds the electric Hamiltonian of the 2 plaquette system
\begin{widetext}
    \begin{equation}
        \begin{aligned}
            H_E&=2(N_r+L)\qty(\frac{N_r+L}{2} + 1)+2(N_l+L)\qty(\frac{N_l+L}{2} + 1)-\frac{L}{2}(L+1)\\
            &\hphantom{{}={}}+\lambda_{-}(l)\lambda_{-}(r)\Lambda_+\frac{L+1}{2}\sqrt{\frac{N_lN_r(N_l+2L+2)(N_r+2L+2)}{4(L+1)^2-1}}\\
            &\hphantom{{}={}}+\lambda_+(l)\lambda_+(r)\Lambda_-\frac{L}{2}\sqrt{\frac{(N_l+1)(N_r+1)(N_l+2L+1)(N_r+2L+1)}{4L^2-1}}\,.
        \end{aligned}
    \end{equation}
\end{widetext}

Let us finish by interpreting the first two terms of this Hamiltonian as the electric Hamiltonian of the two plaquette system in a universe where they are non-interacting. This can be used to study the system with one plaquette away from the physical states with $L=0$. In particular, when considered in conjunction with the results of~\cite{DAndrea:2023qnr}, this shows~\eqref{eq:differential_equation}. 

\section{Conclusions and Outlook}
\label{sec:Conclusions}

In this work we developed a formulation of gauge theories on general graphs composed of half-links, which can be assembled into a general arrangement of vertices connected by links.
We developed the general theory of graph coarsening, of which point splitting and the maximal tree construction are examples of, and showed how this can be used to relate theories on different graphs. 
This was used in the main text to understand a formulation of maximal tree gauge fixing in terms of a flower shaped graph, as well as the point splitting of said graph into a branch shaped graph.
The latter consisted only of trivalent vertices, which made it amenable to treatment using gauge LSH operators. It further contained a single loop for every physical link after the gauge fixing.

We further extended the gauge LSH formulation to general graphs, inspired by a detailed understanding of its group-theoretic properties. This allowed us to obtain a fully gauge fixed formulation of the theory on the branch providing a basis that is digitized by nature. The relationship between the resulting basis of the Hilbert space, which is already digitized by nature, with the magnetic basis explored in \cite{DAndrea:2023qnr} was made explicit. The results found in the magnetic basis then show that in the small coupling limit the degrees of freedom on the loops of the branch are the relevant ones. This suggests a decoupling of the loops from the rest of the branch which would give a systematic method of performing numerical explorations of the low coupling regime on quantum computers. Future work will be done along these lines.

This work further made explicit the description of the dynamics of $\SU$ Yang-Mills theory in the maximal tree gauge using LSH operators. A new ingredient was an LSH expression for inner products of electric operators probing different sectors of a graph. We showed that terms of this form, along with all of the other terms in the Hamiltonian of the theory can be expressed in terms of LSH Wilson line operators defined on corner trajectories.

Finally, the algorithm to construct fully gauge fixed Hamiltonians in this formulation was described. This algorithm was put in place in the explicit examples of one and two plaquette systems.

{\bf Note:} A paper on a related topic appeared recently in~\cite{Grabowska:2024emw}.

\section{Acknowledgements}
The authors would like to thank Dorota Grabowska, Christopher Kane and Jesse Stryker for valuable discussions. 
IMB would particularly like to extend gratitude towards Jesse for help in understanding computational aspects of the LSH formalism.
This work was supported by the DOE, Office of Science under contract DE-AC02-05CH11231, partially through Quantum Information Science Enabled Discovery (QuantISED) for High Energy Physics (KA2401032).
This work was partly supported by the Lawrence Berkeley National Lab LDRD project No. LDRD25-131.
%%%%%%%%%%%%%%%%%%%%%%%%%%%%%%%%%%%%%%%%%%%%%%%%%%%%%%%%%%%%
\bibliography{references.bib}
\bibliographystyle{utphys}
%%%%%%%%%%%%%%%%%%%%%%%%%%%%%%%%%%%%%%%%%%%%%%%%%%%%%%%%%%%%
\onecolumngrid
\newpage
\large
\appendix

The following appendices are meant to complement the main body of the paper in two ways.
On one hand, a complete understanding of our results requires a treatment of HLGTs on general lattices.
Furthermore, such treatment is made easier by introducing notation and conventions which are not standard in the HLGT community at the moment.
These appendices therefore provide a general introduction to this field that takes advantage of this.
On the other, we have relegated some of the more involved calculations performed in the main body of the paper to the appendices, so as to not deviate the reader's attention from the main conceptual results.

\section{Notation and Conventions}

Let us start with a reference guide for the reader on the notation and conventions we will use in this appendix. 

Given a set $A$, we say an element $a$ is in $A$ with the notation $a\in A$. 
If we introduce another set $B$, we can then construct the cartesian product $A\times B$ which is the set of all (ordered) tuples $(a,b)$ where $a\in A$ and $b\in B$. 
On the other hand, the union of two sets is the set $A\cup B$ containing all of the elements from both $A$ and $B$. 
If we want to emphasize that $A$ and $B$ have no elements in common, we usually write the union as $A\sqcup B$.

A function $f:A\rightarrow B$ is a relation that assigns to every $a\in A$ a \textit{unique} element $f(a)\in B$.
When we want to make the rule of the function explicit, we will use the notation
\begin{equation}
    \begin{aligned}
        f:A&\rightarrow B\\
        a&\mapsto f(a)
        \,.
    \end{aligned}
\end{equation}
We denote the set of all such functions by $A^B$.
This notation is inspired by the following remark.
The reader might be familiar with the set $\mathbb{R}^3=\mathbb{R}\times\mathbb{R}\times\mathbb{R}$ of all triples $x=(x_1,x_2,x_3)$ of real numbers.
But a slight switch of notation to $(x(1),x(2),x(3))$ makes it clear that the triple $x$ can equivalently be thought of as a map $x:\{1,2,3\}\rightarrow\mathbb{R}$, i.e. an element $x\in\mathbb{R}^3$ is equivalent to an element $x\in\mathbb{R}^{\{1,2,3\}}$.
This notation is therefore an efficient way of determining a set of functions.
Its usefulness in LGT stems from the fact that the notion often we are dealing with spaces of functions on which no continuity criterion has to be imposed given the discreteness of graphs (these functions are of course the fields).
Any continuity properties are instead achieved dynamically (or not) in the continuum limit.

A function $f:A\rightarrow B$ is said to be injective if every two $a,b\in A$ that are different $a\neq b$, get mapped to two different elements $f(a),f(b)\in B$. Sometimes this is called one-to-one, although a better option is probably two-to-two. A function is said to be surjective if every element $b\in B$ has at least one element $a\in A$ which gets mapped into it $f(a)=b$. 

Given two vector spaces $V$ and $W$, a linear map is a function $T:V\rightarrow W$ which respects the addition and scalar multiplication of these spaces. When dealing with such maps, we will often denote $T(v)$ simply by $Tv$. We define the subspace
\begin{equation}
    \ker T\coloneq \Set{v\in V|Tv=0}
\end{equation}
of $V$ which corresponds to the set of vectors annihilated by $T$. For linear maps, being injective is equivalent to having $\ker T=\{0\}$. We will also require the direct sum $V\oplus W$, which is the vector space of vectors uniquely determined by expressions of the form $v+w$ with $v\in V$ and $w\in W$. Loosely speaking, this is the vector space containing both $V$ and $W$ directions. 
More formally, this is the space $V\times W$ equipped with the appropriate vector space operations suggested by the notation.
In fact, for our purposes we will mainly be interested in inner product spaces. 
For these $V\oplus W$ is the space in which $V$ and $W$ sit orthogonally with respect to one another.
Finally, we also define the dual space $V^*$ to $V$ to be the set of all linear maps $V\rightarrow\mathbb{C}$.

Another notion we will sometimes use is that of a quotient.
Whenever there is a relation $\sim$ on a set $A$, we can construct the quotient set $\faktor{A}{\sim}$. 
Informally, this set can be thought of as a set with ``the same elements as A,'' with the modification that whenever $a,b\in A$ are such that $a\sim b$, when considered as elements of $\faktor{A}{\sim}$, then $a=b$. 
The primary example where we use this is in the definition of the set of links introduced in~\cref{app:graph_theory}.
The set of half-links $H$ has a relation implemented by a function $\sigma:H\rightarrow H$, which assigns to every half-link $h\in H$ the other half-link $\sigma(h)\in H$ to which it is attached. 
In this relation then $h\sim\sigma(h)$. 
An undirected link $e$ is of course determined completely either by $h$ or $\sigma(h)$, once the function $\sigma$ is known.
Therefore an undirected link can be thought of as an element of $\faktor{H}{\sim}$, which we more commonly denote as $\faktor{H}{\sigma}$ to remember that the relation is equivalent to the determination of $\sigma$.

Given functions $f:X\rightarrow Z$ and $g:Y\rightarrow Z$, we can define the set-theoretic pull back
\begin{equation}
    X\times_{f,Z,g}Y=\Set{(x,y)\in X\times Y|f(x)=g(y)}\,.
\end{equation}
The graph of $f$ is an example of such a pullback obtained when $Z=Y$ and $g$ is the identity function on $Y$
\begin{equation}
    \begin{aligned}
        g:Y\rightarrow Y\qc
        y\mapsto y\,.
    \end{aligned}
\end{equation}
In this case we simply denote the pullback by $X\times_f Y$. A particular example will be the set of oriented links, for which this concept gives the convenient notation $H\times_\sigma H$.

Given a group $G$ and a set $X$, an action of the group on the set is a map
\begin{equation}
    \begin{aligned}
        \triangleright:G\times X\rightarrow X\qc (g,x)\mapsto g\triangleright x
        \,,
    \end{aligned}
\end{equation}
which respects the group multiplication
\begin{equation}
    g_1\triangleright(g_2\triangleright x)=g_1g_2\triangleright x
    \,.
\end{equation}
Many group actions will appear throughout these appendices, which we will distinguish by adorning the $\triangleright$ symbol.

\renewcommand{\arraystretch}{1.5}
\setlength{\tabcolsep}{5pt}
\begin{table}[]
    \centering
        \begin{tabular}{|c|p{0.8\textwidth}|}
        \hline
            $\G$ & Graph.\\
            $H$ & Set of half-links.\\
            $h$ & Generic half-link. For explicit examples other latin letters are sometimes used.\\
            $\Gamma$ & Set of vertices.\\
            $p$ & Generic vertex. For explicit examples other latin letters are sometimes used.\\
            $\pi$ & Map that anchors every half-link to its associated vertex.\\
            $\sigma$ & Map that attaches every half-link to its associated vertex.\\
            $\faktor{H}{\sigma}$ & Set of undirected links.\\
            $H\times_\sigma H$ & Set of directed links.\\
            $e=(h_1,h_2)$ & Generic link going from $h_2$ to $h_1$.\\
            $\gamma$ & Generic trajectory. Links can be thought of as special trajectories.\\
            $\gamma_2\gamma_1$ & Trajectory that goes first through $\gamma_1$ and then $\gamma_2$. We maintain the convention of reading orientation of objects from right to left throughout this paper.\\
            $s(\gamma)$ & Starting point of $\gamma$.\\
            $t(\gamma)$ & Target point of $\gamma$.\\
            $E$ & Oriented set of links.\\
            $C$ & Set of corners of the graph.\\
            $\mathcal{C}$ & Oriented set of corners.\\
            $G$ & Structure group. For this paper, mostly $G=\SU$.\\
            $u$ & Classical configuration of Wilson lines.\\
            $G^\Gamma$ & Group of gauge transformations at all vertices of the graph - the gauge group.\\
            $g$ & Generic element of the structure or gauge groups.\\
            $\triangleright_h$ & Action of the structure group at a single half-link $h$.\\
            $R_h$ & Lift to the quantum setting of the action of the structure group at a single half-link $h$. Its generator is the electric operator $E(h)$.\\
            $a,b,c,\dots$ & Used both for $\mathrm{U}(1)$ indices associated to the Abelian Gauss law, as well as sometimes to denote half-links in explicit graphs.\\
            $A,B,C,\dots$ & $\SU$ indices\\
            $U\indices{^A_B}(h_1,h_2)$ & Wilson line operator on the link $(h_1,h_2)$.\\
            $U\indices{^a_b}(h_1,h_2)$ & LSH Wilson line operator on the corner $(h_1,h_2)$.\\
            $R$ & Set of trajectories used when coarsening a graph. Also used as subscript to denote objects after coarsening.\\
            $h_\gamma^{(s)}$, $h_\gamma^{(t)}$ & Pair of half-links corresponding to $\gamma\leftrightarrow (h_\gamma^{(t)},h_\gamma^{(s)})$ in the coarsened graph $\G_R$.\\
            $S$ & Surjection associating a field configuration on the coarsened graph $\G_R$ to each field configuration on the original graph $\G$\\
        \hline
        \end{tabular}
    \caption{Some notation commonly used throughout the appendix.}
    \label{tab:my_label}
\end{table}

\section{Hamiltonian Lattice Gauge Theory on General Graphs}

Let us now delve into the extension of the framework of lattice Hamiltonian Yang-Mills theory, as developed by Kogut and Susskind (KS)~\cite{Kogut1975}, to general graphs.

\subsection{Graph theory}\label{app:graph_theory}

\begin{figure}
    \centering
    \subfloat[]{\includegraphics[width = 0.3\textwidth]{figures/edge}}\qquad
    \subfloat[]{\includegraphics[width = 0.2\textwidth]{figures/loop}}\qquad
    \subfloat[]{\includegraphics[width = 0.3\textwidth]{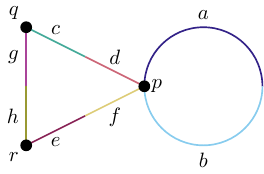}}
    
    \caption{On the left there is the graph determined by $\Gamma=\{p_1,p_2\}$, $H=\{h_1,h_2\}$, $\pi(h_1)=p_1$, $\pi(h_2)=p_2$, and $\sigma(h_1)=h_2$.
    On the middle we have a graph determined by taking $p_1=p_2=p$.
    These are the only two graphs with a single undirected link.
    On the right we have the graph $\Gamma=\{p,q,r\}$, $H=\{a,b,c,d,e,f,g,h\}$, $\pi(a)=\pi(b)=\pi(d)=\pi(f)=p$, $\pi(c)=\pi(g)=q$, $\pi(h)=\pi(e)=r$, $\sigma(a)=b$, $\sigma(c)=d$, $\sigma(e)=f$, and $\sigma(g)=h$.
    The set of undirected links of the latter is $\faktor{H}{\sigma}=\{\{a,b\},\{c,d\},\{e,f\},\{g,h\}\}$.
    The set of links on the other hand is $H\times_\sigma H=\{(a,b),(b,a),(c,d),(d,c),(e,f),(f,e),(g,h),(h,g)\}$.
    At the vertex $q$ the set of corners is $\{(g,g),(c,c),(g,c),(c,g)\}$.
    At the vertex $p$ there are $4+(3!!\times 2)=10$ corners.}
    \label{fig:graph_examples}
\end{figure}

We define a graph $\mathrm{G}$ through~\cite{Costello2011} a finite set of vertices $\Gamma$ and a set of half-links (also known as half-\textbf{e}dges in the mathematical literature) $H$, equipped with an anchoring map $\pi:H\rightarrow \Gamma$ and an attaching map $\sigma:H\rightarrow H$.
\cref{fig:graph_examples} shows a few examples that illustrate the definitions used.
The anchoring map anchors each half-link to its corresponding vertex. 
We will demand that it be surjective, meaning that there are no vertices without half-links attached to it. 
The attachment map attaches each half-link to its partner, with which it will form a full link. 
Accordingly, we will ask that for each $h\in H$ we have $\sigma(\sigma(h))=h$, i.e. $\sigma$ is an involution. 
Also, in the theory of Feynman diagrams one is often interested in graphs with external links. 
Such links can be characterized by half-links $h\in H$ which are their own partner $\sigma(h)=h$. 
For our purposes we will not allow this possibility. 

We will define an undirected link to be a set of the form $\{h,\sigma(h)\}\subseteq H$. 
We will denote the set of all undirected links by $\faktor{H}{\sigma}$. 
By a directed link, we will refer to an ordering of such an orbit $e=(h,\sigma(h))\in H\times H$, representing the link going from $\sigma(h)$ to $h$.
Throughout this paper, whenever we refer to a link without any additional qualification, the reader should assume we are referring to a directed link.
The set of all links will be denoted $H\times_\sigma H$, where the notation used highlights the fact that the set of links is simply the graph of $\sigma$. 
This set is equipped with maps that invert the direction of the link
\begin{align}
        H\times_\sigma H\rightarrow H\times_\sigma H
        \,,\qquad 
        e=(h_1,h_2)\mapsto e^{-1}\coloneq (h_2,h_1)
        \,,
\end{align}
determine its source
\begin{align}
        s:H\times_\sigma H\rightarrow \Gamma
        \,,\qquad
        e=(h_1,h_2)\mapsto s(e)=\pi(h_2)
        \,,
\end{align}
and determine its target
\begin{align}
        t:H\times_\sigma H\rightarrow \Gamma\,, \qquad 
        e=(h_1,h_2)\mapsto t(e)=\pi(h_1)
        \,.
\end{align}
Finally, a corner is a pair of half-links $(h_1,h_2)\in H$ that are anchored at the same vertex $\pi(h_1)=\pi(h_2)$. Let us denote the set of corners by $C$.

\begin{figure}
    \centering
    \subfloat[]{\includegraphics[width = 0.45\textwidth]{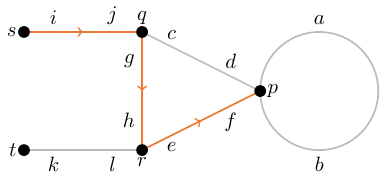}}\qquad
    \subfloat[]{\includegraphics[width = 0.45\textwidth]{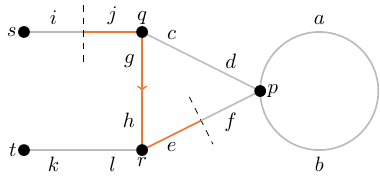}}
    \caption{On the left we show the trajectory $\gamma\in T_3$ of length 3 given by $\gamma(1)=(j,i)$, $\gamma(2)=(h,g)$, and $\gamma(3)=(f,e)$. We have $s(\gamma)=s$ and $t(\gamma)=p$. The inverse trajectory would then be given by $\gamma^{-1}(1)=(e,f)$, $\gamma^{-1}(2)=(g,h)$, and $\gamma^{-1}(3)=(i,j)$. We also have a corner trajectory $\gamma_c$ on the right given by $\gamma_c(1)=(g,j)$, $\gamma_c(2)=(e,h)$. It follows the same path as $\gamma$ except that it doesn't go through $i$ or $f$.} 
    \label{fig:trajectory_example}
\end{figure}

The set of links naturally extends to the set of trajectories $H\times_\sigma H\subseteq T$ (see~\cref{fig:trajectory_example} for an example). 
The set $T_n$ of trajectories of length $n$ is the set of maps $\gamma:\{1,\dots,n\}\rightarrow H\times_\sigma H$ such that for all $i\in\{1,\dots,n-1\}$ we have $t(\gamma(i))=s(\gamma(i+1))$. 
Each link can be identified with a trajectory of length 1, and we can therefore identify $H\times_\sigma H = T_1$. The union of all possible paths will be called $T$
\begin{equation}
    T=\bigsqcup_{n=1}^\infty T_n
    \,.
\end{equation}
Indeed, the inversion, source and target maps extend to all of $T$
\begin{align}
    T_n\rightarrow T_n\,,\qquad &\gamma\mapsto \gamma^{-1}:i\mapsto \gamma^{-1}(i)\coloneq \gamma(n-i+1)^{-1},\\
    s,t:T_n\rightarrow \Gamma\,,\qquad
    &\gamma\mapsto s(\gamma)\coloneq s(\gamma(1))\qand t(\gamma)\coloneq t(\gamma(n))
    \,.
\end{align}
Whenever we have two trajectories $\gamma_1\in T_n$ and $\gamma_2\in T_m$ satisfying $t(\gamma_1)=s(\gamma_2)$, i. e. the starting point of the second path is equal to the end point of the first path, they can be concatenated to a new trajectory $\gamma_2\gamma_1\in T_{n+m}$ that first goes through $\gamma_1$ and then through $\gamma_2$
\begin{equation}
    \gamma_2\gamma_1(i)=\begin{cases}
        \gamma_1(i) & i\in\{1,\dots,n\}\\
        \gamma_2(i-n) & i\in\{n+1,\dots,n+m\}
        \,,
    \end{cases}
\end{equation}
We will further call a trajectory $\gamma\in T_n$ closed if it starts and ends at the same point $s(\gamma(1))=t(\gamma(n))$. 

Much like trajectories are continuous sequences of links, we can also define corner trajectories $\gamma:\{1,\dots,n\}\rightarrow C$ which are continuous sequences of corners. 
An example of this is also shown in~\cref{fig:trajectory_example}.
Much like trajectories, corner trajectories can also be inverted, composed, and have source and target maps. 
The first two are defined by the same equations as for trajectories. 
The source and target are instead defined by the equations
\begin{equation}
    \gamma(1)=(\sigma(s(\gamma)),s(\gamma))\qand \gamma(n)=(t(\gamma),\sigma(t(\gamma)))\,.
\end{equation}
In particular, they are half-links instead of vertices.

Let us wrap up our graph-theoretic definitions by introducing the notion of an orientation. 
This is obtained by choosing a unique orientation for each undirected link. 
It is thus given by a set of links $E\subseteq H\times_\sigma H$ such that, for all links $e\in H\times_\sigma H$, either $e\in E$ or $e^{-1}\in E$, but not both. 
In the literature most of the theory is usually explained by heavily emphasizing the role of $E$. 
However, while choosing an $E$ is useful for computations, the orientation does not have any real physical meaning -- physical observables should be independent of $E$. 
Accordingly, in order to have more conceptual clarity, throughout our discussion we will do our best to deemphasize the role of $E$. 

\subsection{Classical Yang-Mills theory}\label{app:classical_theory}

In order to specify a Yang-Mills theory we need to choose a structure group $G$ (note the difference between this $G$ and the $\G$ we use to denote a graph). 
We will restrict to compact semisimple Lie groups in this work.
The kinematical information of the theory is encoded in Wilson lines defined on the links of the lattice.
The classical field configurations $\fields$ consist of all maps $u:H\times_\sigma H\rightarrow G$ assigning to each link the corresponding Wilson line supported on it. 
Then, for every link $e\in H\times_\sigma H$, the group element $u(e)\in G$ tells us how to perform parallel transport of the matter degrees of freedom from $s(e)$ to $t(e)$. 
Of course, the transport from $t(e)$ to $s(e)$ can be implemented by doing the opposite operation.
Thus, we will further restrict the maps in $\fields$ to satisfy
\begin{equation}\label{eq:consistency}
    u(e^{-1})=u(e)^{-1}\,.
\end{equation}

Note that a field configuration is completely specified by the configuration at each undirected link. 
In other words, the space of field configurations is the Cartesian product of the space of field configurations on each undirected link. 
Furthermore, the configuration of a field on such an undirected link is completely determined by the configuration of a field on one of its orientations.
We then conclude that
\begin{equation}\label{eq:orientation_isomorphism}
    \fields\cong G^E\,,    
\end{equation}
for any orientation $E$. 
We remark againt that, although the choice of such an orientation is useful for computations, it is unphysical and thus we will try to not put emphasis on it for the time being. 
Finally, let us also note that classical field configurations naturally assign Wilson lines to trajectories $\gamma\in T$ as well
\begin{equation}\label{eq:factorization_Wilson}
    u(\gamma)\coloneq \overleftarrow{\prod_{\{e\in\gamma\}}}u(e)\coloneq u(\gamma(n))\cdots u(\gamma(1))\,.
\end{equation}
Observe that the order chosen here reflects the physical interpretation of $u$ as implementing parallel transport. First $u(\gamma(1))$ acts on degrees of freedom at $s(\gamma(1))$ and maps them to degrees of freedom at $t(\gamma(1))$. 
Then $u(\gamma(2))$ acts on degrees of freedom at $t(\gamma(1))=s(\gamma(2))$ and maps them to degrees of freedom at $t(\gamma(2))$.

The fields in $\fields$ are however not all physically distinguishable from one another. 
In order to describe Yang-Mills theory locally, as we will explain more below, the space of fields has an inherent redundancy. 
This redundancy is parametrized by a group element at each vertex of the graph. 
The set $G^\Gamma$ of such assignments is called the gauge group.
This group acts on the field configurations in a way that we define two fields to be physically equivalent if one can be transformed into the other by a gauge transformation.

\begin{figure}
    \centering
    \subfloat[]{\includegraphics[width = 0.3\textwidth]{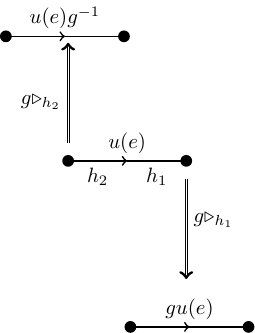}}\hspace{3cm}
    \subfloat[]{\includegraphics[width = 0.3\textwidth]{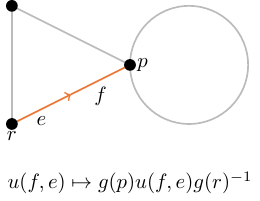}}
    \caption{On the left it is shown how gauge transformations act locally on every half-link of a graph.
    On the right it is shown how full gauge transformations act on the Wilson line supported on $(f,e)$.
    All of the other Wilson lines are similarly affected by the gauge transformation simulaneously.}
    \label{fig:gauge_transformations}
\end{figure}

In order to describe this action, it is useful to, as an intermediate step, define an action of the structure group that is supported on the half-link $h\in H$.
It will be non-trivial only on the two Wilson lines with support at the half-link, acting differently depending on whether the line is coming into or out of the half-link.
Said group action can be succinctly summarized as\footnote{Note how the middle line in this expression is completely determined by the first line $(gu(e^{-1}))^{-1}=u(e)g^{-1}$. 
From now on we will omit these redundancies from our expressions. The reader should however remain alert of these detail when reading ``otherwise'' from now on.}
\begin{equation}
    (g\triangleright_h u)(e)\coloneq \begin{cases}
    gu(e) & e = (h,\sigma(h)),\\
    u(e)g^{-1} & e=(\sigma(h),h)\\
    u(e) & \text{otherwise}\,.
    \end{cases}
\end{equation}
A graphical interpretation of this action can be found in the left of~\cref{fig:gauge_transformations}.
The idea here is that, if $h$ is at the end of the link $e$, then the transformation $g$ acts after the parallel transport has been implemented.
If instead $h$ is at the beginning, the transformation acts before the parallel transport takes place.
The appearance of $g^{-1}$ in this case ensures that the transformed field configuration still satisfies the consistency condition $u(e)^{-1}=u(e^{-1})$.
The action of a full gauge transformation $g\in G^\Gamma$ is obtained by letting the transformation $g(p)$ at each vertex $p\in \Gamma$ act on each of the half-links $h\in \pi^{-1}(\{p\})$ attached to it.
In particular, since every link is attached to two vertices, we simply have

\begin{align}
    (g\triangleright u)(h_1,h_2)\coloneq &(g(\pi(h_1))\triangleright_{h_1}g(\pi(h_2))\triangleright_{h_2} u)(h_1,h_2)\nonumber\\
    =& g(\pi(h_1))u(h_1,h_2)g(\pi(h_2))^{-1}\,.
\end{align}
This action is depicted at the right of~\cref{fig:gauge_transformations}.

\subsection{Quantum Yang-Mills theory}\label{app:quantum_theory}

The Haar measure provides a natural way of integrating over the structure group.
This in turn induces a natural integration measure on the space of field configurations $\fields$.
Explicitly, given an orientation $E$, one can use the isomorphism \eqref{eq:orientation_isomorphism} and set
\begin{equation}
    \dd{u}=\prod_{e\in E}\dd{u(e)}
    \,.
\end{equation}
The resulting measure is independent of $E$ due to the consistency condition \eqref{eq:consistency} and the invariance of the Haar measure under the inversion operation.
In particular we can use this measure to define densities of such field configurations. 

In the quantum theory, the states of the theory are described by wave functions that specify the probability amplitude density for each classical field configuration. 
The Hilbert space of the theory will therefore be
\begin{equation}
    \hilbert\coloneq  L^2\qty(\fields)
    \,.
\end{equation}
Given that one can specify the field configuration independenty at each undirected link, this space factorizes as the tensor product of the Hilbert spaces on each of the undirected links.
Indeed, through a choice of orientation $E$, \eqref{eq:orientation_isomorphism} induces
\begin{equation}\label{eq:factoriazation_Hilbert}
    \hilbert\cong\bigotimes_{e\in E}L^2(G)\,.
\end{equation}
The Hilbert space is spanned by states $\ket{u}$ in which the field is in a definite configuration $u\in\fields$.
More general wave functions are then given by linear combinations
\begin{equation}
    \ket{\psi}=\int_{\fields}\dd{u}\ket{u}\psi(u)\qc\psi(u)=\ip{u}{\psi}\,.
\end{equation}

The state space $\hilbert$ is equipped with a field operator $U$ that measures the field configuration. 
Each trajectory $\gamma\in T$ supports a $G$-valued Wilson line operator
\begin{equation}
    U(\gamma)\ket{u}=u(\gamma)\ket{u}
    \,.
\end{equation}
Given that they are all simultaneously diagonalizable, the matrix elements of this operator all commute with each other (although the full operators in general don't commute unless $G$ is Abelian).
On the other hand, every half-link $h\in H$ supports a (chromo)electric operator\footnote{These are sometimes called electric field operators in the literature given that their role in the Hamiltonian will be very similar to the role of the electric field in its continuum counterpart.
However, this identification can also lead to confusion.
For example, while in our case the electric operator will generate gauge transformations on its support, in the continuum the integral of the covariant derivative of the electric field plays this role.}. 
They are defined through a representation of the group on the Hilbert space, induced naturally from the action of the group on the classical field configurations
\begin{equation}\label{eq:half_link_representation}
    R_h(g)\ket{u}\coloneq \ket{g\triangleright_hu}
    \,.
\end{equation}
The electric operators $E_X(h)$ are then the generators of this representation\footnote{In this paper we will follow the mathematicians convention that Lie algebras of unitary groups are anti-hermitian. 
We however follow the physicists convention of including factors of $i$ to obtain Hermitian operators from the representation of such Lie algebras.}
\begin{equation}
    R_h(e^{\omega X})=e^{-i\omega E_X(h)}\qc X\in\mathfrak{g}\coloneq \text{Lie}(G)
    \,.
\end{equation}
In particular the electric operators bundle into a $\mathfrak{g}^*$-valued operator. 
Explicitly, if $X_1,\dots,X_{\dim G}$ is a basis of $\mathfrak{g}$ and $X^1,\dots,X^{\dim G}$ is a basis of the dual, we have
\begin{equation}
    E(h)=X^aE_{X_a}(h)\equiv X^a E_a(h)
    \,.
\end{equation}

Since the electric operators $-i E(h)$ furnish a representation of the Lie algebra $\mathfrak{g}$ of $G$, they satisfy its commutation relations
\begin{equation}\label{eq:electric_commutation}
    [E_X(h),E_Y(h')]=iE_{[X,Y]}(h)\delta_{h,h'}
    \,.
\end{equation}
The Wilson line operators transform like
\begin{equation}
    R_h(g)^{-1}U(\gamma)R_h(g)=(g\triangleright_h U)(\gamma)
    \,,
\end{equation}
where we have extended the group action on the classical field configurations to their quantum counterparts
\begin{equation}
    (g\triangleright_h U)(\gamma)\ket{u}\coloneq (g\triangleright_h u)(\gamma)\ket{u}
    \,.
\end{equation}
This induces an antirepresentation of the Lie algebra\footnote{Since the action on the Wilson line is from the right, it reverses the order of the product of elements in $G$.
This is why we obtain an anti-representation, with the crucial minus sign in \eqref{eq:wilson_transformation}}
\begin{equation}\label{eq:wilson_transformation}
    \begin{aligned}
        &[-iE_X(h),U(h_1,h_2)]=-X U(h_1,h_2)\delta_{h_1,h}+U(h_1,h_2)X\delta_{h_2,h}
    \,.
    \end{aligned}
\end{equation}

At the quantum level, gauge transformations act by a representation $R$ of the gauge group $G^\Gamma$ on $\hilbert$
\begin{equation}
    R(g)\ket{u}=\ket{g\triangleright u}
    \,.
\end{equation}
The generator of these is given by the Gauss operator at each vertex $p\in\Gamma$
\begin{equation}\label{eq:gauss_operator}
    G(p)=\sum_{h\in\pi^{-1}(\{p\})}E(h)
    \,.
\end{equation}
The physical Hilbert space is then identified as the space of states $\mathfrak{h}(\mathrm{G})$ that are gauge invariant, i.e. annihilated by the Gauss operators
\begin{equation}
    \mathfrak{h}(\mathrm{G})=\bigcap_{p\in\Gamma}\ker G(p)
    \,.
\end{equation}
These correspond to wave functions that are invariant under gauge transformations
\begin{equation}
    \ip{g\triangleright u}{\psi}=\mel{u}{R(g^{-1})}{\psi}=\ip{u}{\psi}
    \,.
\end{equation}

\begin{figure}
    \centering
    \subfloat[]{\includegraphics[width = 0.3\textwidth]{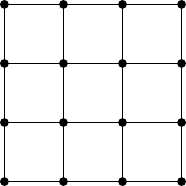}}\hspace{3cm}
    \subfloat[]{\includegraphics[width = 0.3\textwidth]{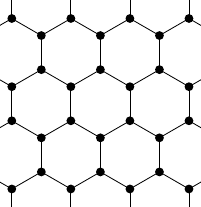}}
    \caption{Examples of admissible spacetime lattices. They are distinguished by the fact that they have elementary plaquettes.
    On the left we have a square lattice, whose elementary plaquettes are the smallest squares.
    On the right picture we have a honeycomb lattice, whose elementary plaquettes are given by the smallest hexagons.}
    \label{fig:spacetime_lattices}
\end{figure}

The Hamiltonian of the system depends on the choice of a coupling constant $g^2$, an invariant inner product $\ev{\cdot,\cdot}$ on $\mathfrak{g}$, and the interpretation of the graph $\G$. 
For the groups that we are interested in, we can take as the inner product
\begin{equation}
    \ev{X,Y}=-2\tr(XY)\,,
\end{equation}
with the trace computed in the fundamental representation\footnote{A different choice of inner product ammounts to a redefinition in the coupling constant.}.
On the other hand, for now, let us think of $\Gamma$ as a graph superimposed in space. 
Such spatial graphs are distinguished for having a set of closed loops $\mathcal{P}\subseteq T$ known as plaquettes, such as the ones in~\cref{fig:spacetime_lattices}. 
Whenever there is a plaquette $\gamma\in\mathcal{P}$, the loop with the reversed orientation is also a plaquette $\gamma^{-1}\in\mathcal{P}$. 
The magnetic part of the Hamiltonian is given by the sum of Wilson loops
\begin{equation}\label{eq:magnetic_hamiltonian}
    H_B\coloneq \sum_{\gamma\in \mathcal{P}}W(\gamma)\qc W(\gamma)\coloneq \tr(U(\gamma))
    \,.
\end{equation}
In particular it gives standard operator on $\mathcal{H}(\G)$ not a $G$-valued one.
On the other hand, the electric Hamiltonian is given by
\begin{equation}\label{eq:electric_hamiltonian}
    H_E\coloneq \sum_{\{h_1,h_2\}\in \faktor{H}{\sigma}}E^2(h)
    \,.
\end{equation}
In here $E^2(h)$ is the Casimir operator with respect to our inner product of the representation at either of the two half links $h\in\{h_1,h_2\}$.
It is conveniently described in any basis as
\begin{equation}
    E^2(h)\coloneq E_a(h)E^a(h)
    \,.
\end{equation}
$E_a$ denotes the components of an operator $E_a X^a$ in $\mathfrak{g}^*$, and we can use the inner product to raise its indices and make it an operator in $\mathfrak{g}$.
We will explain in the next section why the choice of half-link $h\in\{h_1,h_2\}$ does not matter in the Hamiltonian.

The full Hamiltonian of the theory is given by \eqref{eq:hamiltonian}
\begin{equation}\label{eq:ks_hamiltonian}
    H=\frac{g^2}{2}H_E+\frac{1}{2g^2}H_B
    \,.
\end{equation}
This is controlled by a dimensionless coupling constant $g^2$. 
For strong coupling $g^2\gg 1$, the theory is in the electric regime, where the behaviour is dominated by the electric Hamiltonian. 
At weak coupling $g^2\ll 1$, the theory is in the magnetic regime. 
Performing computations in a magnetic basis~\cite{DAndrea:2023qnr} is better suited in this regime. 

\subsection{SU(2) prepotentials on general graphs}\label{app:prepotential}

\subsubsection{Peter-Weyl theorem}

We are now ready to introduce the localization of the degrees of freedom from links to half-links.
This is most easily done by first focusing on a single link in light of \eqref{eq:factoriazation_Hilbert}. 
We will thus restrict to the graph shown on ~\cref{fig:single_link}. 
Let us for the moment further orient the link $e=(h_1,h_2)$. 
Then the field configurations can be identified with Wilson lines along said link and the Hilbert space $L^2(G)$ is given by wave functions describing the probability amplitude densities for such a line. 
In particular, this Hilbert space is a birepresentation of $G$ according to its action on each of the half-links
\begin{align}
        R_{h_1}(g_1)R_{h_2}(g_2)\ket{u}=\ket{g_1ug_2^{-1}}\quad \Leftrightarrow\quad 
        \mel{u}{R_{h_1}(g_1)R_{h_2}(g_2)}{\psi}=\ip{g_1^{-1}ug_2}{\psi}
        \,.
\end{align}

The technique we will use stems from the Peter-Weyl theorem
\begin{equation}
    L^2(G)\cong\bigoplus_{r\in \hat{G}}(V_r\otimes V_r^*)
    \,,
\end{equation}
with $\hat{G}$ an exhaustive collection of the inequivalent irreducible unitary representations of $G$, $V_r$ the $r$ representation and $V_r^*$ its dual. 
Let us understand this isomorphism in detail. 
The isomorphism is given by interpreting an elementary operator $\op{\psi}{\phi}\in V_r\otimes V_r^*$ in the $r$ representation as a wave function given by\footnote{We remark that on a finite dimensional Hilbert space, such as $V_r$, every operator can be written in terms of its matrix elements, so that the space of operators is equal to $V_r\otimes V_r^*$.
However, a tensor product of two Hilbert spaces is, in itself, a Hilbert space.
Thus, the space of operators on $V_r$ is, in itself, a Hilbert space.
In here we will be using that that Hilbert space can be identified as ``the Hilbert space of chromoelectric flux $r$'' in the Yang-Mills theory.}
\begin{equation}\label{eq:peter_weyl}
    \bra{u}\qty(\op{\psi}{\phi})\coloneq \sqrt{\dim V_r}\mel{\phi}{r(u)}{\psi}
    \,.
\end{equation}
The  localization to half-links
is obtained because this is an equality of birepresentations
\begin{align}
        \bra{u}\qty(r(g_2)\op{\psi}{\phi}r(g_1^{-1}))&=
        \sqrt{\dim V_r}\mel{\phi}{r(g_1^{-1})r(u)r(g_2)}{\psi}\nonumber\\
        &=\bra{g_1^{-1}ug_2}\qty(\op{\psi}{\phi})\nonumber\\
        &=\bra{u}R_{h_2}(g_2)R_{h_1}(g_1)\qty(\op{\psi}{\phi})
        \,.
\end{align}
In particular, the first factor $V_r$ contains the gauge redundancy of the starting point $h_2$ while the second factor $V_{r}^*$ contains the gauge redundancy of endpoint $h_1$.
This localization suggests an enlargement of the Hilbert space to
\begin{equation}\label{eq:abelian_enlargement}
    L^2(G)\subseteq \mathcal{H}_{\SB}\otimes\mathcal{H}_{\SB}^*\qc \mathcal{H}_\SB\coloneq \bigoplus_{r\in\hat{G}}V_r
    \,.
\end{equation}
Each of the two $\mathcal{H}_\SB$ factors is thus supported on a single half-link.
The original Hilbert space is then obtained by the introduction of new \textbf{Abelian} Gauss' laws demanding that physical vectors are spanned by tensor products of vectors in the same irreducible representation.

\subsubsection{Schwinger bosons and SU(2) duality}

The next step in the prepotential formulation is the introduction of Schwinger bosons. 
We will reduce our discussion at this stage to $G=\SU$, but the approach can be generalized to different gauge groups (see, e.g.,~\cite{Anishetty2009, Anishetty:2009nh, Kadam:2022ipf} for discussions of the prepotential formulation with $G=\mathrm{SU}(3)$). 
The space $\mathcal{H}_{\SB}$ can be thought of as the Hilbert space of two bosonic species, the Schwinger bosons described by a doublet of operators $a^A$, $A=1,2$ satisfying two copies of the canonical commutation relations (CCR) algebra
\begin{equation}
    [a^A,a^\dagger_B]=\delta^A_B
    \,.
\end{equation}
At this point we will adopt the conventions that upper indices correspond to the fundamental $\SU$ representation, lower indices correspond to the antifundamental, and the Einstein summation convention contracting upper and lower indices.
Indeed, the space of Schwinger bosons carries a representation of the Lie algebra $\su$ given by
\begin{equation}\label{eq:schwinger_representation}
    X\mapsto a^\dagger X a(=a^\dagger_A X\indices{^A_B}a^B)
    \,.
\end{equation}
The isomorphism is then implemented through the identification 
\begin{equation}\label{eq:spin_vs_schwinger}
    \begin{aligned}
        \ket{j,m}\equiv\ket{n_1,n_2}\coloneq \frac{\qty(a^\dagger_1)^{n_1}\qty(a^\dagger_2)^{n_2}}{\sqrt{n_1!n_2!}}\ket{0}\qc
        n_1 = j+m\qc n_2 = j-m
        \,,
    \end{aligned}
\end{equation}
with $\ket{0}$ the bosonic vacuum of the theory.
On the left hand side of the $\equiv$ sign, we are expressing states in a ``spin'' convention, where the first entry determines the total spin of the representation and the second its $z$-component.
On the right hand side, we instead represent the state in using an ``occupation'' number convention, where the first entry determines the total number of $a^\dagger_1$ bosons and the second the total number of $a^\dagger_2$ bosons.
Although ambiguous, we hope that the reader can identify which convention is being used through the context and the variable names inside of the kets.

One of the fundamental simplifying features of $\SU$ is that all representations are self-dual $V_j\cong V_j^*$. 
This can be made manifest by noting that we can raise indices using the totally antisymmetric symbol $\epsilon^{AB}$. 
We will define $\epsilon_{AB}=-\epsilon^{AB}$ so that $\epsilon_{AB}\epsilon^{BC}=\delta^{C}_A$.
By convention we will take $\epsilon^{12}=1$ and the contraction of $\SU$ doublets to be\footnote{The reader more familiar with index manipulation in general relativity might find this odd.
In our case, since we are raising and lowering indices with an antisymmetric symbol, we have $\psi_A\phi^A=-\psi^A\phi_A$.}
\begin{equation}
\psi\phi\coloneq \psi_A\phi^A=\psi_A\epsilon^{AB}\phi_B
\,.    
\end{equation}
We can then write the explicit formula of the isomorphism of every representation and its dual by
\begin{equation}\label{eq:self_dual}
    \begin{aligned}
        V_j^*&\rightarrow V_j\\
        \bra{0}\frac{\qty(a^1)^{j+m}\qty(a^2)^{j-m}}{\sqrt{(j+m)!(j-m)!}}&\mapsto \frac{\qty(a^{\dagger 1})^{j+m}\qty(a^{\dagger 2})^{j-m}}{\sqrt{(j+m)!(j-m)!}}\ket{0}\\
        &\Leftrightarrow\\
        \bra{j,m}&\mapsto(-1)^{j-m}\ket{j,-m}\\
        &\Leftrightarrow\\
        \bra{n_1,n_2}&\mapsto(-1)^{n_2}\ket{n_2,n_1}
        \,.
    \end{aligned}
\end{equation}
This will allow us to implement the prepotential formulation without the choice of an orientation. 
Indeed, now we have an isomorphism
\begin{equation}\label{eq:disorientation}
    L^2(\fields)\cong\bigoplus_{r\in \hat{G}}V_r\otimes V_r\subseteq \mathcal{H}_{\SB}\otimes \mathcal{H}_{\SB}
    \,,
\end{equation}
where the first factor is supported in $h_1$ and the second in $h_2$, given by
\begin{equation}
    \begin{aligned}
        \bra{u}(\ket{j,m_1}\otimes\ket{j,m_2})=(-1)^{j+m_1}\mel{j,-m_1}{r_j(u(h_1,h_2))}{j,m_2}
        \,.
    \end{aligned}
\end{equation}

Putting these ingredients together we obtain the prepotential formulation of the Hilbert space
\begin{equation}
    \mathfrak{H}(\G)=\mathcal{H}_\SB^{\otimes H}\supseteq\mathcal{H}(\G)
    \,,
\end{equation}
obtained by tensoring the Schwinger boson Hilbert space on each half-link. 
The action of $\su$ on each half-link \eqref{eq:schwinger_representation} extends the definition of the electric operators to the Hilbert space of this formulation
\begin{equation}
    -iE_X(h)\coloneq a^\dagger(h) X a(h)
    \,.
\end{equation}
The index structure of the Schwinger bosons reflects the commutator with these
\begin{equation}\label{eq:schwinger_transformation}
    \begin{aligned}
        \qty[-iE_X(h_1),a^A(h_2)]&=-X\indices{^A_B} a^B(h_2)\delta_{h_1h_2}\qand\\
        \qty[-iE_X(h_1),a^\dagger_A(h_2)]&=a^\dagger_B(h_2)X\indices{^B_A} \delta_{h_1h_2}
        \,.
    \end{aligned}
\end{equation}
The commutation relations with other operators can be read from the placement of its physical indices and their support. 

While we have technically enlarged the Hilbert space, it is now very simple to characterize the gauge invariant states. 
Indeed, states in this Hilbert space are spanned by monomials in the creation operators\footnote{We note that not all of the half-links in this notation have to be different.
The same can be said for the indices.}
\begin{equation}
    a^\dagger_{A_1}(h_1)\cdots a^\dagger_{A_n}(h_n)\ket{0}
    \,.
\end{equation}
The transformation properties of such monomials are easily read from their index structure.
Furthermore, the gauge invariant linear combinations can be immediately constructed by demanding that the resulting state satisfies the property that, at each vertex, all of the half-links attached to it have indices that are fully contracted. 

\subsubsection{Loop-String-Hadron operators}
\label{sec:LSHOps}

A particularly convenient way of describing these gauge invariant states is obtained by collecting the creation and annihilation operators into a single operator
\begin{equation}\label{eq:wilson}
    \begin{aligned}
        A\indices{^A_a}(h)&\coloneq \begin{cases}
            a^{\dagger A}(h) & a=+\\
            a^{A}(h) & a=-
            
        \end{cases}=\begin{pmatrix}
            a_2^\dagger(h) & a^1(h) \\
            -a_1^\dagger(h) & a^2(h)
        \end{pmatrix}\indices{^A_a}\,.
    \end{aligned}
\end{equation}
In terms of these, the commutation relations become
\begin{equation}\label{eq:wilson_commutations}
    [A\indices{^A_a}(h_1),A\indices{^B_b}(h_2)]=\epsilon^{AB}g_{ab}\delta_{h_1,h_2}\qc g=\mqty(0 & -1\\ -1 & 0)
    \,.
\end{equation}
This $g$ is the metric of $1+1$ Minkowski space in lightcone coordinates, indicating that the $a$ index is in the vector representation of this Poincaré group.
Since we are working with complex representations, this is equivalent to the vector representation of $\mathrm{SO}(2)=\mathrm{U}(1)$, which is the Abelian gauge symmetry associated with the Abelian Gauss law.
In this representation we can choose the convention $a=+$ objects transform like $z=x+iy$ while $a=-$ objects transform like $z=x-iy$, so that it acts by
\begin{equation}
    A\indices{^A_a}(h)\mapsto e^{ia\phi}A\indices{^A_a}(h)\,.
\end{equation}
Thus the associated charge counts the number of Schwinger bosons created by the operator.

With this we can form the loop operators
\begin{equation}
    \mathcal{L}_{ab}(h_1,h_2)\coloneq A\indices{_{Aa}}(h_1)A\indices{^A_b}(h_2)
    \,,
\end{equation}
obtained by contraction of the physical indices.
Therefore, as long as $\pi(h_1)=\pi(h_2)$, i.e. that $(h_1,h_2)$ is a corner, this loop operator is gauge invariant. 
On the other hand, the $\mathrm{U}(1)$ charge of such an operator is $a$ on $h_1$ and $b$ on $h_2$ (and they add up if $h_1=h_2$). 
These behave simply under change of order of the edges
\begin{equation}
    \mathcal{L}_{ab}(h_1,h_2)=-\mathcal{L}_{ba}(h_2,h_1) + 2g_{ab}\delta_{h_1,h_2}\,,
\end{equation}
and conjugation
\begin{equation}
    \mathcal{L}_{ab}(h_1,h_2)^\dagger = -ab\mathcal{L}^{ba}(h_2,h_1)\,,
\end{equation}
and form a closed Lie algebra\footnote{The table of commutation relations found in~\cite{Raychowdhury:2019iki} can be summarized in a similar fashion if one creates a $\mathbb{Z}_2$-graded version of the field $A$ that takes fermionic values on vertices and bosonic on half-links.}
\begin{equation}
    \begin{aligned}
        \qty[\mathcal{L}_{ab}(h_1,h_2),\mathcal{L}_{cd}(h_3,h_4)] = &\, g_{ad}\delta_{h_1,h_4}\mathcal{L}_{cb}(h_3,h_2) -g_{bc}\delta_{h_2,h_3}\mathcal{L}_{ad}(h_1,h_4)\\
        &+g_{bd}\delta_{h_2,h_4}\mathcal{L}_{ac}(h_1,h_3)-g_{ac}\delta_{h_1,h_3}\mathcal{L}_{db}(h_4,h_2)\,.
    \end{aligned}
\end{equation}
Let us look at some particular examples of such operators that will be relevant for us.
For $h_1\neq h_2$, the operator
\begin{equation}\label{eq:loop_creation}
    \mathcal{L}_{++}(h_1,h_2)=a^\dagger(h_1)a^\dagger(h_2)=-\mathcal{L}_{++}(h_2,h_1)
    \,,
\end{equation}
precisely provides the gauge invariant linear combinations of creation operators. 
It is thus useful to define a set of oriented corners $\mathcal{C}\subseteq C$ which is a subset of the set of all corners such that for each corner $(h_1,h_2)$ at a vertex $p\in\Gamma$ either $(h_1,h_2)\in\mathcal{C}$, $(h_2,h_1)\in\mathcal{C}$ but not both. 
In other words, this is an orientation of the corners of the graph.
We then have
\begin{equation}\label{eq:basis_unnormalized}
    \begin{aligned}
        \Span\qty{\left.\prod_{c\in \mathcal{C}}\mathcal{L}_{++}^{n(c)}(c)\ket{0} \right|n:\mathcal{C}\rightarrow\mathbb{N}}=\bigcap_{p\in\Gamma,X\in\mathfrak{g}}\ker G_X(p)
        \,.
    \end{aligned}
\end{equation}

If $h_1=h_2=h$, the only non-trivial operator we obtain is the number operator
\begin{equation}
    N(h)\coloneq \mathcal{L}_{+-}(h,h)=a^\dagger(h)a(h)
    \,.
\end{equation}
In terms of the identification \eqref{eq:spin_vs_schwinger}, this operator measures $2j=n_1+n_2$.
Accordingly, the Casimir of this representation is
\begin{equation}\label{eq:su2_casimir}
    \vb{E}^2(h)\coloneq \frac{N(h)}{2}\qty(\frac{N(h)}{2}+1)
    \,. 
\end{equation}
In this equation we have chosen the vector of Pauli matrices $i\vb*{\sigma}/2$ as a basis of $\su$. This is orthonormal and yields the electric operators
\begin{equation}
    \vb{E}(h)=E_{i\vb*{\sigma}/2}(h)=-\frac{1}{2}a^\dagger(h)\vb*{\sigma}a(h)\,.
\end{equation}

In light of the Abelian enlargement \eqref{eq:abelian_enlargement}, the physical Hilbert space must be further constrained to satisfy $N(h_1)=N(h_2)$ for every undirected link $\{h_1,h_2\}\in \faktor{H}{\sigma}$. 
In other words, we must introduce the Abelian constraints
\begin{equation}\label{eq:abelian_gauss_law}
    G(\{h_1,h_2\})\coloneq |N(h_1)-N(h_2)|
    \,,
\end{equation}
in terms of which the physical Hilbert space is now given by the set of all quantum numbers $N(h)$ subject to the constraint $N(h_1) - N(h_2) = 0$
\begin{equation}
    \mathfrak{h}(\Gamma)=\qty(\bigcap_{p\in\Gamma}\ker G(p))\cap\qty(\bigcap_{\{h_1,h_2\}\in\faktor{H}{\sigma}}\ker G(\{h_1,h_2\}))
    \,.
\end{equation}
The first intersection, corresponding to the invariant states under the non-Abelian Gauss' law, is resolved by creating states through the loop operators. 
The second, corresponding to the invariant states under the Abelian one, by cycling through every undirected link $\{h_1,h_2\}\in\faktor{H}{\sigma}$ and ensuring the numbers of loop creation operators with support on $h_1$ and $h_2$ coincides.

With these operators in mind, it is instructive to look at particular examples of the commutation relations.
Let us start looking at the relations on a single corner $(h_1,h_2)$ with $h_1\neq h_2$.
First, we note that
\begin{equation}
    [N(h_1),\mathcal{L}_{ab}(h_1,h_2)]=\delta_{a+}\mathcal{L}_{+b}(h_1,h_2)-\delta_{a-}\mathcal{L}_{-b}(h_1,h_2)\,,
\end{equation}
so that the number operator as expected measures the number of creations or annihilations at the half-link.
Furthermore, unlike a harmonic oscillator, the commutator of creation and annihilation operators is not proportional to the identity
\begin{equation}\label{eq:creation_annihilation_on_site}
    [\mathcal{L}_{--}(h_1,h_2),\mathcal{L}_{++}(h_1,h_2)]=N(h_1)+N(h_2)+2\,.
\end{equation}
This is of course a consequence of the fact that our loop operators are bilinears in the Schwinger boson operators.
Finally, let us remark that the mixed operators commute with the creation and annihilation operators
\begin{equation}\label{eq:mixed_commutators}
    \begin{aligned}
        {[}\mathcal{L}_{+-}(h_1,h_2),\mathcal{L}_{++}(h_1,h_2){]}&=0\\
        {[}\mathcal{L}_{+-}(h_1,h_2),\mathcal{L}_{--}(h_1,h_2){]}&=0
    \end{aligned}
\end{equation}

It is also interesting to study the commutation relations obtained when one introduces another corner $(h_3,h_1)$ with $h_3\neq h_1$ and $h_3\neq h_2$.
In that case we have
\begin{equation}\label{eq:half_overlapping}
    [\mathcal{L}_{ab}(h_1,h_2),\mathcal{L}_{cd}(h_3,h_1)]=-g_{ad}\mathcal{L}_{bc}(h_2,h_3)\,.
\end{equation}
In particular, if the indices at the overlapping half-link match, the operators commute.
Otherwise, we obtain commutators like
\begin{equation}\label{eq:creation_annihilation}
    [\mathcal{L}_{--}(h_1,h_2),\mathcal{L}_{++}(h_3,h_1)] = \mathcal{L}_{-+}(h_2,h_3)\,.
\end{equation}
This show that switching flux from $h_3$ to $h_2$ can be achieved by anlihilation of flux on $(h_3,h_1)$ and creation of flux on $(h_1,h_2)$.

\subsubsection{Wilson line operators}
\label{sec:Wilson_line_operators}

Let us finish by discussing the construction of the Wilson line operators $U\indices{^A_B}(h_1,h_2)$ in terms of prepotentials.
For this, let us take advantage of the fact that the $\mathrm{U}(1)$ and the $\SU$ representations appearing in $A$ have the same dimension to define
\begin{equation}
    A\indices{^{\dagger a}_A}(h)\coloneq (A\indices{^A_a}(h))^\dagger=\begin{pmatrix}
            a^2(h)& -a^1(h) \\
            a_1^\dagger(h) & a_2^\dagger(h)
        \end{pmatrix}\indices{^a_A}\,.
\end{equation}
with the index placement reflecting the transformation behavior of this operator.
Explicitly, one can check that
\begin{equation}\label{eq:dagger_A}
    A\indices{^{\dagger a}_A}(h)=aA\indices{_A^a}(h)\,,
\end{equation}
raising and lowering indices using $g_{ab}$.
With this one can then prove that
\begin{equation}
    (A(h) A^\dagger(h))\indices{^A_B}\coloneq A(h)\indices{^A_a}A^\dagger(h)\indices{^a_B}=\delta^A_B(N(h)+1)\,.
\end{equation}

This suggests the introduction of a new Wilson line operator on each half-link given by
\begin{equation}\label{eq:unitary_half_link}
    U(h)\indices{^A_a} \coloneq  \frac{1}{\sqrt{N(h)+1}}A(h)\indices{^A_a}\,,
\end{equation}
which then satisfies
\begin{equation}
    U(h)\indices{^A_a}U^\dagger(h)\indices{^a_B}=\delta^A_B\,.
\end{equation}
In fact, this operator is ``unitary,'' for 
\begin{equation}
    U^\dagger(h)\indices{^a_A} U(h)\indices{^A_b}=\delta^a_b\,.
\end{equation}
This can be checked by noticing that
\begin{equation}
    A^\dagger(h)\indices{^a_A} A(h)\indices{^A_b}=(N(h)+1+a)\delta^a_b\,,
\end{equation}
and that 
\begin{equation}\label{eq:N_charge}
    \mathcal{O}f(N(h))=f(N(h)-o)\mathcal{O}\,
\end{equation}
with $o$ the $\mathrm{U}(1)$ charge of the operator $\mathcal{O}$ at $h$.

Now consider an edge $(h_1,h_2)$. We wish to obtain a unitary operator $U\indices{^A_B}(h_1,h_2)$ out of the operators $U\indices{^A_a}(h_1)$ and $U\indices{^{\dagger b}_A}(h_2)$. 
This can then be done by multiplying the two operators together and contracting the $a$ and $b$ indices (note that these operators commute since they are localized in different half-links). 
This naive approach would however be invariant under the $U(1)$ action acting simultaneously on $h_1$ and $h_2$ because there are no more $U(1)$ indices. 
In other words, the operator would commute with $N(h_1)+N(h_2)$. 
This is because of the placement of the indices - whenever one of the operators creates a boson on a half-link, the other destroys one on the other half-link. 
Instead, the Abelian Gauss law demands that physical operators commute with $N(h_1)-N(h_2)$, so that whenever a boson is created in a half-link, another boson should be created on the other.
Thus, instead of contracting the $a$ and $b$ indices with one on top and one below, we should contract them with the same placement.
In order to keep consistency with the Einstein summation convention, we can then introduce a matrix $t$ that does that
\begin{equation}
    U\indices{^A_B}(h_1,h_2) = U\indices{^A_a}(h_1)t\indices{^a_b}U\indices{^{\dagger b}_B}(h_2)\,.
\end{equation}
Any off-diagonal matrix would do to obtain something that satisfies the Abelian Gauss law.
If we further want a unitary result, we should choose a unitary $t$, which would then guarantee
\begin{equation}
    \begin{aligned}
        U\indices{^A_B}(h_1,h_2)U\indices{^{\dagger B}_C}(h_1,h_2)=U\indices{^{\dagger A}_B}(h_1,h_2)U\indices{^C_D}(h_1,h_2)=\delta^A_C\,.
    \end{aligned}
\end{equation}
A final restriction, which fixes the form of this matrix is that the result should have unit determinant.
At this stage, the meaning of this statement is not clear, for the elements of $U\indices{^A_B}(h_1,h_2)$ are themselves operators.
The meaning is acquired by the fact that the components commute, so that the requirement that 
\begin{equation}
    1 = U\indices{^1_1}(h_1,h_2)U\indices{^2_2}(h_1,h_2)-U\indices{^1_2}(h_1,h_2)U\indices{^2_1}(h_1,h_2)\,,
\end{equation}
becomes independent of the operator ordering.
This is proven in~\cref{sec:Wilson_line_operators_commute}.
Using the techniques developed there, one can show that the determinant condition is equivalent to taking $t$ to have unit determinant.
This therefore reduces the possible choice of $t$ down to
\begin{equation}
    t\indices{^a_b}=\mqty(0 & e^{i\theta}\\
    -e^{-i\theta} & 0)\indices{^a_b}.
\end{equation}

In general, one can choose a different $\theta$ for every link, so that $t\indices{^a_b}=t\indices{^a_b}(h_1,h_2)$.
One however would like to impose the condition that $U(h_1,h_2)^{-1}(=U(h_1,h_2)^\dagger)=U(h_2,h_1)$.
We must then have
\begin{equation}
    t(h_1,h_2)^\dagger=t(h_2,h_1).
\end{equation}
Given our specific form for $t$, we conclude that $t(h_2,h_1)=-t(h_1,h_2)$.

In practice, for definiteness one can orient the graph.
Along the positive oriented links one can absorb the $\theta$ dependence in the $\mathrm{U}(1)$ gauge invariance.
This amounts to chosing
\begin{equation}
    t\indices{^a_b}=\mqty(0 & 1\\
    -1 & 0)\indices{^a_b}.
\end{equation}
for those edges and its conjugate for the negatively oriented ones. 
For this choice we have $T^{ab}=\delta^{ab}$.
Thus, on the positively oriented links we can take
\begin{equation}
    \begin{aligned}
        U\indices{^A_B}(h_1,h_2)=\frac{1}{\sqrt{N(h_1) + 1}} A\indices{^A_a}(h_1) A\indices{_{Ba}}(h_2) \frac{1}{\sqrt{N(h_2) + 1}}\,,
    \end{aligned}
\end{equation}
and its negative for the negatively oriented links.
Even though the $a$'s are not one on top and one below, a summation is implied in this formula.
This is done to minimize the factors of $\delta^{ab}$ appearing in our formulae.

Not being gauge invariants, it is not surprising that generic Wilson line operators themselves are not immediately expressible in terms of LSH operators.
Nevertheless, gauge invariant combinations built out of them can.
In order to do this, it is useful to define LSH Wilson lines.
Unlike, the usual Wilson lines, these are supported on corners instead of links.
Accordingly, instead of transforming under the $\SU$ copies at the source and target of a link, the transform under the $\mathrm{U}(1)$ copies at the source and target of the corner.
They are obtained by contracting in a $\SU$ invariant way the indices of the Wilson line operators on the half-links of the corner.
Thus, given a corner $(h_1,h_2)$ we define
\begin{equation}\label{eq:vertex_factor}
    \begin{aligned}
        U\indices{^a_b}(h_1,h_2)=U\indices{^{\dagger a}_A}(h_1)U\indices{^A_b}(h_2)=\begin{cases}
            a\frac{1}{\sqrt{N(h_2)+1}}\mathcal{L}\indices{^a_b}(h_1,h_2)\frac{1}{\sqrt{N(h_1)+1}} & h_1\neq h_2\,\\
            \delta^a_b & h_1 = h_2\,.
        \end{cases}
    \end{aligned}
\end{equation}
These are known as vertex factors in~\cite{Raychowdhury:2019iki}.

As an example of their use, consider a pair of corners $(h_1,h_2)$ and $(h_3,h_4)$ with $\pi(h_3)=\pi(h_2)$. 
Then we have
\begin{equation}
    \begin{aligned}
        U\indices{^A_B}(h_1,h_2)U\indices{^B_C}(h_3,h_4)=U\indices{^A_a}(h_1)t\indices{^a_b}(h_1,h_2)U\indices{^b_c}(h_2,h_3)t\indices{^c_d}(h_3,h_4)U\indices{^{\dagger d}_C}(h_4)\,.
    \end{aligned}
\end{equation}
More generally, let $\gamma$ be a trajectory. 
Let us define the trajectory $\gamma_c$ as follows. 
Denote $\gamma(i)=(a(i),b(i))$. If the length of $\gamma$ is $n$, $\gamma_c$ is the corner trajectory of length $n-1$ defined by
\begin{equation}
    \gamma_c(i)=(b(i+1),a(i))\,.
\end{equation}
An example of this is given in ~\cref{fig:trajectory_example}.
We can then define
\begin{equation}
    \begin{aligned}
        U\indices{^a_b}(\gamma_c)=&\,U\indices{^a_{b_1}}(\gamma_c(n-1))t\indices{^{b_1}_{a_2}}(\gamma(n-1))
        \, U\indices{^{a_2}_{b_2}}(\gamma_c(n-2))t\indices{^{b_2}_{a_3}}(\gamma(n-2))\times \ldots\\
        &\times U\indices{^{a_{n-2}}_{b_{n-2}}}(\gamma_c(2))t\indices{^{b_{n-2}}_{a_{n-1}}}(\gamma(2))\, U\indices{^{a_{n-1}}_{b}}(\gamma_c(1))\,.
    \end{aligned}
\end{equation}
With this we then have
\begin{equation}\label{eq:open_Wilson_line}
    \begin{aligned}
        U\indices{^A_B}(\gamma)=U\indices{^A_a}(a(n))t\indices{^a_b}(\gamma(n))U\indices{^b_c}(\gamma_c)t\indices{^c_d}(\gamma(1))U\indices{^{\dagger d}_B}(b(1))\,.
    \end{aligned}
\end{equation}
If $\gamma$ is closed, we can then consider the gauge invariant $\tr U(\gamma)$. This amounts to setting $B=A$ in the equation above. Furthermore, if $\gamma$ only goes through $a(n)$ once, we can push the first factor all the way to the end and recover yet another corner Wilson line operator. We then have
\begin{equation}\label{eq:loops_lsh}
    \begin{aligned}
        \tr U(\gamma)=&t\indices{^a_b}(\gamma(n))U\indices{^b_c}(\gamma_c)t\indices{^c_d}(\gamma(1))U\indices{^d_a}(b(1),a(n))\\
        =&t\indices{^a_b}(\gamma(n))U\indices{^b_a}(\tilde{\gamma})\,,
    \end{aligned}
\end{equation}
where $\tilde{\gamma}$ is the corner trajectory that first goes through $(b(1),a(n))$ and then through $\gamma_c$.
This is now fully expressed in the context of LSH operators.

Let us finish this section with a few comments on the application of this formulation.
First of all, the condition that $\gamma$ only goes through $a(n)$ once might seem rather restrictive at first.
In reality, since $\tr U(\gamma)$ is invariant under shifts of the starting (which is also the finishing) point of $\gamma$, this is not the case.
Thus, in order to use this formula, we just need there to exist at least a single half-link that $\gamma$ goes through once. 
Second, let us remark that
\begin{equation}
    \begin{aligned}
        t\indices{^a_b}(h_1,h_2)U\indices{^b_c}(h_2,h_3)=\pm\delta^{ab}\frac{1}{\sqrt{N(h_3)+1}}\mathcal{L}_{bc}(h_2,h_3)\frac{1}{\sqrt{N(h_2)+1}}\,,
    \end{aligned}
\end{equation}
where the sign is determined by whether $(h_1,h_2)$ is positively oriented or not.
Therefore, in practical applications the $t$ matrices are there only to keep track of the sign, and one can immediately write everything in terms of loop operators with lower indices.

\subsubsection{Proof of commutative property of Wilson line operator components}
\label{sec:Wilson_line_operators_commute}

In this section we show that the components of the Wilson line operator commute.
To streamline this computation, let us introduce some notation that we will use taking advantage of the fact that for this the half-links $h_1$ and $h_2$ will be fixed.
First, we will suppress the labels associated to these.
In order to avoid confusion due to this, we will set
\begin{equation}
    A\indices{^A_a}(h_1)=A\indices{^A_a}\qand A\indices{^A_a}(h_2)=B\indices{^A_a}\,.
\end{equation}
We will also restrict to the subspace satisfying the Abelian Gauss law, so that
\begin{equation}\label{eq:loop_number}
    \mathcal{L}_{+-}(h_1,h_1)=\mathcal{L}_{+-}(h_2,h_2)=\mathcal{L}_{+-}=-\mathcal{L}_{+-}-2=N\,.
\end{equation}
Finally, we will introduce the notation
\begin{equation}
    \Phi(a)\coloneq \frac{1}{\sqrt{N+a+1}}\,.
\end{equation}

With this notation in place, let us define a rescaled commutator that will be more comfortable to work with
\begin{equation}
    C\indices{^A_B^C_D}\coloneq \Phi(0)^{-1}[U\indices{^A_B},U\indices{^C_D}]\Phi(0)^{-1}\,.
\end{equation}
Our main goal will be to show that this vanishes on the subspace satisfying the Abelian Gauss law.
Let us write
\begin{equation}
    U\indices{^A_B}=T^{ae}\Phi(0)A\indices{^A_a}B\indices{_{Be}}\Phi(0)\qc T^{ae}\coloneq \sum_{b=\pm}bt\indices{^a_b}g^{be}\,,
\end{equation}
using \eqref{eq:unitary_half_link} and \eqref{eq:dagger_A}. 
The only property we will need of the matrix $T^{ab}$ is that it is diagonal since both $t\indices{^a_e}$ and $g^{be}$ are off-diagonal.
With this we have that
\begin{equation}
\begin{aligned}
    \Phi(0)^{-1}U\indices{^A_B}U\indices{^C_D}\Phi(0)^{-1}
    &=T^{ae}T^{cf}A\indices{^A_a}B_{Be}\Phi(0)^2A\indices{^C_c}B_{Df}\\
    &=\sum_a\Phi(-a)^2T^{ae}T^{cf}A\indices{^A_a}A\indices{^C_c}B_{Be}B_{Df}\,.
\end{aligned}
\end{equation}
In the last line we used the fact that the $A$'s and $B$'s commute since their support does not overlap, \eqref{eq:N_charge}, and made the sum over $a$ explicit since we broke the summation convention. Then the commutator is equal to
\begin{equation}
    \begin{aligned}
        &C\indices{^A_B^C_D}=\sum_{a}\Phi(-a)^2T^{ae}T^{cf}\qty(A\indices{^A_a}A\indices{^C_c}B_{Be}B_{Df}-A\indices{^C_a}A\indices{^A_c}B_{De}B_{Bf})\,.
    \end{aligned}
\end{equation}

Next, we note that the term in parenthesis, which is quartic in Schwinger bosons, can be reduced to a bilinear with functions of $N$ as coefficients. 
For this one adds and subtracts $A\indices{^C_a}A\indices{^A_c}B_{Be}B_{Df}$, so that the term in parenthesis reduces to the sum
\begin{equation}
    \begin{aligned}
        &(A\indices{^A_a}A\indices{^C_c}-A\indices{^C_a}A\indices{^A_c})B_{Be}B_{Df}+A\indices{^C_a}A\indices{^A_c}(B_{Be}B_{Df}-B_{De}B_{Bf})\,.
    \end{aligned}
\end{equation}
We claim that the terms in parenthesis are functions of $N$. 
Focusing on the one in the first row, the $A\leftrightarrow C$ antisymmetry guarantees that
\begin{equation}
    A\indices{^A_a}A\indices{^C_c}-A\indices{^C_a}A\indices{^A_c}=\epsilon^{AC}(A\indices{^1_a}A\indices{^2_c}-A\indices{^2_a}A\indices{^1_c})=\epsilon^{AC}(A\indices{_{2a}}A\indices{^2_c}+A\indices{_{1a}}A\indices{^1_c})=\epsilon^{AC}\mathcal{L}_{ac}\,.
\end{equation}
Including the analogous result for the second row and using the diagonality of the $T^{ae}$ matrices, the commutator takes the form
\begin{equation}
    C\indices{^A_B^C_D}=\epsilon^{AC}M_{BD}(B)+\epsilon_{DB}M^{CA}(A)\,,
\end{equation}
where
\begin{equation}
    M^{CA}(A)=\sum_a\Phi(-a)^2 T^{ae}T^{cf}\mathcal{L}_{ef}A\indices{^C_a}A\indices{^A_c}\,.
\end{equation}

We further claim that $M^{CA}$ is itself a function of the number operator. 
For this one can start by noting that the $a=c$ terms vanish identically because $\mathcal{L}_{ef}$ would vanish. 
Therefore, there are only two terms in the contractions of $M^{CA}$. 
Furthermore, both are proportional to $T^{++}T^{--}$. 
Dividing this common factor out, the $a=+$ term is
\begin{equation}
    \Phi(-1)^2\mathcal{L}_{+-}A\indices{^C_+}A\indices{^A_-}=A\indices{^C_+}A\indices{^A_-}\,.
\end{equation}
In the last equal sign we used \eqref{eq:loop_number}.
Similarly, the $a=-$ term is
\begin{equation}
    \Phi(+1)^2\mathcal{L_{-+}}A\indices{^C_-}A\indices{^A_+}=-A\indices{^C_-}A\indices{^A_+}\,.
\end{equation}
Therefore
\begin{equation}
\begin{aligned}
    \frac{M^{CA}(A)}{T^{++}T^{--}}&=A\indices{^C_+}A\indices{^A_-}-A\indices{^C_-}A\indices{^A_+}\\
    &=A\indices{^C_+}A\indices{^A_-}-A\indices{^A_+}A\indices{^C_-}+\epsilon^{CA}\\
    &=\epsilon^{CA}(N+1)\,.
\end{aligned}
\end{equation}

Inserting these relations we conclude that
\begin{equation}
    C\indices{^A_B^C_D}=\epsilon^{AC}\epsilon_{DB}(N+1)+\epsilon_{DB}\epsilon^{CA}(N+1)=0\,,
\end{equation}
concluding our proof.

The same technique employed in this proof can be used to show that
\begin{equation}
    \Phi(0)^{-1}\det U\Phi(0)^{-1}=T^{++}T^{--}(N+1)\,.
\end{equation}
Thus, the unit determinant condition demands that
\begin{equation}
    T^{++}T^{--}=1.
\end{equation}
This is equivalent to demanding that $t$ has unit determinant.

\section{Relating Theories on Different Graphs}\label{app:graphs}

In this section we will attempt to formalize the graph manipulations required in the main body of the paper. In~\cref{app:classical_graphs} and~\cref{app:quantum_graphs}, we give a general description of graph manipulations and whether they yield kinematically equivalent physical representations. 
The equivalences are made explicit in these sections as well. The former deals with the classical setting. 
The latter with the quantum mechanical one. In~\cref{app:examples} these general theorems are put in practice for the two manipulations required in the main body of the paper. 
In~\cref{app:point_splitting}, virtual point splitting is discussed. This is fundamental in transforming petals into leaves and flowers into branches. 
We show that graphs can always be manipulated to become fully trivalent. 
We then give proofs for the normalization of the states and LSH operators given in~\cite{Raychowdhury:2019iki}. 
In~\cref{app:maximal_tree}, we discuss the flower construction in the context of maximal tree gauge fixing. 
Finally,~\cref{app:dynamics} lifts the kinematical equivalences to dynamical ones by showing how the ingredients for the engineering of Hamiltonians are translated through these manipulations. This is first done in the context of the original Kogut-Susskind formulation~\cref{app:ks_dof}. 
In the context of maximal tree gauge fixing, we simply restate and give proofs for the results found in~\cite{DAndrea:2023qnr} in our context. 
We also provide analogous results for the case of point splitting. 
We then show how to rewrite the ingredients in the language of LSH in~\cref{app:lsh_dof}.

\subsection{Classical aspects}\label{app:classical_graphs}

In this paper we will find ourselves multiple times with the following situation.
Consider a set of trajectories $R\subseteq T$ on $\G$.
Then one can define a new graph $\G_R$ as follows.
The vertices of $\G_R$ will be given by the start and endpoints of the trajectories in $R$
\begin{equation}
    \begin{aligned}
        \Gamma_R\coloneq &\Set{p\in \Gamma|\exists\gamma\in R:s(\gamma)=p\qq{or} t(\gamma)=p}\\
        =&\Set{s(\gamma)|\gamma\in R}\cup\Set{t(\gamma)|\gamma\in R}
        \,.
    \end{aligned}
\end{equation}
The half-links of $\G_R$ will be given by two copies of $R$.
This means that we will define two half-links $h_\gamma^{(s)}$ and $h_\gamma^{(t)}\in H_R$ to each trajectory $\gamma\in R$.
If one wants to, it is useful to have the picture in mind that $h_\gamma^{(s)}$ corresponds to the first half of $\gamma$ while $h_\gamma^{(t)}$ to the second half, so that the link $(h_\gamma^{(t)},h_\gamma^{(s)})$ in $\G_R$ is simply the trajectory $\gamma$.
The attachment map is defined by the declaring that each of these pairs is attached to one another, i.e.
\begin{equation}
    \sigma_R(h_\gamma^{(s)})=h_\gamma^{(t)}\qand\sigma_R(h_\gamma^{(t)})=h_\gamma^{(s)}
    \,.
\end{equation}
Finally, the anchoring map $\pi_R$ is given by
\begin{equation}
    \pi_R(h_\gamma^{(s)})=s(\gamma)\qand\pi_R(h_\gamma^{(t)})=t(\gamma)
    \,.
\end{equation}

It is useful to note that $\G_R$ is actually independent of the orientation of the trajectories in $R$.
However, in order to make the notation as simple as possible, we have decided to not make this independence manifest. 

Whenever we find ourselves in this situation, every field configuration on $\G$ naturally induces a field configuration on $\G_R$, according to the intution presented above when defining the half-links.
This defines a map
\begin{equation}
    S:\fields\rightarrow \mathcal{E}(\G_R)
    \,,
\end{equation}
given by identifying the Wilson line on the link $(h_\gamma^{(t)},h_\gamma^{(s)})$ with that on $\gamma$
\begin{equation}
    S(u)(h_\gamma^{(t)},h_\gamma^{(s)})\coloneq u(\gamma)
    \,,
\end{equation}
for all $\gamma\in R$.
Let us spell out some of the properties of this map.

First of all, we may wonder whether $S$ is surjective.
This need not be the case if the trajectories in $R$ are not different enough.
For example, if there is a $\gamma\in R$ such that $\gamma^{-1}\in R$ as well, then the field configurations in the image of $S$ have Wilson lines on $(h_\gamma^{(t)},h_\gamma^{(s)})$ and $(h_{\gamma^{-1}}^{(t)},h_{\gamma^{-1}}^{(s)})$ that are inversely related to one another.
However, in all of the examples that we will be interested in, the set of trajectories $R$ satisfies a property that will guarantee the surjectivity of $S$: 
for each trajectory $\gamma\in R$, there is a link $\gamma(i)$ that is not traversed, in any direction, by any other trajectory in $R$\footnote{This assumption is not true in the $R$ one would choose to recover the results of~\cite{mathur_canonical_2015}. 
Still, one can show that in that case $S$ is surjective and the rest of our proofs carry through.}.
Of course, there may be several links that satisfy this property for each trajectory in $R$.
Let us however choose one of them, defining a map
\begin{equation}
    f:R\rightarrow E
    \,.
\end{equation}
Explicitly, this map has the properties:
\begin{enumerate}
    \item for each $\gamma\in R$ of length $n$, there is an $i\in\{1,\dots, n\}$, such that $\gamma(i)=f(\gamma)$,
    \item for each $\gamma\in R$, there is no $\gamma'\in R$ for which there is an $i\in\{1,\dots,n\}$, with $n$ the length of $\gamma'$, for which $f(\gamma)=\gamma'(i)$ or $f(\gamma)=\gamma'(i)^{-1}$.
\end{enumerate}

We will show that a choice of such an $f$ guarantees the surjectivity of $S$ by constructing from it a right inverse
\begin{equation}
    \iota:\fieldsR\rightarrow \fields
    \,,
\end{equation}
to $S$.
The choice of notation chosen for it is meant to remind the reader that, given that $\iota$ has a left inverse, namely $S$, it must be injective.
Physically, one should thing of $\iota$ as a \emph{choice} of interpretation of each field configuration on $\G_R$ as a field configuration on $G$.
Note that, while $S$ is canonically defined from the choice of trajectories $R$, the embedding $\iota$ is not.

The definition of $\iota$ relies on utilizing each $f(\gamma)$ to store the value of the Wilson line on $\gamma$.
Given that $f(\gamma)$ is not traversed by any other trajectories, there are no conflicts in this procedure.
The rest of the Wilson lines can then be set to the identity.
Explicitly, consider a $u\in \G_R$.
Define
\begin{equation}
    \iota(u)(e)\coloneq \begin{cases}
        u(h_\gamma^{(t)},h_\gamma^{(s)}) & e = f(\gamma)\qc\gamma\in R,\\
        1 & \text{otherwise}
        \,.
    \end{cases}
\end{equation}
The conditions that $f$ satisfy guarantee that $\iota(u)$ is well-defined and that $S(\iota(u))=u$. 

Both $\G$ and $\G_R$ are acted on by their own gauge groups $G^\Gamma$ and $G^{\Gamma_R}$, respectively.
Given that $\Gamma_R\subseteq \Gamma$, there is a surjective map $S:G^\Gamma\rightarrow G^{\Gamma_R}$ between these gauge groups simply given by restricting the gauge transformations on $\G$, so that
\begin{equation}
    S(g)(p)\coloneq g(p)
    \,,
\end{equation}
for all $p\in \Gamma_R$.
We have used the same name for it as for the map $S:\fields\rightarrow\fieldsR$, given that they play similar roles.
Any confusion should be dispelled by looking at the argument of the function.

The map $S$ is equivariant with respect to these group actions
\begin{equation}
    S(g\triangleright u)=S(g)\triangleright S(u)
    \,.
\end{equation}
In order to see this, one simply notes that
\begin{equation}
    \begin{aligned}
        S(g\triangleright u)(h_\gamma^{(t)},h_\gamma^{(s)})
        =&\,(g\triangleright u)(\gamma)=g(t(\gamma))u(\gamma)g(s(\gamma))^{-1}\\
        =&\,g(\pi_R(h_\gamma^{(t)}))S(u)(h_\gamma^{(t)},h_\gamma^{(s)})g(\pi_R(h_\gamma^{(s)}))^{-1}\\
        =&\,(S(g)\triangleright S(u))(h_\gamma^{(t)},h_\gamma^{(s)})
        \,, 
    \end{aligned}
\end{equation}
for all $\gamma\in R$.
This property is a reflection of the naturalness of $S$.
On the other hand, $\iota$, being non-canonical, does not satisfy an analogous property in general.
It is curious to note that in both of the cases that we will consider in this paper, it turns out that $\iota$ is indeed equivariant.
We will however not use this fact and therefore will not dwell on this.

It will be useful from now on to go to a slightly more abstract scenario and consider the generic case where we have two classical systems with configuration spaces given by $X$ and $Y$, on which the groups $G$ and $H$ act respectively.
Suppose we have a surjective map $S:X\rightarrow Y$, which is equivariant with respect to a surjective map $S:G\rightarrow H$.
Let $\iota:Y\rightarrow X$ be a right inverse for $S:X\rightarrow Y$.
Let us show that $[S]:\faktor{X}{G}\rightarrow\faktor{Y}{H}$ is an isomorphism if and only if every $x\in X$ is $G$-equivalent to $\iota(S(x))$.
Note that the surjectivity of $S$ already guarantees that $[S]$ is surjective.
Therefore $[S]$ being an isomorphism is equivalent to $[S]$ being injective.

On the one hand, assume that $[S]$ is injective.
Take $x\in X$.
Then we have that $S(x)=S(\iota(S(x)))$.
Thus, given that $[S]$ is injective, we must have that $x\sim \iota(S(x))$.
On the other hand, now assume that every $x\in X$ is $G$-equivalent to $\iota(S(x))$.
Further, assume that we have $x_1,x_2\in X$ such that $S(x_1)$ and $S(x_2)$ are $H$-equivalent.
To show injectivity of $[S]$, we need to show that $x_1$ and $x_2$ are $G$-equivalent.
The key to see this is to note that, the equivalence of $S(x_1)$ and $S(x_2)$, as well as the surjectiveness of $S:G\rightarrow H$, guarantees there is a $g\in G$ such that
\begin{equation}
    S(x_2)=S(g)\triangleright S(x_1)=S(g\triangleright x_1)
    \,,
\end{equation}
so that, without loss of generality, we can assume that $S(x_2)=S(x_1)$.
In that case, the equivalence of $x_1$ and $x_2$ is clear since
\begin{equation}
    x_1\sim\iota(S(x_1))=\iota(S(x_2))\sim x_2
    \,.
\end{equation}

Going back to our specific case of interest, this means that the theories on $\G$ and $\G_R$ are kinematically equivalent as long as every field $u\in\fields$ is gauge equivalent to $\iota(S(u))$. 

\subsection{Quantum aspects}\label{app:quantum_graphs}

Thinking back to our general scenario, the map $S:X\rightarrow Y$ induces a linear map $S^*:L^2(Y)\rightarrow L^2(X)$ which assigns to every wave function $\ket{\psi}\in L^2(Y)$ a wave function $S^*\ket{\psi}\in L^2(X)$.
As mentioned above, in the cases that we will be interested in, $S$ will always be surjective.
This guarantees that $S^*$ is injective.
Indeed, the only $\ket{\psi}\in L^2(Y)$ such that $S^*\ket{\psi}=0$, is $\ket{\psi}=0$.
To see this, consider $y\in Y$.
Then, since $S$ is surjective, there is an $x\in X$ such that $S(x)=y$.
We conclude that
\begin{equation}
    \ip{y}{\psi}=\ip{S(x)}{\psi}=\mel{x}{S^*}{\psi}=0
    \,.
\end{equation}
Another face of the above is to note that the map $\iota^*:L^2(X)\rightarrow L^2(Y)$ is now a left inverse to $S^*$, for
\begin{equation}
    \mel{y}{\iota^*S^*}{\psi}=\ip{S(\iota(y))}{\psi}=\ip{y}{\psi}
    \,.
\end{equation}

The equivariance of $S$ further guarantees that $S^*$ restricts to a map between gauge invariant wave functions $S^*:L^2(Y)^H\rightarrow L^2(X)^G$, where
\begin{equation}
    \begin{aligned}
        L^2(X)^G\coloneq \{&\ket{\psi}\in L^2(X)|\ip{g\triangleright x}{\psi}=\ip{x}{\psi} \qq{for all} x\in X \qand g\in G\}
        \,.
    \end{aligned}
\end{equation}
This is because if $\ket{\psi}\in L^2(Y)^H$, then for all $x\in X$ and $g\in G$, we have
\begin{equation}
    \begin{aligned}
        \mel{g\triangleright x}{S^*}{\psi}&=\ip{S(g\triangleright x)}{\psi}=\ip{S(g)\triangleright S( x)}{\psi}\\
        &=\ip{S(x)}{\psi}=\mel{x}{S^*}{\psi}
        \,,
    \end{aligned}
\end{equation}
so that indeed $S^*\ket{\psi}\in L^2(X)^G$.

In fact, whenever the induced map
\begin{equation}
    [S]:\faktor{X}{G}\rightarrow\faktor{Y}{H}
\end{equation}
is an isomorphism, we will have that $S^*:L^2(Y)^H\rightarrow L^2(X)^G$ is an isomorphism as well.
Of course, the restriction is injective because the full $S^*$ was.
We thus only need to show that it is surjective.
For this, let us assume we have a gauge invariant $\ket{\psi}\in L^2(X)^G$.
Then we will argue that $\iota^*\ket{\psi}$ is gauge invariant and satisfies $S^*\iota^*\ket{\psi}=\ket{\psi}$.

Tackling gauge invariance first, let $y\in Y$ and $h\in H$.
Consider
\begin{equation}
    \mel{h\triangleright y}{\iota^*}{\psi}=\ip{\iota(h\triangleright y)}{\psi}
    \,.
\end{equation}
By surjectivity of $S$, we have $x\in X$ and $g\in G$ such that $S(x)=y$ and $S(g)=h$.
Then, equivariance of $S$ guarantees that 
\begin{equation}
    \iota(h\triangleright y)=\iota(S(g)\triangleright S(x))=\iota(S(g\triangleright x))
    \,.
\end{equation}
Furthermore, since $[S]$ is an isomorphism, we have
\begin{equation}
    \iota(h\triangleright y)=\iota(S(g\triangleright x))\sim g\triangleright x\sim \iota(S(x))=\iota(y)
    \,.
\end{equation}
Therefore, by gauge invariance of $\ket{\psi}$, we have
\begin{equation}
    \mel{h\triangleright y}{\iota^*}{\psi}=\ip{\iota(y)}{\psi}=\mel{y}{\iota^*}{\psi}
    \,,
\end{equation}
proving that $\iota^*\ket{\psi}$ is gauge invariant.

Now, we need to show that we indeed have $S^*\iota^*\ket{\psi}=\ket{\psi}$.
For this, let $x\in X$.
Then
\begin{equation}
    \mel{x}{S^*\iota^*}{\psi}=\ip{\iota(S(x))}{\psi}
    \,.
\end{equation}
Then, the bijectivity of $[S]$ guarantees that $\iota(S(x))\sim x$, while the gauge invariance of $\ket{\psi}$ then gives
\begin{equation}
    \mel{x}{S^*\iota^*}{\psi}=\ip{x}{\psi}
    \,.
\end{equation}

In other words, the classical equivalence described in the previous section lifts to a quantum mechanical one.

\subsection{Examples}\label{app:examples}

\subsubsection{Virtual point splitting}\label{app:point_splitting}

\begin{figure}
    \centering
    \includegraphics[width = 0.7\textwidth]{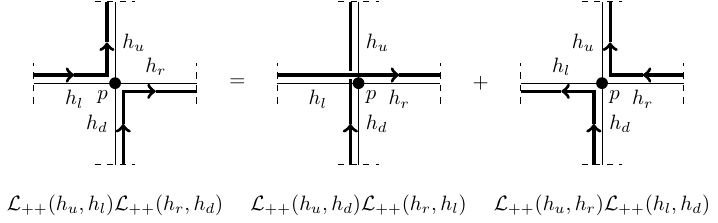}
    \caption{Mandelstam constraint between gauge invariant states obtained from the loop operators \eqref{eq:loop_creation}.}
    \label{fig:mandelstam}
\end{figure}

A general difficulty associated to applying the LSH formulation in more than one dimension is that the gauge invariant states constructed with the loop operators \eqref{eq:loop_creation} are not orthogonal to one another, even after choosing a preferred orientation for each corner. 
So, while they span the whole gauge invariant Hilbert space, they satisfy linear relations known as Mandelstam constraints. 
Indeed, consider a graph $\G$ with a vertex $p\in\Gamma$ of valency $4$, such as the one shown in~\cref{fig:mandelstam}. 
Due to the identity~\cite{Raychowdhury}
\begin{equation}
\label{eq:mandelstam}
    \begin{aligned}
        \epsilon^{AB}\epsilon^{CD}&(=\delta^{AC}\delta^{BD}-\delta^{AD}\delta^{BC}\\
        &=\delta^{AC}\delta^{BD}-\delta^{AB}\delta^{CD}+\delta^{AB}\delta^{CD}-\delta^{AD}\delta^{BC})\\
        &=\epsilon^{AD}\epsilon^{CB}+\epsilon^{AC}\epsilon^{BD}
        \,,  
    \end{aligned}
\end{equation}
we have the linear relation of~\cref{fig:mandelstam}. 
It is clear what the issue is: the configuration with a unit of gauge flux on every half-link of this vertex has three distinct interpretations in terms of the 6 non-trivial corners in a four-valent vertex.

\begin{figure}
    \centering
    \includegraphics[width = 0.6\textwidth]{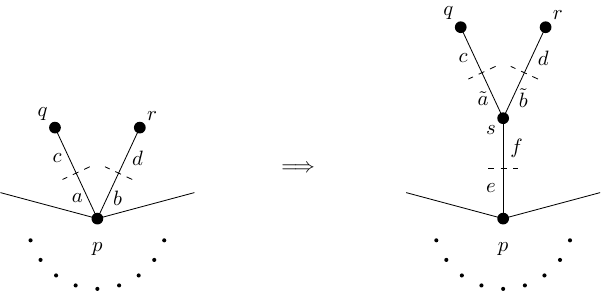}
    \caption{On the left we have a vertex $p$ of valency greater than 3 with half-links $a$ and $b$ attached to it.
    We further have the half-links $c=\sigma(a)$ and $d=\sigma(b)$, attached respectively to vertices $q$ and $r$.
    On the right we show a point splitting of said vertex.
    A new virtual vertex $\tilde{s}$ has been produced, as well as the half-links $e$, $f$, $\tilde{a}$ and $\tilde{b}$.}
    \label{fig:point_splitting_edges}
\end{figure}

This problem of course extends to higher valency vertices. 
If, on the other hand, we had a graph whose vertices had valency less than or equal to three, Mandelstam constraints would not be present, since no analog of~\cref{eq:mandelstam} can be written.
In this case, the loop operators would truly furnish a basis of the gauge invariant Hilbert space. 
Fortunately, there is a generic method that one can employ to lower the valency of a vertex $p\in \G$ while not changing the theory~\cite{Anishetty2018a}. 
The idea is to take two half-links $a$ and $b$ emerging from $p$ and collapse their bases, leaving us with a new virtual unoriented link $\{e,f\}$ and vertex $s$. 
From this new vertex stems two further half-links $\tilde{a}$ and $\tilde{b}$, which are the remainders of the original half-links $a$ and $b$. 
Let us denote the resulting graph by $\tilde{\G}$. 
This construction is shown in~\cref{fig:point_splitting_edges}. 
It is known in the literature as virtual point splitting due to the intuition that the construction corresponds to splitting the point $p$ into $p$ and $s$, which is different from the interpretation we presented above. 
In any case, the two graphs $\G$ and $\tilde{\G}$ differ in two ways. 
On the one hand, the latter has the extra vertex $s$. 
On the other, the half-links $a$ and $b$ of the former get replaced by the half-links $\tilde{a}$, $\tilde{b}$, $f$, and $e$ in the latter. 
The vertices and half-links in common between the two graphs will be identified along with the structures supported in them.

\begin{figure}
    \centering
    \includegraphics[width = 0.6\textwidth]{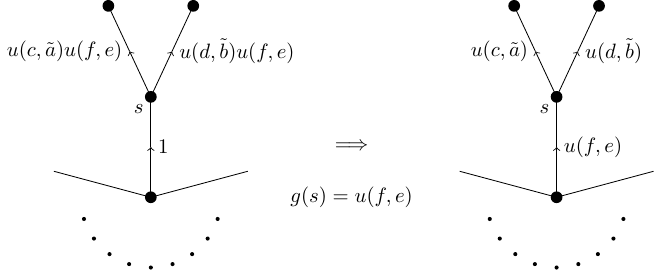}
    \caption{On the right we have a field configuration $u\in\mathcal{E}(\tilde{\G})$.
    On the left we have $\iota(S(u))$.
    The two are equivalent by applying a gauge transformation $u(f,e)$ at $s$.}
    \label{fig:point_splitting_gauge_fixing}
\end{figure}

We can prove that the physics on the graphs $\G$ and $\tilde{\G}$ are equivalent using the techniques we developed in the previous sections.
Namely, consider the paths 
\begin{equation}
    R=\{((c,\tilde{a}),(f,e)),((d,\tilde{b}),(f,e))\}
    \,.
\end{equation}
Then $G=\tilde{G}_R$.
The corresponding map $S$ is surjective, for each trajectory in $R$ has a corresponding link (either $(c,\tilde{a})$ or $(d,\tilde{b})$) for which the other trajectory doesn't go through.
This gives us the map $\iota$ obtained by setting
\begin{equation}
    \begin{aligned}
        &\iota(u)(f,e)=1\qc\iota(u)(c,\tilde{a})=u((c,\tilde{a}),(f,e))\qc\iota(u)(d,\tilde{b})=u((d,\tilde{b}),(f,e))
        \,,
    \end{aligned}
\end{equation}
for all $u\in\mathcal{E}(\tilde{\G})$.
Thus, all that we are left with checking is that $u\sim\iota(S(u))$, for all $u\in\mathcal{E}(\tilde{\G})$.
This is obtained by noting that $u=g\triangleright\iota(S(u))$, where $g$ is equal to the identity for every vertex different than $s$ and is equal to $u(f,e)$ at $s$.
This is shown in~\cref{fig:point_splitting_gauge_fixing}.

When applying point splitting we are usually in a setting where we understand the dynamics on $G=\tilde{\G}_R$.
Accordingly, the Wilson line operators can be immediately translated to operators on $\tilde{\G}$.
Similarly, the electric operators $E(c)$ and $E(d)$ can be immediately replaced by their counterparts in $\tilde{\G}$.
The counterparts to the operators $E(a)$ and $E(b)$ are however more delicate.
This issue is not usually addressed in the literature since $E(a)^2=E(c)^2$ on $\G$ and thus either can be used to build the Hamiltonian on $\tilde{G}$.
But in our applications we will need to have explicit control over the operator $E(a)$ in order to build the Hamiltonian. The way this is done will be discussed in~\cref{app:ks_dof}.

Repeated application of this procedure yields graphs that only have trivalent vertices.
With this we can now normalize the states appearing in \eqref{eq:basis_unnormalized} to recover the basis discussed in~\cite{Raychowdhury:2019iki}.
Given that these states are tensor products of states localized at each site, we need only normalize each of these.
Let us the consider a vertex to which three half-edges $a$, $b$, and $c$ are attached.
This situation is shown in~\cref{fig:trivalent_vertex}.
Let us pick as a set of oriented corners $\mathcal{C}=\{(a,b),(b,c),(c,a)\}$.
We then need to compute the vacuum expectation value 
\begin{equation}
    \begin{aligned}
        &C(n(a,b),n(b,c),n(c,a))\\
        &\coloneq \bra{0}\mathcal{L}_{--}^{n(a,b)}(a,b)\mathcal{L}_{--}^{n(b,c)}(b,c)\mathcal{L}_{--}^{n(c,a)}(c,a)\mathcal{L}_{++}^{n(a,b)}(a,b)\mathcal{L}_{++}^{n(b,c)}(b,c)\mathcal{L}_{++}^{n(c,a)}(c,a)\ket{0}\,.
    \end{aligned}
\end{equation}

In order to compute this expectation value, let us first study the state
\begin{equation}
    \mathcal{L}_{--}(a,b)\mathcal{L}_{++}^{n(a,b)}(a,b)\mathcal{L}_{++}^{n(b,c)}(b,c)\mathcal{L}_{++}^{n(c,a)}(c,a)\ket{0}\,.
\end{equation}
Since the LSH vacuum is annihilated by all annihilation operators, we can replace the operator above by a commutator. 
Then applying the Leibniz rule of derivations, we see that said commutator is the sum of two commutators.
Accordingly, our state of interest is the sum of the states
\begin{equation}\label{eq:former_vector}
    [\mathcal{L}_{--}(a,b),\mathcal{L}_{++}^{n(a,b)}(a,b)]\mathcal{L}_{++}^{n(b,c)}(b,c)\mathcal{L}_{++}^{n(c,a)}(c,a)\ket{0}\,,
\end{equation}
and
\begin{equation}\label{eq:latter_vector}
    \mathcal{L}_{++}^{n(a,b)}(a,b)[\mathcal{L}_{--}(a,b),\mathcal{L}_{++}^{n(b,c)}(b,c)\mathcal{L}_{++}^{n(c,a)}(c,a)]\ket{0}\,.
\end{equation}
We will now show that the latter of these two vanishes.

Let us recall the commutator \eqref{eq:creation_annihilation}.
With this and the Leibniz rule we conclude that
\begin{equation}
    \begin{aligned}
        &[\mathcal{L}_{--}(a,b),\mathcal{L}_{++}^{n}(b,c)]=\sum_{k=1}^{n}\mathcal{L}_{++}^{k-1}(b,c)\mathcal{L}_{-+}(a,c)\mathcal{L}_{++}^{n-k}(b,c)\,.
    \end{aligned}
\end{equation}
But in light of \eqref{eq:half_overlapping}, all of these terms are actually the same and we obtain the simpler result
\begin{equation}
    [\mathcal{L}_{--}(a,b),\mathcal{L}_{++}^{n}(b,c)]=n\mathcal{L}_{++}^{n-1}(b,c)\mathcal{L}_{-+}(a,c)\,.
\end{equation}
This operator in particular annihilates the vacuum. 
Along with further applications of the Leibniz rule and \eqref{eq:mixed_commutators}, one concludes that \eqref{eq:latter_vector} vanishes.

Let us now study the contribution from \eqref{eq:former_vector}. 
The Leibniz rule in conjunction with \eqref{eq:creation_annihilation} and \eqref{eq:N_charge} shows that
\begin{equation}\label{eq:annihilation_of_excitation}
    \begin{aligned}
        \qty[\mathcal{L}_{--}(a,b),\mathcal{L}_{++}^{n}(a,b)]=&\sum_{k=1}^n\mathcal{L}_{++}^{k-1}(a,b)(N(a)+N(b)+2)\mathcal{L}_{++}^{n-k}(a,b)\\=
        &\sum_{k=1}^n\mathcal{L}_{++}^{n-1}(a,b)(N(a)+N(b)+2+2(n-k))\\=
        &\, n\mathcal{L}_{++}^{n-1}(a,b)(N(a)+N(b)+2+n-1)\,.\\
    \end{aligned}
\end{equation}
Therefore
\begin{equation}
\begin{aligned}
    &[\mathcal{L}_{--}(a,b),\mathcal{L}_{++}^{n(a,b)}(a,b)]\mathcal{L}_{++}^{n(b,c)}(b,c)\mathcal{L}_{++}^{n(c,a)}(c,a)\\
    &=n(a,b)\mathcal{L}_{++}^{n(a,b)-1}(a,b)\mathcal{L}_{++}^{n(b,c)}(b,c)\mathcal{L}_{++}^{n(c,a)}(c,a)(N(a)+N(b)+2+n(a,b)-1+n(b,c)+n(c,a))\,.
\end{aligned}
\end{equation}
With this result we can now conclude that
\begin{equation}
    \begin{aligned}
        C(n(a,b),n(b,c),n(c,a))=n(a,b)\left(n(a,b)+n(b,c)+n(c,a)+1\right)C(n(a,b)-1,n(b,c),n(c,a))\,.
    \end{aligned}
\end{equation}

This is the fundamental recursion relation from which $C$ can be recovered. Repeated applications of it yield
\begin{equation}
    \begin{aligned}
        C(n(a,b),n(b,c),n(c,a))=n(a,b)!\frac{(n(a,b)+n(b,c)+n(c,a)+1)!}{(n(b,c)+n(c,a)+1)!}C(0,n(b,c),n(c,a))\,.
    \end{aligned}
\end{equation}
Given that $C$ is symmetric under permutations of its entries, and $C(0,0,0)=1$, we conclude that
\begin{equation}
    \begin{aligned}
        &C(n(a,b),n(b,c),n(c,a))=\\
        &n(a,b)!\frac{(n(a,b)+n(b,c)+n(c,a)+1)!}{(n(b,c)+n(c,a)+1)!}n(b,c)!\frac{(n(b,c)+n(c,a)+1)!}{(n(c,a)+1)!}n(c,a)!(n(c,a)+1)!\\
        &=n(a,b)!n(b,c)!n(c,a)!(n(a,b)+n(b,c)+n(c,a)+1)!\,. 
    \end{aligned}
\end{equation}
Therefore the states
    \begin{equation}\label{eq:orthonormal_basis}
        \ket{n}\coloneq \frac{\mathcal{L}_{++}^{n(a,b)}(a,b)\mathcal{L}_{++}^{n(b,c)}(b,c)\mathcal{L}_{++}^{n(c,a)}(c,a)}{\sqrt{C(n(a,b),n(b,c),n(c,a))}}\ket{0}\,,
    \end{equation}
give an orthonormal basis.

As in~\cite{Raychowdhury:2019iki}, we note that the action of the LSH operators on these states is rather simple. 
Indeed, they always amount to a change in $n$ and a change in normalization. Observe that
\begin{equation}
\begin{aligned}
    \frac{C(n(a,b)+1,n(b,c),n(c,a))}{C(n(a,b),n(b,c),n(c,a))}=\left(n(a,b)+1)(n(a,b)+n(b,c)+n(c,a)+2\right)\,.
\end{aligned}
\end{equation}
Therefore, the introduction of the operators
\begin{equation}
\begin{aligned}
    N(a,b)=\frac{1}{2}(N(a)+N(b)-N(c))\qc N(a,b,c)=N(a,b)+N(b,c)+N(c,a)+1\,,
\end{aligned}    
\end{equation}
allows us to express
\begin{equation}
    \mathcal{L}_{++}(a,b)=\Lambda_+(a,b)\sqrt{(N(a,b,c)+1)(N(a,b)+1)}
\end{equation}
In here $\Lambda_{+}(a,b)$ is the operator that raises and lowers the corresponding quantum number without changing the normalization of the state.
We can further use \eqref{eq:annihilation_of_excitation} in conjunction with the fact that
\begin{equation}
\begin{aligned}
    \frac{C(n(a,b)-1,n(b,c),n(c,a))}{C(n(a,b),n(b,c),n(c,a))}=\frac{1}{n(a,b)(n(a,b)+n(b,c)+n(c,a)+1)}\,,
\end{aligned}
\end{equation}
to show that
\begin{equation}
    \mathcal{L}_{--}(a,b)=\Lambda_{-}(a,b)\sqrt{N(a,b)N(a,b,c)}.
\end{equation}
Finally, we can use \eqref{eq:creation_annihilation} to show that
\begin{equation}
    \mathcal{L}_{-+}(a,b)=\Lambda_-(c,a)\Lambda_+(b,c)\sqrt{N(c,a)(N(b,c)+1)}\,.
\end{equation}
and therefore
\begin{equation}
    \mathcal{L}_{+-}(a,b)=-\Lambda_-(b,c)\Lambda_+(c,a)\sqrt{N(b,c)(N(c,a)+1)}\,.
\end{equation}

\subsubsection{Maximal tree gauge fixing}\label{app:maximal_tree}

In this section we discuss the idea of maximal tree gauge fixing, which identifies a set trajectories in a general graph whose Wilson lines can be set to the identity by choosing appropriate gauge transformations at the vertices. 

A tree is a connected graph which has no loops. 
Now, consider a graph $\mathrm{G}$ that is connected and for which every pair of vertices is connected by a single link at most, i.e. for all $e_1,e_2\in H\times_\sigma H$ if $s(e_1)=s(e_2)$ and $t(e_1)=t(e_2)$, we then have that $e_1=e_2$. 
We will further demand that there are no closed trajectories of length 1. 
A maximal tree on such a graph $\G$ is a tree that is a subgraph of $\G$ and is maximal, in the sense that the inclusion of any additional link of $\G$ would produce a loop. 
In the mathematical literature these are more often called spanning trees, for they are equivalently described as trees that have all of the vertices $\G$. 
Let us denote the set of half-links of the maximal tree by $M\subseteq H$. 
An example is shown in~\cref{fig:maximal_tree}.

Now, let us choose a vertex $o\in\G$ and an orientation $E_M$ for all the links that are not in the maximal tree.
For each $e\in E_M$ define a path $\gamma_e$ that starts and ends at $o$ as follows.
Let $\gamma_{\rightarrow}$ be the only path that goes from $o$ to $s(e)$ along the maximal tree without backtracking (if it wasn't unique then composing it with the inverse of another one would give a loop along the maximal tree).
Similarly, let $\gamma_{\leftarrow}$ be the corresponding path going from $t(e)$ to $o$.
Then define $\gamma_e = \gamma_\leftarrow\cdot e\cdot\gamma_\rightarrow$.
Let $R$ denote the set of these trajectories.
The resulting $\G_R$ looks like a flower of loops starting and ending at $o$.
As we will now show, the physics on $\G$ and $\G_R$ are equivalent.

The map $S$ is clearly surjective, for $\gamma_e$ is the only trajectory in $R$ going through $e$.
Given a field configuration $u\in\fields$, we have
\begin{equation}
    \iota(S(u))(e)=\begin{cases}
        1 & e\in M\times_\sigma M\\
        u(\gamma_e) & e\in H\times_\sigma H\setminus M\times_\sigma M
        \,.
    \end{cases}
\end{equation}
We are left with showing that $u\sim\iota(S(u))$.
This is the result of the maximal tree gauge-fixing construction we will describe next.

First let $q$ be a vertex in $\G$ which is only attached to one link in the maximal tree.
Next, go along the path $\gamma$ defined by the maximal tree from $o$ to $q$.
At each step $i$, perform the gauge transformation $u(\gamma(i))^{-1}$ at $t(\gamma(i))$.
The resulting gauge transformation leaves the identity at $\gamma(i)$ and doesn't disrupt this gauge fixing condition for the previous $\gamma(j)$.
Furthermore, note that after this step the Wilson line at $\gamma(i+1)$ is equal to
\begin{equation}
    u(\gamma(i+1))u(\gamma(i))u(\gamma(i-1))\cdots u(\gamma(1))
    \,.    
\end{equation}
Repeating this procedure for all the possible $q$'s shows this gauge fixing procedure.
Every link $e$ will pick up a factor of $u(\gamma_\rightarrow)$ on the right for the path that goes through $s(e)$ and a factor of $u(\gamma_\leftarrow)$ on the left for the one going through $t(e)$.

\subsection{Dynamics}\label{app:dynamics}

\subsubsection{Kogut-Susskind degrees of freedom}\label{app:ks_dof}

So far we have concluded that, as long as every trajectory in $R$ has a link that is not traversed by any other trajectory in any direction and every field configuration $u\in\fields$ is gauge equivalent to $\iota(S(u))$, then the theories on $\G$ and $\G_R$ are equivalent at the kinematic level.
In order to make this equivalence useful however, we need to explain how to translate the Hamiltonian of one graph into a Hamiltonian for the other
\begin{equation}
    HS^*=S^*H_R
    \,.
\end{equation}
In the case of point splitting we will need to translate the dynamical information in $\G_R$ to a Hamiltonian on $\G$.
In the case of the maximal tree construction we will need to do the opposite.
To the best of our knowledge, there is not a unique construction that works for any coarsening.
We will therefore only answer it in these two cases.
Some general lessons will however be explained as we go along.

Due to the way we have set up the equivalence, it is simplest to understand the relationship between the Wilson line operators on $\G$ and on $\G_R$.
Namely, we have
\begin{equation}
\begin{aligned}
    \hphantom{{}={}}\mel{u}{S^*U(h_\gamma^{(t)},h_\gamma^{(s)})}{\psi}
    &=\mel{S(u)}{U(h_\gamma^{(t)},h_\gamma^{(s)})}{\psi}\\
    &=S(u)(h_\gamma^{(t)},h_\gamma^{(s)})\ip{S(u)}{\psi}\\
    &=u(\gamma)\mel{u}{S^*}{\psi}=\mel{u}{U(\gamma)S^*}{\psi}
    \,,
\end{aligned}
\end{equation}
i.e.
\begin{equation}\label{eq:magnetic_hamiltonian_point_splitting}
    S^*U(h_\gamma^{(t)},h_\gamma^{(s)})=U(\gamma)S^*
    \,.
\end{equation}
This allows us to translate a magnetic Hamiltonian from $\G_R$ to $\G$ immediately.
One simply replaces the Wilson line operators on a link in $\G_R$ by the Wilson line operator supported on its corresponding path in $\G$.
We have thus solved the problem of the magnetic Hamiltonian under point splitting.
It is however not yet obvious that this solves the problem of the magnetic Hamiltonian under the maximal tree construction.
In order to do this we need to show that in that case the Hamiltonian can be built solely from Wilson lines supported on the trajectories in the corresponding $R$.

In order to do this, we note that every plaquette $\gamma\in\mathcal{P}$ in $\G$, a cubic lattice, has a non-empty set $E_\gamma\subseteq E_M$ of links that it goes through that are outside of the maximal tree.
For every $e\in E_\gamma$ let $\sigma_e$ be $1$ if $\gamma$ goes along it with the orientation chosen when constructing the flower or $-1$ if it goes against it.
We then have
\begin{equation}\label{eq:magnetic_hamiltonian_maximal_tree}
    \tr U(\gamma)=\tr U\qty(\prod_{e\in E_\gamma} \gamma_e^{\sigma_e})\,,
\end{equation}
where the product is oriented using the ordering in which $\gamma$ travels through the links.
The operators on the right hand side are all supported on paths in $R$.
They can therefore be replaced by operators on the corresponding links of $\G_R$.
This recovers the solution found in~\cite{DAndrea:2023qnr} to the problem of relating the magnetic Hamiltonian of a cubic lattice to that of its corresponding flower.

Let us go back to the general case once more.
Relating the electric operators between $\G$ and $\G_R$ requires us to relate the group actions on the half-links of $\G$ with those of $\G_R$.
Here the concept of parallel transport arises naturally as follows.
Fix a half-link $h\in H$ and a trajectory $\gamma\in T_n$ (which could potentially be in $R$).
We can then wonder how $\triangleright_h$ behaves on $\gamma$.
Of course $(g\triangleright_h u)(\gamma)=u(\gamma)$ if $\gamma$ does not go through $h$.
On the other hand, if $\gamma$ passes once through $(h,\sigma(h))$ we have
\begin{equation}
    (g\triangleright_h u)(\gamma)=u(\gamma_2)gu(h,\sigma(h))u(\gamma_1)=u(\gamma_2)gu(\gamma_2)^{-1}u(\gamma)
    \,,
\end{equation}
where we have written $\gamma=\gamma_2\cdot(h,\sigma(h))\cdot\gamma_1$.
This immediately implies that
\begin{equation}
    (u(\gamma_2)^{-1}gu(\gamma_2)\triangleright_h u)(\gamma)=gu(\gamma)
    \,.
\end{equation}
This prompts us to the following definition. 

Let $h\in H$ and $\gamma\in T_n$ be such that $\pi(h)=t(\gamma)$.
We then define $\circ_{\gamma,h}:G\times \fields\rightarrow\fields$ by 
\begin{equation}
    g\circ_{\gamma,h} u=u(\gamma)gu(\gamma)^{-1}\triangleright_h u
    \,.
\end{equation}
With this definition we have
\begin{equation}
    (g\circ_{\gamma_2^{-1},h} u)(\gamma)=gu(\gamma)
\end{equation}
in our previous setting.
In other words, if $\gamma\in R$,
\begin{equation}\label{eq:equivariance_maximal_tree}
    S(g\circ_{\gamma_2^{-1},h} u)(h_\gamma^{(t)},h_\gamma^{(s)})=g\triangleright_{h_\gamma^{(t)}}S(u)(h_\gamma^{(t)},h_\gamma^{(s)})
    \,.
\end{equation}
Now, assume that the path $\gamma_2$ is common to all trajectories in $R$ which go through $(h,\sigma(h))$
Then the generators of $\circ_{\gamma_2^{-1},h}$ and $\triangleright_{h_\gamma^{(t)}}$ are related in the quantum theory.
So, let us take some time to explore the general properties of $\circ_{\gamma,h}$.

First of all, the choice of notation reflects that $\circ_{\gamma,h}$ need not be a left group action.
Of particular importance to virtual point splitting and the maximal tree construction, is the case in which $\gamma$ and $h$ are disjoint.
In this case we are guaranteed that $\circ_{\gamma,h}$ is, indeed, a left group action.
We can then associate to it an Electric operator.
This mimics the construction of the Electric operators on half-links.
Namely, one first lifts the group action to the quantum level
\begin{equation}
    \begin{aligned}
        R_{\gamma,h}(g)\ket{u}\coloneq &\ket{g\circ_{\gamma,h} u}=\ket{u(\gamma)gu(\gamma)^{-1}\triangleright_h u}\\
        =&R_{h}(u(\gamma)gu(\gamma)^{-1})\ket{u}
        \,,
    \end{aligned}
\end{equation}
and then defines the electric operator to be its generator
\begin{equation}
    e^{-i\omega E_X(\gamma, h)}\coloneq R_{\gamma, h}(e^{\omega X})\qc X\in\mathfrak{g}
    \,.
\end{equation}

The relationship between the parallel transported operators $E(\gamma,h)$ and the operators $E(h)$ is clear by noting that
\begin{equation}
    \begin{aligned}
        e^{-i\omega E_X(\gamma,h)}\ket{u}&=R_h(u(\gamma)e^{\omega X}u(\gamma)^{-1})\ket{u}\\
        &=R_h(e^{\omega \Ad_{u(\gamma)}X})\ket{u}\\
        &=e^{-i\omega E_{\Ad_{u(\gamma)}(X)}(h)}\ket{u}
        \,,
    \end{aligned}
\end{equation}
where $\Ad$ denotes the adjoint representation of the structure group $G$ on its Lie algebra.
In other words
\begin{equation}
    E_X(\gamma,h)\ket{u}=E_{\Ad_{u(\gamma)}(X)}(h)\ket{u}
    \,.
\end{equation}
On a basis of $\mathfrak{g}$, this becomes
\begin{equation}
    E_a(\gamma,h)\ket{u}=\Ad\indices{_{u(\gamma)}^b_a}E_b(X)(h)\ket{u}
    \,.
\end{equation}
Multiplying by $X^a$, the dual basis of $\mathfrak{g}$ under the Killing form, we can further rewrite this in matrix form
\begin{equation}
    E(\gamma,h)\ket{u}=u(\gamma)^{-1}E(h)u(\gamma)\ket{u}
    \,.
\end{equation}
In here we used the fact that
\begin{equation}
    \begin{aligned}
        X^a\Ad\indices{_{u(\gamma)}^b_a}&=X^a\ev{X^b,\Ad_{u(\gamma)}(X_a)}\\
        &=X^a\ev{\Ad_{u(\gamma)^{-1}}X^b,X_a}\\
        &=\Ad_{u(\gamma)^{-1}}(X^b)\\
        &=u(\gamma)^{-1}X^bu(\gamma)
        \,,
    \end{aligned}
\end{equation}
due to the $\Ad$-invariance of the Killing form.
Finally, this can be lifted to an operator equation since $\gamma$ and $h$ do not overlap
\begin{equation}\label{eq:parallel_transport_operator}
    E(\gamma,h) = U(\gamma)^{-1}E(h)U(\gamma)
    \,.
\end{equation}
It is simple to remember the placement of the $U$'s, which guarantee that $E(\gamma,h)$ transform under the adjoint representation of the structure group at $s(\gamma)$.

Not only does this expression provide an explicit construction of the parallel transport in terms of other operators that are simpler to implement, but it also guarantees that $E(\gamma,h)^2=E(h)^2$ due to the $Ad$-invariance of the Killing form
\begin{equation}
    K_{ab}=K_{cd}\Ad\indices{_g^c_a}\Ad\indices{_g^d_b}
    \,.
\end{equation}
This is also a good place to comment that, another case of general interest is when $\gamma$ touches $h$ only once at the very end $\gamma(n)_2=h$.
This is precisely the case of interest when trying to relate $E(h)$ and $E(\sigma(h))$.
As it turns out, in that case instead of a left group action, we will have a right group action.
Thus, these two operators will be related to each other by parallel transport up to a sign.
In this case, since we do have an overlap, we do not have an analogue of \eqref{eq:parallel_transport_operator}. Interestingly however, it is still true that parallel transport preserves the quadratic Casimir. 

Now, we are ready to consider our specific examples.
First, let us treat the simpler example of virtual point splitting.
We have that
\begin{equation}
    S(g\triangleright_h u)=g\triangleright_h S(u)
\end{equation}
for the half-links $h$ in ~\cref{fig:point_splitting_edges} which retained their name after point splitting (meaning $c$, $d$ and the ones that are outside of the scope of the picture).
This implies that
\begin{equation}
    S^*E(e)=E(e)S^*
    \,,
\end{equation}
for those links.
This leaves us with the operator $E(a)$ (and, of course, the operator $E(b)$ to which the same procedure can be applied) in order to translate the electric information from the original graph $\tilde{G}_R$ to the one after point splitting $\tilde{G}$.
So, we start by computing
\begin{equation}
    g\triangleright_a S(u)=S(g\circ_{(f,e),\tilde{a}} u)
    \,.
\end{equation}
This is checked by evaluating both sides on every link of $\tilde{G}_R$.
In particular, the group action on both sides is trivial on all links except for $(c,a)$, where the equality can be checked explicitely.
We then have
\begin{equation}\label{eq:parallel_transport_electric}
    S^*E(a)=E((f,e),\tilde{a})S^*
    \,,
\end{equation}
establishing the equivalence of the electric content in the case of point splitting.

Second, let us focus on the maximal tree construction.
For every vertex $p\in \Gamma$, let $\gamma_p\in T$ be the path starting at the marked vertex $o\in V$ and finishing at $p$ along the maximal tree. 
Using this path we will be able to understand $E(h)^2$ for all of the half-edges $h\in \pi^{-1}(p)$ anchored at $p$ except for the ones that $\gamma$ goes through. 
Fortunately, for every link there is at least one half-link $h$ for which $\gamma_{\pi(h)}$ does not go through $h$ (namely, for each link on the maximal tree there is exactly one, and for each link outside of the maximal tree, both suffice). 
Let us then choose such a half-link for every link and define the set of these to be $H_{\text{dyn}}$. 
We will refer to these as the set of ``dynamical'' half-links.
The sum in \eqref{eq:electric_hamiltonian} can be performed over this set.

Now, given $h\in H_\text{dyn}$, we can consider $E(\gamma_{\pi(h)},h)$, which satisfies
\begin{equation}
    \vb{E}(h)^2=\vb{E}(\gamma_{\pi(h)},h)^2.
\end{equation}
The latter however satisfies much simpler properties on $G_R$. Indeed, given $\gamma\in R$, there are only four possibilities:
\begin{enumerate}
    \item $\gamma$ and $h$ do not overlap.
    \item $\gamma$ goes along $(h,\sigma(h))$ once and doesn't go through $(\sigma(h),h)$.
    \item $\gamma$ goes along $(\sigma(h),h)$ once and doesn't go through $(h,\sigma(h))$, and
    \item $\gamma$ goes both along $(h,\sigma(h))$ and $(\sigma(h),h)$, each once.
\end{enumerate}
Even more specially, because of the uniqueness of paths along maximal trees, all $\gamma\in R$ that go through $(\sigma(h),h)$ share the same trajectory $\gamma_{\pi(h)}$ leading up to it.
Let us denote the set of such paths by $R_{h,\leftarrow}$.
Similarly, all $\gamma\in R$ that go through $(h,\sigma(h))$ share the same trajectory after going through it, namely, $\gamma_{\pi(h)}^{-1}$.
Let us denote the set of such paths by $R_{h,\rightarrow}$.
These remarks lead us to consider
\begin{equation}
    \begin{aligned}
        S(g\circ_{\gamma_{\pi(h)}, h}u)(h_\gamma^{(t)}h_\gamma^{(s)})=\begin{cases}
            S(u)(h_\gamma^{(s)}h_\gamma^{(t)}) & \text{case 1.} \\
            (g\triangleright_{h_\gamma^{(t)}} S(u))(h_\gamma^{(t)}h_\gamma^{(s)}) & \text{case 2.}\\
            (g\triangleright_{h_\gamma^{(s)}} S(u))(h_\gamma^{(t)}h_\gamma^{(s)}) & \text{case 3.}\\
            (g\triangleright_{h_\gamma^{(t)}}\triangleright g\triangleright_{h_\gamma^{(s)}} S(u))(h_\gamma^{(t)}h_\gamma^{(s)}) & \text{case 4.}
            \,,
        \end{cases}
    \end{aligned}
\end{equation}
by comparison with \eqref{eq:equivariance_maximal_tree}.
In other words, the composition of $S$ after acting with $\circ_{\gamma_h,\gamma}$ is the same as the composition of $g\triangleright_{h_\gamma^{(t)}}$ for all $\gamma\in R_{h,\rightarrow}$ and $g\triangleright_{h_\gamma^{(s)}}$ for all $\gamma\in R_{h,\leftarrow}$.
Therefore, we have
\begin{equation}
    E(\gamma_h,h)S^* = S^*\qty(\sum_{\gamma\in R_{h,\leftarrow}}E(h_\gamma^{(s)})+\sum_{\gamma\in R_{h,\leftarrow}}E(h_\gamma^{(t)}))
    \,.
\end{equation}
In particular, in the case of the maximal tree construction, this gives us a way of translating the quadratic Casimir appearing in the Hamiltonian of a cubic lattice
\begin{equation}\label{eq:electric_hamiltonian_maximal_tree}
    \begin{aligned}
        \vb{E}(h)^2S^*&=\vb{E}(\gamma_h,h)^2S^*\\
        &=S^*\qty(\sum_{\gamma\in R_{h,\leftarrow}}\vb{E}(h_\gamma^{(s)})+\sum_{\gamma\in R_{h,\rightarrow}}\vb{E}(h_\gamma^{(t)}))^2
        \,.
    \end{aligned}
\end{equation}
This shows how to translate the electric Hamiltonian from that lattice to the flower. In this way we then recover the result shown in~\cite{mathur_canonical_2015, DAndrea:2023qnr}.

\subsubsection{LSH degrees of freedom}\label{app:lsh_dof}

In light of equation \eqref{eq:magnetic_hamiltonian_maximal_tree}, which allows for the transition of the magnetic Hamiltonian from the spatial lattice to the flower, and equation \eqref{eq:magnetic_hamiltonian_point_splitting}, which allows for the transition of the magnetic Hamiltonian from flowers to branches, the magnetic Hamiltonian in our formulation will correspond to Wilson loops on branches. The description of these in the LSH formalism has already been discussed in \eqref{eq:loops_lsh}. 

Let us then focus on the electric part. The transition from the spatial lattice to flower, requires \eqref{eq:electric_hamiltonian_maximal_tree}. 
Therefore, we need to understand the inner product between two electric operators.
Furthermore, the point splitting process required to transition from flowers to branches, requires us to understand this inner product for electric operators that have been parallel transported back to a common vertex \eqref{eq:parallel_transport_electric}.

\begin{figure}
    \centering
    \includegraphics[width=0.6\linewidth]{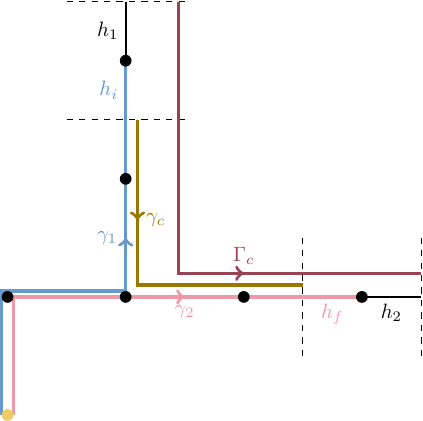}
    \caption{Trajectories $\gamma_1$ and $\gamma_2$ are depicted in red and blue respectively. They have a common source at the orange vertex $s(\gamma_1)=s(\gamma_2)$. A corner trajectory $\gamma_c$ is also shown in purple. The Wilson line operator on the trajectory $\gamma_2\gamma_1^{-1}$ is equivalent to a the Wilson line operator on a trajectory $\gamma$ that first goes through $h_i$, then through $\gamma_c$, and finally through $h_f$. Another corner trajectory $\Gamma_c$ is shown in brown. This trajectory is obtained by first going through $(h_1,h_i)$, then through $\gamma_c$, and finally through $(h_2,h_f)$. This can equivalently be described as first going through $h_1$, then through $\gamma$, and finally through $h_2$.}
    \label{fig:electric_inner_product}
\end{figure}

Consider then the inner product
\begin{equation}
    \vb{E}(\gamma_1,h_1)\cdot \vb{E}(\gamma_2,h_2)\,
\end{equation}
with $\gamma_1$ and $\gamma_2$ two trajectories that start at the same vertex $s(\gamma_1)=s(\gamma_2)$ and end at $t(\gamma_1)=\pi(h_1)$ and $t(\gamma_2)=\pi(h_2)$. 
We will further assume that the two trajectories do not go throught $h_1$ or $h_2$, so that, as operators, the Wilson line and electric operators appearing in this quantity commute.
This geometry is depicted in~\cref{fig:electric_inner_product}.
We can therefore manipulate this quantity as one would any other Lie algebra manipulation.
In particular, the invariance of the inner product chosen guarantees that one can shift the parallel transport on $h_1$ all the way to $h_2$.
Thus, defining $\gamma=\gamma_2\gamma_1^{-1}$, this inner product can be rewritten as
\begin{equation}\label{eq:parallel_transport_simplified}
    \vb{E}(h_1)\cdot\vb{E}(\gamma,h_2).
\end{equation}
Observe that $\gamma$ then implements the parallel transport of $h_2$ to $h_1$.
In applications it is useful to note that this operator only depends on $U(\gamma)$. 
Therefore, if the trajectory $\gamma$ backtracks the corresponding Wilson operators will cancel each other.
By eliminating these we can replace the trajectory $\gamma$ by a simpler one.
This is the trajectory mentioned in the caption of~\cref{fig:electric_inner_product}.

On $\SU$ with our choice of inner product, we have the orthonormal basis $i\sigma_i/2$ constructed out of Pauli matrices. 
In this basis, we can extract the coefficients of $X=X_ii\sigma_i/2\in\su$ by
\begin{equation}
    X_i=-\tr(i\sigma_i X)\,.
\end{equation}
Then
\begin{equation}
    \begin{aligned}
        E(h_1)_i=E_{i\sigma_i/2}(h_1)=-\frac{1}{2}a^\dagger(h_1)\sigma_i a(h_1)=-\frac{1}{2}A_{A+}(h_1)A\indices{^B_-}(h_1)\sigma\indices{_i^A_B}\,,
    \end{aligned}
\end{equation}
and 
\begin{equation}
    \begin{aligned}
        E(\gamma,h_2)_i=-\frac{1}{2}A_{A+}(h_2)A\indices{^B_-}(h_2)\,\sigma\indices{_j^A_B}\qty(-\tr(i\sigma_i U(\gamma)^{-1}i\sigma_j/2U(\gamma)))\,.
    \end{aligned}
\end{equation}
The trace can explicitly be written in components as 
\begin{equation}
    U\indices{^B_C}(\gamma^{-1})U\indices{^D_A}(\gamma)\sigma\indices{_i^A_B}\sigma\indices{_j^C_D}\,.
\end{equation}
Thus the inner product is the contraction of the tensor
\begin{equation}
    A_{A+}(h_1)A\indices{^B_-}(h_1)A_{C+}(h_2)A\indices{^D_-}(h_2)U\indices{^E_F}(\gamma^{-1})U\indices{^G_H}(\gamma)\,,
\end{equation}
with
\begin{equation}
\label{eq:sigmaContraction}
    \frac{1}{8}\sigma\indices{_i^A_B}\sigma\indices{_j^C_D}\sigma\indices{_i^H_E}\sigma\indices{_j^F_G}
    = 
    \frac{1}{2}\delta^A_E\delta^H_B\delta^C_G\delta^F_D
-\frac{1}{4} \delta^A_B\delta^H_E\delta^C_G\delta^F_D
-\frac{1}{4} \delta^A_E\delta^H_B\delta^C_D\delta^F_G
+ \frac{1}{8}\delta^A_B\delta^H_E\delta^C_D\delta^F_G\,,
\end{equation}
where the expression in terms of Kronecker deltas can be obtained from the Fierz identity of the Pauli matrices. 

For example, the fact that $t(\gamma)=s(\gamma^{-1})=\pi(h_2)$ implies that the index $G$ can only be contracted with the $C$ or the $F$ index. 
This suggests a reordering of the product of operators into
\begin{equation}\label{eq:first_operator}
    A_{A+}(h_1)U\indices{^E_F}(\gamma^{-1})A\indices{^D_-}(h_2)A_{C+}(h_2)U\indices{^G_H}(\gamma)A\indices{^B_-}(h_1)\,.
\end{equation}
In doing so we however switched the order of the $h_2$ factors, so that we have to include the additional operator
\begin{equation}\label{eq:second_operator}
    -\delta^D_CA_{A+}(h_1)U\indices{^E_F}(\gamma^{-1})U\indices{^G_H}(\gamma)A\indices{^B_-}(h_1)\,.
\end{equation}
This additional operator is however simpler to study, since gauge invariance demands that $G$ and $F$ be contracted.
But then the Wilson lines are multiplied and lead to a $\delta^E_F$.
Therefore the four possible contractions must be proportional to $-N(h_1)$.

Let us now make these contractions explicit.
The term coming from \eqref{eq:second_operator} is
\begin{equation}
    -\qty(\frac{1}{2}-\frac{2}{4}-\frac{2}{4}+\frac{4}{8})N(h_1)=0\,,
\end{equation}
where the 4 terms in the bracket correspond to the 4 possible combinations of Kronecker deltas in~\cref{eq:sigmaContraction}, such that this term vanishes.
Now we can perform the contractions of the non-zero contribution \eqref{eq:first_operator}. 
The last three combinations of Kronecker deltas include contractions of the Wilson lines. 
Thus, they are $\gamma$ independent. 
Therefore gauge invariance demands that they are functions of $N(h_1)$ and $N(h_2)$. 
In fact, because of the order that we wrote the operators in, these contractions are proportional to $N(h_1)(N(h_2)+2)$.
Performing these contractions explicitly, we have
\begin{equation}
\begin{aligned}
    \qty(-\frac{1}{4}-\frac{1}{4}+\frac{2}{8})N(h_1)(N(h_2)+2)=-\frac{1}{4}N(h_1)(N(h_2)+2)\,.
\end{aligned}
\end{equation}
We are then left with the first combination of Kronecker deltas, which gives
\begin{equation}\label{eq:final_contraction}
    \frac{1}{2}A_{A+}(h_1)U\indices{^A_D}(\gamma^{-1})A\indices{^D_-}(h_2)A_{C+}(h_2)U\indices{^C_B}(\gamma)A\indices{^B_-}(h_1)\,.
\end{equation}

This is a product of two terms that can be written in the LSH formulation. To see this, note that
\begin{equation}
    \begin{aligned}
        A_{A+}(h_2)U\indices{^A_B}(\gamma)A\indices{^B_-}(h_1)=A_{A+}(h_2)U\indices{^A_a}(h_f)t\indices{^a_b}(\gamma(n))U\indices{^b_c}(\gamma_c)t\indices{^c_d}(\gamma(1))U\indices{^{\dagger d}_B}(h_i)A\indices{^B_-}(h_1)\,.
    \end{aligned}
\end{equation}
Using~\eqref{eq:N_charge} one can write
\begin{equation}
    U\indices{^B_-}(h)=\frac{1}{\sqrt{N(h)+1}}A\indices{^B_-}(h)=A\indices{^B_-}(h)\frac{1}{\sqrt{N(h)}}\,.
\end{equation}
Thus, we can set
\begin{align}
    A_{A+}(h_2)U\indices{^A_a}(h_f) &= \sqrt{N(h_2)}U\indices{^-_a}(h_2,h_f)\nonumber\\
    U\indices{^{\dagger d}_B}(h_i)A\indices{^B_-}(h_1)& =U\indices{^d_-}(h_i,h_1)\sqrt{N(h_1)}
    \,.
\end{align}
We therefore find
\begin{equation}
    \begin{aligned}
        A_{A+}(h_2)U\indices{^A_B}(\gamma)A\indices{^B_-}(h_1)&=\sqrt{N(h_2)}U\indices{^-_a}(h_2,h_f)t\indices{^a_b}(\gamma(n))U\indices{^b_c}(\gamma_c)t\indices{^c_d}(\gamma(1))U\indices{^d_-}(h_i,h_1)\sqrt{N(h_1)}\\
        &=\sqrt{N(h_2)}U\indices{^-_-}(\Gamma_c)\sqrt{N(h_1)}\,,
    \end{aligned}
\end{equation}
where $\Gamma_c$ is the corner trajectory defined by first going through $h_1$, then through $\gamma$, and finally through $h_2$, as shown in~\cref{fig:electric_inner_product}.
Putting these results together, we conclude that
\begin{equation}
    \begin{aligned}
        \vb{E}(\gamma_1,h_1) \cdot\vb{E}(\gamma_2,h_2)=\frac{1}{2}\sqrt{N(h_1)}U\indices{^-_-}(\Gamma_c^{-1})N(h_2)U\indices{^-_-}(\Gamma_c)\sqrt{N(h_1)}-\frac{N(h_1)}{2}\qty(\frac{N(h_2)}{2}+1)\,.
    \end{aligned}
\end{equation}
One can further see that while $U(\Gamma_c)$ creates a Schwinger boson on $h_2$ and annihilates one on $h_1$, $U(\Gamma_c)$ annihilates one on $h_2$ and creates one on $h_1$.
Therefore, this can be simplified to \eqref{eq:electric_inner_product}
\begin{equation}
    \begin{aligned}
        \vb{E}(\gamma_1,h_1)\cdot\vb{E}(\gamma_2,h_2)=U\indices{^-_-}(\Gamma_c^{-1})U\indices{^-_-}(\Gamma_c)\frac{N(h_1)}{2}(N(h_2)+1)-\frac{N(h_1)}{2}\qty(\frac{N(h_2)}{2}+1)
        \,.
    \end{aligned}
\end{equation}
In this way we have succeeded at describing the electric Hamiltonian solely in terms of LSH operators as well.

\section{Derivation of the Wave Function}
\label{app:wavefunction}

As shown in~\cref{eq:LSHStates}, the basis states of a leaf (or a loop in maximal tree gauge fixing) are given by 
\begin{align}                   
    \ket{n,l,m}\coloneq \ell_{++}^n\mathscr{L}_{++}^l(a_1^\dagger(d))^{l+m}(a_2^\dagger(d))^{l-m}\ket{0}
    \,,
\end{align}
where $n$ counts the internal chromoelectric flux while $l$ the external.

To derive the corresponding wave functions, we first note that we can use the binomial theorem in order to expand the product of the loop operators in terms of Schwinger Bosons
\begin{equation}
\begin{aligned}
    \ell^n_{++}\mathscr{L}^l_{++}&=\sum_{k=0}^n\sum_{p,q=0}^l(-1)^{n+l+k+p+q}\binom{n}{k}\binom{l}{p}\binom{l}{q}\, a^\dagger_1(a)^{k+p}a^\dagger_2(a)^{n+l-k-p}a^\dagger_1(b)^{n+l-k-q} a^\dagger_2(b)^{k+q}\\
    &\hphantom{{}={}}\times a^\dagger_1(c)^{l+q-p}a^\dagger_2(c)^{l+p-q}
    \,.
\end{aligned}
\end{equation}
Next, we note that, other than the combinatorial factors, the dependence on $p$ and $q$ is only through the combinations $k+p$ and $k+q$.
We can thus shift these variables so as to isolate the $k$-dependence
\begin{equation}
    \begin{aligned}
        \ell^n_{++}\mathscr{L}^l_{++}&=\sum_{p,q}\qty[\sum_k(-1)^k\binom{n}{k}\binom{l}{p-k}\binom{l}{q-k}](-1)^{n+l+p+q}\times\\
        &\hphantom{{}={}}a^\dagger_1(a)^{p}a^\dagger_2(a)^{n+l-p}a^\dagger_1(b)^{n+l-q} a^\dagger_2(b)^{q}a^\dagger_1(c)^{l+q-p}a^\dagger_2(c)^{l+p-q}
        \,.
    \end{aligned}
\end{equation}
When switching the order of summation, the limits become complicated and we have therefore left them implicit in this formula.
They are however completely determined by the requirement that no factorials of negative numbers are taken within the combinatorial factors.

At this stage we recognize that the term in parenthesis looks extremely similar to a Clebsch-Gordan coefficient (e.g. section 8.2.4 in~\cite{varshalovich_quantum_1988}). Introducing these coefficients then allows us to eliminate the sum over $k$
\begin{equation}
    \begin{aligned}
        \ell^n_{++}\mathscr{L}^l_{++}&= \sum_{p,q}(-1)^{n+l+p+q}\binom{n+l}{n}^{-1}\sqrt{\binom{n+2l+1}{n}\binom{n+l}{p}\binom{n+l}{n+l-q}\binom{2l}{2l+p-q}}\\
        &\hphantom{{}={}}\times\ip{\frac{n+l}{2},\frac{n+l}{2}-p;\frac{n+l}{2},q-\frac{n+l}{2}}{l,q-p}a^\dagger_1(a)^{p}a^\dagger_2(a)^{n+l-p}
        a^\dagger_1(b)^{n+l-q} a^\dagger_2(b)^{q}
        \\
        &\hphantom{{}={}}\times a^\dagger_1(c)^{l+q-p}a^\dagger_2(c)^{l+p-q}
        \,.
    \end{aligned}
\end{equation}
The remaining binomial coefficients that appear when doing this greatly simplify against the normalization factors in \eqref{eq:sb_electric}, so that when acting on the vacuum state, this operator creates
\begin{equation}
    \begin{aligned}
        \ell^n_{++}\mathscr{L}^l_{++}\ket{0}&=\sum_{p,q}(-1)^{n+l+p+q}n!l!\sqrt{\binom{n+2l+1}{n}\binom{2l}{2l+p-q}}\\
        &\hphantom{{}={}}\times\ip{\frac{n+l}{2},\frac{n+l}{2}-p;\frac{n+l}{2},q-\frac{n+l}{2}}{l,q-p}\\
        &\hphantom{{}={}}\times\Ket{\frac{n+l}{2},p-\frac{n+l}{2}}_a\Ket{\frac{n+l}{2},\frac{n+l}{2}-q}_b
        a^\dagger_1(c)^{l+q-p}a^\dagger_2(c)^{l+p-q}
        \,.
    \end{aligned}
\end{equation}

With this expression in hand, we are now in a position to calculate the wave functions. 
With our gauge fixing condition, the Wilson line $u_1$ on the $(a,b)$ link is oriented along the $z$-axis.
When evaluating the state above on such a Wilson line via \eqref{eq:electric_to_magnetic}, each term in the sum will be proportional to the matrix element
\begin{equation}
    \mel{\frac{n+l}{2},\frac{n+l}{2}-p}{r_{\frac{n+l}{2}}(u_1)}{\frac{n+l}{2},\frac{n+l}{2}-q}
    \,,
\end{equation}
which vanishes unless $p=q$ in this gauge.
Let us thus focus on this subsector of the sum.
In it the binomial coefficients can be factored out of the sum, which in turn becomes a resolution of the identity on the space obtained by coupling two spin $(n+l)/2$ representations.
Thus, up to proportionality constants, we obtain that
\begin{equation}
    \ell^n_{++}\mathscr{L}^l_{++}\ket{0}\propto\ket{l,0}_n a^\dagger_1(c)^{l+q-p}a^\dagger_2(c)^{l+p-q}+\cdots
    \,.
\end{equation}

The state $\ket{l,0}_n$ is the total spin $l$ state with $0$ component in the $z$ direction in this coupled space
\begin{equation}
    \begin{aligned}
        \ket{l,0}_n&=\sum_p\Ket{\frac{n+l}{2},p-\frac{n+l}{2}}_a\Ket{\frac{n+l}{2},\frac{n+l}{2}-p}_b\ip{\frac{n+l}{2},\frac{n+l}{2}-p;\frac{n+l}{2},p-\frac{n+l}{2}}{l,0}
        \,.
    \end{aligned}
\end{equation}
This evaluates to
\begin{equation}
    \begin{aligned}
        \ip{e^{i\omega J_z}}{l,0}_n&\propto\sqrt{n+l+1}\sum_{m=-\frac{n+l}{2}}^{\frac{n+l}{2}}e^{i(\omega-\pi) m}\ip{\frac{n+l}{2},m;\frac{n+l}{2},-m}{l,0}
        \,,
    \end{aligned}
\end{equation}
up to an overall sign.
Due to the symmetry properties of the Clebsch-Gordan coefficients, we can study this function through the purely real function defined by
\begin{equation}
    \begin{aligned}
        \phi_{n,l}(\omega)&\coloneq  i^n\sqrt{n+l+1}\sum_{m=-\frac{n+l}{2}}^{\frac{n+l}{2}}e^{i(\omega-\pi) m}\ip{\frac{n+l}{2},m;\frac{n+l}{2},-m}{l,0}
        \,,
    \end{aligned}
\end{equation}
to which it is proportional.

For this we need need to now compute the remaining wave function on the stem
\begin{equation}
    \begin{aligned}
        \mel{u_2}{(a_1^\dagger(c)a_2^\dagger(c))^la_1^\dagger(d)^{l+m}a_2^\dagger(d)^{l-m}}{0}\propto \ip{u_2}{l,0;l,m}\propto\mel{l,-m}{r_l(u_2)}{l,0}
        \,,
    \end{aligned}
\end{equation}
Fortunately, this is simple to express in terms of the Euler angles\footnote{We will use the $zyz$ convention for Euler angles.} $(\alpha,\theta,\phi)$ of $u_2$ (see e.g. (37) in~\cite{varshalovich_quantum_1988})
\begin{equation}
    \mel{l,0}{r_l(u_2)}{l,m}\propto Y^l_m(\theta,\phi)
    \,.
\end{equation}
However, recall that $u_2$ was any rotation that mapped the axis $\hat{n}$ of $u$ to the $z$-axis.
If the spherical coordinates of $\hat{n}$ are $(\theta,\phi)$, then the rotations that do this are those that have Euler angles $(\alpha,\theta,\phi)$, with $\alpha$ parametrizing this $\mathrm{U}(1)$ worth of such rotations.
This then finishes the proof of \eqref{eq:petal_function}.

\end{document}